%% file: main.tex
\input{topmatter}

\input{aux/macros}

\input{titlepage}

\allowdisplaybreaks[2]
\begin{document}
\maketitle

\eject

\input{sections/sec_1_intro}
\input{sections/sec_2_simplest}
\input{sections/sec_3_nonanomalous}
\input{sections/sec_4_anomalous_VOA}
\input{sections/sec_5_higgs_branches}
\input{sections/sec_6_discussion}


\acknowledgments

The authors would first and foremost like to thank Mario Martone, who initiated this study in the first instance with the first author. They would also like to thank Tomoyuki Arakawa, Niklas Garner, Julius Grimminger, and Palash Singh for helpful conversations and suggestions related to this work. CB is supported in part by ERC Consolidator Grant \#864828 ``Algebraic Foundations of Supersymmetric Quantum Field Theory'' (SCFTAlg), and by the STFC consolidated grant ST/X000761/1. The work of HK is supported by the Clarendon Fund Scholarship and the Eddie Dinshaw Scholarship at Balliol College.


\vfill\eject

\appendix
\input{sections/appendix_a}
\input{sections/appendix_b}
\input{sections/appendix_c}
\input{sections/appendix_d}


\bibliographystyle{aux/JHEP}
\bibliography{aux/biblio}
\end{document}

%% file: topmatter.tex
\documentclass[a4paper,dvipsnames,11pt]{article}
\RequirePackage{pdf14}
\RequirePackage{fix-cm}

\usepackage{./aux/jheppub}
\usepackage{amsthm}
\usepackage{graphicx}
\usepackage{verbatim}
\usepackage{amsmath,amssymb,amsfonts,amsthm,amscd}

\newtheorem*{theorem*}{Theorem}
\newtheorem*{remark}{Remark}
\newtheorem{conjecture}{Conjecture}
\newtheorem{prop}{Proposition}

\usepackage{mathtools}
\usepackage{bm}
\usepackage{bbm}
\usepackage{mathrsfs} 
\usepackage{lscape} 
\usepackage{cancel}
\usepackage{enumitem}
\usepackage{tcolorbox}
\usepackage{multirow} 
\usepackage{colortbl} 
\usepackage{array,hhline,arydshln,multirow}
\usepackage{floatrow}

\usepackage[export]{adjustbox}
\usepackage[normalem]{ulem}
\usepackage{rotating} 
\usepackage[parfill]{parskip}

\usepackage{pifont,dsfont}
\usepackage{array,setspace,mathrsfs,yfonts,dsfont,amscd,euscript}
\usepackage{relsize,suffix,mathtools,cancel,bm,bbm}

\usepackage{graphicx,tikz}
\usepackage[framemethod=TikZ]{mdframed} 
\usepackage{tikz-cd} 

\usepackage{xcolor}

\definecolor{amaranth}{rgb}{0.9, 0.17, 0.31}
\definecolor{coolblack}{rgb}{0.0, 0.18, 0.39}
\definecolor{gold(web)(golden)}{rgb}{1.0, 0.84, 0.0}
\definecolor{deepcarmine}{rgb}{0.66, 0.13, 0.24}
\definecolor{mutedred}{RGB}{240,45,45}

%% file: aux/macros.tex
\def\nn{\nonumber}

\def\bM{\begin{matrix}}
\def\eM{\end{matrix}}
\newcommand{\bpm}{\begin{pmatrix}}
\newcommand{\epm}{\end{pmatrix}}
\newcommand{\bsm}{\begin{smallmatrix}}
\newcommand{\esm}{\end{smallmatrix}}
\newcommand{\bspm}{\left(\begin{smallmatrix}}
\newcommand{\espm}{\end{smallmatrix}\right)}
\newcommand{\beq}{\begin{equation}}
\newcommand{\eeq}{\end{equation}}

\def\hat{\widehat}

\def\^{\wedge}

\def\Tr{{\rm Tr}}

\def\so{\mathfrak{so}}

\def\sl{\mathfrak{sl}}

\def\su{\mathfrak{su}}

\def\sp{\mathfrak{sp}}

\def\cC{{\mathcal C}}

\def\cM{{\mathcal M}}

\def\cN{{\mathcal N}}

\def\cS{{\mathcal S}}

\def\cT{{\mathcal T}}

\def\cW{{\mathcal W}}

\def\Z{\mathbbm{Z}}
\def\a{{\alpha}}

\def\b{{\beta}}
\def\g{{\gamma}}

\def\d{{\delta}}
\def\D{{\Delta}}

\def\s{{\sigma}}

\def\w{{\omega}}
\def\Om{{\Omega}}
\def\ie{\emph{i.e.}}

\def\cf{\emph{cf.}}

\usepackage{tikz}
\usetikzlibrary{arrows,shapes.arrows,decorations.markings}
     \tikzset{>=triangle 90}
     \tikzstyle{bbc}=[draw,circle,fill=black,scale=.75]
     \tikzstyle{rc}=[circle,fill=red,scale=.6]
     \tikzstyle{wc}=[draw,circle,scale=.75]

\def\hat{\widehat}

\def\dim{{\rm dim}}

\def\Tr{{\rm Tr}}

\def\a{{\alpha}}

\def\b{{\beta}}
\def\g{{\gamma}}

\def\d{{\delta}}

\def\D{{\Delta}}

\def\x{{\xi}}

\def\s{{\sigma}}

\def\w{{\omega}}

\def\bh{{\boldsymbol h}}

\def\af{\mathfrak{a}}
\def\bff{\mathfrak{b}}
\def\cf{\mathfrak{c}}

\def\df{\mathfrak{d}}

\def\ff{\mathfrak{f}}
\def\gf{\mathfrak{g}}

\def\hf{\mathfrak{h}}

\def\jf{\mathfrak{j}}

\def\spf{\mathfrak{sp}}

\def\cC{{\mathcal C}}

\def\cD{{\mathcal D}}

\def\cI{{\mathcal I}}

\def\cM{{\mathcal M}}

\def\cN{{\mathcal N}}

\def\cS{{\mathcal S}}

\def\cT{{\mathcal T}}

\def\cW{{\mathcal W}}

\def\E{\mathbb{E}}

\def\Z{\mathbb{Z}} 

\def\beq{\begin{equation}}
\def\eeq{\end{equation}}
\def\nn{\nonumber}
\newcommand{\bpmat}{\begin{pmatrix}}
\newcommand{\epmat}{\end{pmatrix}}
\newcommand{\bsmat}{\begin{smallmatrix}}
\newcommand{\esmat}{\end{smallmatrix}}
\newfloatcommand{capbtabbox}{table}[][\FBwidth]

\def\ADtwo{{\rm AD}(\cf_2)}

%% file: titlepage.tex
\title{Twisted \texorpdfstring{$A_{2N}$}{A(2N)} Argyres--Douglas theories: vertex algebras, Higgs branches, and double covers}

\author[a,b]{Christopher Beem}
\author[a]{and Harshal Kulkarni}

\affiliation[a]{Mathematical Institute, University of Oxford\\Andrew Wiles Building, Woodstock Road, Oxford, OX2 6GG, United Kingdom}
\affiliation[b]{School of Natural Sciences, Institute for Advanced Study\\Einstein Drive, Princeton, NJ 08540, United States of America}
\abstract{
We investigate the associated vertex operator algebras and Higgs branches of two closely related families of generalised Argyres--Douglas theories, arising from the twisted $A_{2N}$ and twisted $D_{N+1}$ series, respectively. We propose that the $A$-series theories are realised as nilpotent Higgsings of the $D$-series, and we assemble a variety of pieces of evidence for this proposal. A consequence is that the associated vertex operator algebras of the $A$-series are finite extensions of affine Kac--Moody vertex algebras at (non-boundary) admissible levels, rather than those Kac--Moody algebras themselves, and that their Higgs branches are in many cases double covers of nilpotent orbit closures. The $\mathbb{Z}_2$ gaugings of the $A$-series theories then furnish examples of unitary SCFTs whose associated vertex operator algebras are precisely affine Kac--Moody algebras at non-boundary admissible levels. We propose that these vertex algebras are equipped with a non-standard $\mathfrak{R}$-filtration, and comment on compatibility with graded unitarity.
}

%% file: sections/sec_1_intro.tex

\section{\label{sec:intro}Introduction}

Generalised Argyres--Douglas (AD) theories \cite{Argyres:1995jj,Argyres:1995xn,Xie:2012hs} constitute a vast and by now reasonably well-mapped landscape of four-dimensional $\cN=2$ superconformal field theories (SCFTs). While these theories are intrinsically strongly coupled, a significant amount of their (supersymmetry-protected) physical data remains accessible. Their Coulomb branch spectra, central charges, and flavour symmetries can be read off algorithmically from the Hitchin systems that define them; furthermore, their associated vertex operator algebras (VOAs) \cite{Beem:2013sza} are in many cases conjectured to be essentially as simple as possible: affine Kac--Moody VOAs at \emph{admissible} levels or quantum Drinfel'd--Sokolov reductions thereof \cite{Xie:2016evu,Song:2017oew,Xie:2019yds}.

However, various finer aspects of these theories remain to be understood better. In many instances, the precise identity of the associated VOA remains unknown or conjectural---often based on knowledge of its central charge and the levels of its affine current subalgebras---and the geometry of the Higgs branch of vacua (expected to match the \emph{associated variety} of the VOA \cite{Beem:2017ooy}) is not always explicit. Moreover, the various connections between theories by Higgs branch renormalisation group flows (mirrored by Drinfel'd--Sokolov reductions at the level of the VOA) are not completely mapped out. The aim of this paper is to supply further detail in some of these areas for two specific (infinite) families of generalised Argyres--Douglas theories.

The families under consideration are associated to irregular Hitchin systems of types $A_{2N}$ and $D_{N+1}$ with a full regular puncture, an irregular puncture, and a $\mathbb{Z}_2$ automorphism twist line \cite{Wang:2018gvb}. Both carry a $\cf_N$ flavour symmetry, though this symmetry is subject to a nontrivial global Witten anomaly in the $A_{2N}$ case \cite{Witten:1982fp, Tachikawa:2018rgw}. While the associated VOAs were conjectured in \cite{Wang:2018gvb} to be exactly the affine Kac--Moody VOAs $V_k(\sp(2N))$ (largely motivated by the Sugawara relation holding for the Virasoro central charges), we argue that this picture is incomplete. Specifically, while the identification holds for the non-anomalous family, we propose that the VOA of the anomalous family is instead a \emph{finite extension} of $V_{k}(\sp(2N))$ by a single module of conformal weight $h=\tfrac12(3+2m)$ in the fundamental representation. This statement follows from a new isomorphism that we propose between the anomalous theories and certain nilpotent Higgsings of parent non-anomalous theories.

These families also provide a natural laboratory for exploring the appearance of admissible-level affine Kac--Moody VOAs in unitary SCFTs. The levels arising in the non-anomalous family are principal \emph{boundary admissible}, whereas those of the anomalous family are \emph{coprincipal admissible} but fail to be boundary levels. Given that boundary admissibility is a recurring feature of affine Kac--Moody VOAs arising from unitary four-dimensional theories \cite{ArabiArdehali:2025fad,Beem:2026lkq}, the anomalous theories---or more precisely them and their $\mathbb{Z}_2$ gaugings---furnish an interesting set of examples. In particular, the graded unitarity analysis of \cite{ArabiArdehali:2025fad} suggests that the $\mathfrak{R}$-filtration on $V_k(\sp(2N))$ in these cases cannot be the ``standard'' filtration introduced \emph{loc. cit.}, and indeed we find that the filtration inherited from the extended VOA differs as required.

To fix notation, we will follow a variant of the conventions of \cite{Wang:2018gvb}. We denote the $\cf_N$ flavour symmetries of the twisted $A_{2N}$ and twisted $D_{N+1}$ theories by $\cf^{\rm anom}_N$ and $\cf_N$, respectively, to distinguish between the presence or absence of the global Witten anomaly. Throughout, the irregular puncture in the defining Hitchin system is assumed to be of regular semisimple type, and we restrict ourselves to the ``class I'' theories of \cite{Wang:2018gvb} (a brief review is provided in Appendix \ref{app:irregular_puncture_review}). These theories are characterised by the absence of exactly marginal couplings and the fact that the irregular puncture introduces no additional mass parameters/flavour symmetries. The irregular singularities for the Higgs fields in the Hitchin systems are given by,\footnote{Here we parametrise the power of $z$ appearing in these irregular singularities differently from \cite{Wang:2018gvb}. We use $m$, where a parameter denoted by $k$ is used \emph{loc. cit.} and in Appendix \ref{app:irregular_puncture_review}. The two conventions are related by $k = -4N - 2 + 3 + 2m$ and $k = -2N - 2 + 3 + 2m$ for the $A_{2N}$ and $D_{N+1}$ cases, respectively.}
\begin{equation}\label{Hitchinparam}
\begin{alignedat}{3}
    &{\rm twisted}~A_{2N}:\qquad &&\Phi=\frac{T}{z^{1+\frac {3+2m}{4N+2}}}+\dots~, \qquad &&\gcd(3+2m,2+4N)=1~, \\
    &{\rm twisted}~D_{N+1}:\qquad &&\Phi=\frac{T}{z^{1+\frac {3+2m}{2N+2}}}+\dots~, \qquad &&\gcd(3+2m,2+2N)=1~, \\
\end{alignedat}
\end{equation}
with $m \in \mathbb{Z}_{\geqslant -1}$. We denote these theories by $\mathrm{AD}(\cf^{\rm anom}_{N},m)$ and $\mathrm{AD}(\cf_{N},m)$, respectively.

A number of general properties of these families were determined in \cite{Wang:2018gvb}, some of which we summarise here. The flavour central charges $k_{4d}$ of the $\cf_N$ flavour symmetries are related to the levels $k$ of the associated affine Kac--Moody currents by $k_{4d}=-2k$, where
\begin{equation}\label{ADlevels}
\begin{alignedat}{3}
    &\mathrm{AD}(\cf^{\rm anom}_{N},m)\,&:&\qquad k=k^{\rm anom}_{N,m}=-h^\vee_{\cf_N}+\frac{h_{\cf_N}+1}{2(3+2m)}~, \qquad &&m \in \mathbb{Z}_{\geqslant -1}~, \\
    &\mathrm{AD}(\cf_{N},m)&:&\qquad k=k_{N,m}=-h^\vee_{\cf_{N}}+\frac{h^\vee_{\cf_N}}{3+2m}~, \qquad &&m \in \mathbb{Z}_{\geqslant -1}~.
\end{alignedat}
\end{equation}
The $c_{4d}$ Weyl anomaly coefficients are given by
\begin{equation} \label{ccent}
\begin{alignedat}{3}
    &\mathrm{AD}(\cf^{\rm anom}_{N},m)\,&:&\qquad c_{4d} = \tfrac{1}{12}N(4N+4Nm+4m+5)~, \\
    &\mathrm{AD}(\cf_{N},m) &:&\qquad c_{4d} = \tfrac{1}{6}N(2N+1)(m+1)~.
\end{alignedat}    
\end{equation}
The Virasoro central charge of the associated VOA is $c=-12c_{4d}$ as usual. In all cases, the levels and central charges in \eqref{ADlevels} and \eqref{ccent} are consistent with the Sugawara relation,
\begin{equation}
    c = \frac{k\,\dim\cf_N}{k+h^\vee_{\cf_N}}~,
\end{equation}
where $\dim\cf_N=N(2N+1)$, $h^\vee_{\cf_N}=N+1$, and $h_{\cf_N}=2N$.

The $a_{4d}$ anomaly coefficient can be determined from $c_{4d}$ and the Coulomb branch operator dimensions via the Shapere--Tachikawa relation \cite{Shapere:2008zf}. The Coulomb branch spectrum was determined in \cite{Wang:2018gvb}, leading to the following anomaly coefficients,
\begin{equation} \label{acent}
\begin{alignedat}{2}
    &\mathrm{AD}(\cf^{\rm anom}_{N},m)\,&:&\qquad 24a_{4d} = \frac{N (22N + 16Nm^2 + 38Nm + 16m^2 + 43m + 28)}{3+2m}~, \\
    &\mathrm{AD}(\cf_{N},m) &:&\qquad 24a_{4d} = \frac{N(2N+1)(m+1)(8m+11)}{3+2m}~.
\end{alignedat}    
\end{equation}
It is precisely this data and the satisfaction of the Sugawara relation that leads naturally to the conjecture that the associated VOAs are the affine Kac--Moody VOAs $V_{k^{\rm anom}_{N,m}}(\sp(2N))$ and $V_{k_{N,m}}(\sp(2N))$. In the language of Appendix \ref{app:KM_admissible}, the levels $k_{N,m}$ are principal boundary admissible, while $k^{\rm anom}_{N,m}$ are coprincipal admissible but not boundary levels.

\subsection{Summary of results}

In the remainder of this paper, we argue that the identification of \cite{Wang:2018gvb} is correct for the non-anomalous theories, but that the anomalous $\mathrm{AD}(\cf^{\rm anom}_{N},m)$ theories give rise to VOAs which are finite extensions of the coprincipal admissible affine Kac--Moody VOAs $V_{k^{\rm anom}_{N,m}}(\sp(2N))$. Concretely, we propose that $\mathrm{AD}(\cf^{\rm anom}_{N},m)$ is obtained by Higgsing the non-anomalous $\mathrm{AD}(\cf_{N+m+1},m)$ theory with respect to the nilpotent orbit $\mathds{O}_{[2+2m,1^{2N}]}$ of $\sp(2(N+m+1))$, so that its VOA is obtained from $V_{k_{N+m+1,m}}(\sp(2(N+m+1)))$ by the corresponding quantum Drinfel'd--Sokolov reduction. The characterisation of this VOA as a finite extension then follows from known results on these reductions \cite{Arakawa:2021ogm}. Our proposal satisfies checks at the level of anomalies, central charges, and the dimension of the Higgs branch. Furthermore, for the $m=0$ and $m=1$ series, we bootstrap closed-form expressions for the OPEs of the strong generators of these VOAs for general $N$. On requiring compatibility with graded unitarity, the bootstrap in particular picks out a unique value for the level $k$, which matches the flavour level $k^{\rm anom}_{N,m}$.

As a point of reference, we observe that for $m=-1$, the levels \eqref{ADlevels} reduce to
\begin{equation}
k^{\rm anom}_{N,-1} = -\frac{1}{2}~, \qquad k_{N,-1} = 0~,
\end{equation}
which corresponds to the theory of $N$ free hypermultiplets and to the trivial theory, respectively. For $m \in \mathbb{Z}_{\geqslant 0}$ the theories are interacting. Recall that the VOA of a free hypermultiplet is the symplectic boson VOA \cite{Beem:2013sza}, and that $N$ copies of this VOA---denoted $\mathrm{Sb}^{\otimes N}$---form a finite extension of $V_{-\frac{1}{2}}(\sp(2N))$ \cite{feingold1985classical}. More precisely, it was shown \emph{loc. cit.} that
\begin{equation} \label{eq:freehyperfiniteext}
    \mathrm{Sb}^{\otimes N} \cong V_{-\frac{1}{2}}(\sp(2N)) \oplus L_{-\frac{1}{2}}(\sp(2N); \varpi_1)~,
\end{equation}
where $L_{-\frac{1}{2}}(\sp(2N); \varpi_1)$ is a simple $V_{-\frac{1}{2}}(\sp(2N))$-module whose conformal highest weight space is in the fundamental representation of the horizontal $\sp(2N)$ algebra. It follows that a $\mathbb{Z}_2$ gauging of $N$ free hypermultiplets has exactly the affine Kac--Moody VOA $V_{-\frac{1}{2}}(\sp(2N))$ as its associated VOA. However, this is not an example of the general Higgsing proposal since the non-anomalous $\mathrm{AD}(\cf_{N},-1)$ theory is trivial.

We find that this pattern generalises to the entire $\mathrm{AD}(\cf^{\rm anom}_{N},m)$ family for $m\geqslant0$: in each case the VOA is an extension of $V_{k^{\rm anom}_{N,m}}(\sp(2N))$ by a single module of conformal weight $h=\tfrac{1}{2}(3+2m)$ with highest weights in the fundamental of $\sp(2N)$. There is then a $\mathbb{Z}_2$ gauging of $\mathrm{AD}(\cf^{\rm anom}_{N},m)$ that gives the associated VOA $V_{k^{\rm anom}_{N,m}}(\sp(2N))$ itself.

Along the way, we also identify a larger set of alternative realisations of the anomalous theories as nilpotent Higgsings. By uncovering certain new $\cW$-algebra isomorphisms at admissible levels, we propose that the ``class I'' twisted $A_{\rm even}$, twisted $A_{\rm odd}$, and twisted $D$-type theories can all be realised as nilpotent Higgsings of parent ``class I'' \emph{untwisted} $D$-type theories. A similar statement relates a twisted $E_6$ theory to the non-anomalous $\mathrm{AD}(\cf_{1},2)$ theory. Such equivalences add some cohesion to the landscape of generalised Argyres--Douglas theories with non-simply laced flavour symmetries, furthering the kind of simplification advocated for in \cite{Beem:2023ofp}.

We further perform a detailed study of the Higgs branches of these theories (as associated varieties of the corresponding VOAs). The associated varieties of admissible-level affine Kac--Moody VOAs have been determined by Arakawa \cite{Arakawa:2010ni}, so by way of the Higgs branch conjecture of \cite{Beem:2017ooy} we have immediate access to the Higgs branches of the non-anomalous theories. For the anomalous theories, through their realisation as nilpotent Higgsings of non-anomalous theories, we can characterise the anomalous-theory Higgs branches as Slodowy slice intersections with nilpotent orbit closures. It is interesting, though, to understand these geometries relative to the associated varieties of their affine Kac--Moody subalgebras. Indeed, we find that the Higgs branch $\cM_H(\mathrm{AD}(\cf^{\rm anom}_{N},m))$ is isomorphic to the associated variety $X_{V_{k^{\rm anom}_{N,m}}(\sp(2N))}$ for $m \geqslant N-1$, but is instead a ramified double cover of the latter for $m<N-1$. We first establish this result in general, and then illustrate it through several examples. This geometric analysis is corroborated in an interesting way by a study of various DS reductions of the extended VOA.

\medskip

The organisation of this paper is as follows. In Section \ref{sec:finite_extensions} we introduce and study the simplest interacting anomalous theory, $\mathrm{AD}(\cf^{\rm anom}_{2},0)$, describe its realisation as a minimal nilpotent Higgsing of $\mathrm{AD}(\cf_{3},0)$, and verify this against the global Witten anomaly and known central charge values. We further identify its associated VOA as a finite extension of an affine Kac--Moody VOA, and relate its Higgs branch to the associated variety of the current subalgebra. In Section \ref{sec:nonanomalous} we treat the non-anomalous family in general, compute vacuum characters and their modular linear differential equations, and recall certain collapsing level statements that relate these theories through Higgs branch RG flows. In Section \ref{sec:anomalVOA} we state and test the general Higgsing conjecture for the anomalous theories, identify the associated VOAs as finite extensions, and bootstrap the $m=0$ and $m=1$ series parametrically in $N$. We also propose and discuss additional equivalences in the landscape of generalised Argyres--Douglas SCFTs with non-simply laced flavour symmetries. In Section \ref{sec:anomalHiggs} we study the Higgs branches of the anomalous theories and establish their relationship to the associated varieties of the corresponding affine Kac--Moody subalgebras. We further investigate the ramification structure of this relation by purely geometric as well as vertex algebra methods. We conclude in Section \ref{sec:discussion}, where we also record some observations on the $\mathfrak{R}$-filtration of the $\mathbb{Z}_2$ gaugings. Four appendices contain background and technical details: Appendix \ref{app:irregular_puncture_review} reviews irregular punctures in twisted class $\cS$, Appendix \ref{app:KM_admissible} reviews admissible levels and associated varieties, Appendix \ref{app:asymptotic} reviews asymptotic data for affine Kac--Moody modules, and Appendix \ref{app:HBramifdetails} presents the details of the quiver analysis used in Section \ref{sec:anomalHiggs}.

%% file: sections/sec_2_simplest.tex

\section{\label{sec:finite_extensions}The simplest twisted \texorpdfstring{$A_{2N}$}{A(2N)} theory}

As a first instance of the general picture advocated in this paper, we revisit the theory $\mathrm{AD}(\cf^{\rm anom}_{2},0)$, which was studied in some detail in \cite{Kaidi:2021tgr} under the name ``$\ADtwo$''. There it was proposed that this theory can be realised on the Coulomb branch of the superconformal $G_2$ gauge theory with four hypermultiplets transforming in the $\mathbf{7}$ of $\gf_2$, and it was further argued to admit a realisation in class $\cS$ of type $\af_4$ with a single $\mathbb{Z}_2$-twisted regular puncture of type $[1^4]$ (\ie, a full twisted puncture) together with an irregular puncture of regular semisimple type. In our conventions, the latter construction is precisely $\mathrm{AD}(\cf^{\rm anom}_{2},0)$, and it is straightforward to confirm that the CFT data of $\mathrm{AD}(\cf^{\rm anom}_{2},0)$ match those of the $\ADtwo$ theory of \cite{Kaidi:2021tgr} (see Table \ref{CcADc2}).

The proposal of the present section is that $\mathrm{AD}(\cf^{\rm anom}_{2},0)$ admits a third description: it is the theory obtained from the non-anomalous $\mathrm{AD}(\cf_{3},0)$ by nilpotent Higgsing with respect to the minimal nilpotent orbit $\mathds{O}_{[2,1^4]}$. We first collect what is known about the parent theory, then test the proposal with anomaly matching, Schur index, and Higgs branch checks.

The associated VOA of $\mathrm{AD}(\cf_{3},0)$ has previously been conjectured to be the affine Kac--Moody VOA $V_{-\frac83}(\sp(6))$, and Table \ref{CcADc2} confirms that the central charge $c=-12c_{4d}$ and the flavour level $k=-k_{4d}/2$ of $\mathrm{AD}(\cf_{3},0)$ are consistent with this identification. The level $k_{3,0}=-\frac83$ is \emph{principal boundary admissible} (see Appendix \ref{app:KM_admissible} for a review of admissible levels), and consequently the vacuum character takes a particularly simple form \cite{kac2016remark}, 
\begin{equation}\label{VCC3}
    \chi_0(V_{-\frac{8}{3}}(\cf_3))=\left(\frac{\eta(q^3)}{\eta(q)}\right)^{21}~,
\end{equation}
where $\eta(q) = q^{\frac1{24}}(q;q)_\infty$ is the Dedekind eta function and $(q;q)_\infty=\prod_{n=1}^\infty(1-q^n)$ is the $q$-Pochhammer symbol. It is also convenient to express this character as a plethystic exponential,
\begin{equation}\label{SchurAD3no}
    \chi_0(V_{-\frac83}(\cf_3))={\rm P.E.}\left[\frac{1}{1-q}\sum_{i\in 3\mathbb{N}}\left(21q^{1+i}-21 q^{3+i}\right)\right]~.
\end{equation}
\begin{figure}
\capbtabbox[15cm]{%
\renewcommand{\arraystretch}{1.1}
\begin{tabular}{|c|c|c|} 
    \hline
    \multicolumn{2}{|c|}{$\mathrm{AD}(\cf^{\rm anom}_{2},0)$}& $\mathrm{AD}(\cf_{3},0)$\\
    \hline\hline
    $\D_{\rm CB}$ &   $(\frac43,\frac{10}3)$  & $(\frac{4}{3},\frac{8}{3},\frac{10}{3})$  \\
    $24a_{4d}$   &   $48$    &   $77$ \\    
    $12c_{4d}$   &   $26$    &   $42$ \\
    $\ff_k$ &   $\sp(4)_{\frac{13}3}$ & $\sp(6)_{\frac{16}3}$ \\ 
    $\dim_{\mathbb{H}}\cM_H$ &   $4$ &   $7$ \\
    $h$     &   $2$     &   $0$ \\
    $T({\bf2}\bh)$  &   $1$ &   $0$\\
    \hline\hline
\end{tabular}
}{%
  \caption{\label{CcADc2} Relevant conformal data for $\mathrm{AD}(\cf^{\rm anom}_{2},0)$ and $\mathrm{AD}(\cf_{3},0)$. Here $h$ denotes the quaternionic dimension of the extended Coulomb branch (ECB) and $T({\bf2}\bh)$ is the quadratic index of the ${\bf2}\bh$ representation of the flavour symmetry algebra $\ff$.}%
}
\end{figure}

Affine Kac--Moody algebras at admissible levels are quasi-lisse, so the vacuum character satisfies a finite-order modular linear differential equation (MLDE) \cite{Arakawa:2016hkg}, from which one can extract the high-temperature limit of the index and thereby the $a$ central charge of the putative associated $\cN=2$ SCFT \cite{ArabiArdehali:2015ybk,DiPietro:2014bca,Beem:2017ooy}. We find by direct computation that the character \eqref{VCC3} is annihilated by a fourth-order untwisted modular linear differential operator (MLDO)
\begin{equation}
    \cD_{\mathrm{AD}(\cf_{3},0)}\coloneqq D_q^{(4)}-1110\E_4(q)D_q^{(2)}+11340\E_6(q) D_q^{(1)}+11025(\E_4)^2~,
\end{equation}
where the $\E_i$ are the weight-$i$ Eisenstein series and we have adopted the notation of \cite{Beem:2017ooy}. From this operator we read off the conformal weights of the remaining components of the vector-valued modular form,
\begin{equation}
    h\in\left\{-\frac73,-2,-\frac53\right\}~,
\end{equation}
and since the minimal weight $h_{\rm min}$ among these is related to the four-dimensional Weyl anomaly coefficients according to \cite{Beem:2017ooy}
\begin{equation}
    h_{\rm min}= 2 a_{4d} - \frac{5}{2} c_{4d}~,
\end{equation}
we obtain $24a_{4d}=77$, in precise agreement with Table \ref{CcADc2}. Finally, the associated variety of this VOA---and hence the Higgs branch of the non-anomalous theory---is the nilpotent orbit closure $\overline{\mathds{O}}_{[3^2]}$, of quaternionic dimension $\dim_{\mathbb{H}}\cM_H = 7$ (see Appendix \ref{app:KM_admissible} for the associated varieties of admissible-level $\sp(2N)$ affine Kac--Moody algebras). 

Two elementary consistency checks of our proposal are now immediate. The commutant in $\sp(6)$ of the $\sl(2)$ associated to the orbit $\mathds{O}_{[2,1^4]}$ is $\sp(4)$, which matches the flavour symmetry of $\mathrm{AD}(\cf^{\rm anom}_{2},0)$. Also, the orbit $\mathds{O}_{[2,1^4]}$ has quaternionic dimension three, so the Higgs branch of the Higgsed theory has quaternionic dimension  $\dim_{\mathbb{H}}\cM_H = 4$, in agreement with Table \ref{CcADc2}. More refined checks come from anomaly matching, to which we turn next.

\subsection{Anomaly matching}

\subsubsection{\label{subsubsec:witten-matching-simplest}Witten anomaly matching}

We begin with the $\mathbb{Z}_2$-valued Witten anomaly. The UV theory $\mathrm{AD}(\cf_{3},0)$ is non-anomalous, so the Witten anomaly of the IR SCFT must cancel against that of the Nambu--Goldstone multiplets that arise from the spontaneous breaking of the flavour symmetry. The latter contribution is determined by decomposing the Nambu--Goldstone modes into representations of the $\sl(2)$-embedding $\Lambda:\sl(2) \hookrightarrow \gf$ and of the residual flavour symmetry, as explained in detail in \cite{Tachikawa:2015bga} and reviewed in \cite{Couzens:2023kyf}. For $\gf = \sp(6)$ and $\mathds{O}_{\rm nil} = \mathds{O}_{[2,1^4]}$, the fundamental representation decomposes as
\begin{equation}
    \mathbf{6}_{\sp(6)} \longrightarrow \left(\mathbf{1}_{\sl(2)} \otimes \mathbf{4}_{\sp(4)}\right) \oplus \left(\mathbf{2}_{\sl(2)} \otimes \mathbf{1}_{\sp(4)}\right)~,
\end{equation}
and consequently,
\begin{equation}
    \mathbf{adj}_{\sp(6)} \longrightarrow (\mathbf{2}_{\sl(2)} \otimes \mathbf{4}_{\sp(4)}) \oplus \left(\mathbf{1}_{\sl(2)} \otimes \mathbf{adj}_{\sp(4)} \right) \oplus (\mathbf{adj}_{\sl(2)} \otimes \mathbf{1})~.
\end{equation}
The Nambu--Goldstone multiplets span the tangent space to the nilpotent orbit of the moment map expectation values, and hence correspond to the states of non-highest weight with respect to the embedded $\sl(2)$: one fundamental $\mathbf{4}$ of $\sp(4)$ and two singlets. The Weyl fermions in these Nambu--Goldstone multiplets therefore contribute $1 \bmod 2 = 1$ to the $\mathbb{Z}_2$-valued Witten anomaly, so the Higgsed theory must itself be anomalous, as $\mathrm{AD}(\cf^{\rm anom}_{2},0)$ indeed is.

\subsubsection{Central charge anomaly matching}

The flavour level of the IR SCFT follows from 't~Hooft anomaly matching,
\begin{equation}
   k_{\rm IR} = I_{\sp(4) \hookrightarrow \sp(6)} k_{3,0} - k_{\rm NG}~,
\end{equation}
where $k_{\rm NG} = -\sum_i T_2(\mathfrak{R}_i)/2$ is the contribution of the Nambu--Goldstone multiplets, with $T_2(\mathfrak{R}_i)$ the Dynkin index of the representation $\mathfrak{R}_i$ and $I_{\sp(4)\hookrightarrow\sp(6)}$ the embedding index. With $k_{\rm NG} = -\frac{1}{2}$ from the decomposition above, one finds
\begin{equation}
    k_{\rm IR} = -\frac{13}{6} = k^{\rm anom}_{2,0}~,
\end{equation}
which is the flavour level of $\mathrm{AD}(\cf^{\rm anom}_{2},0)$. 

Next, Higgs branch (or $\Tr\, r^3$) anomaly matching requires
\begin{equation}
    24(a_{\rm UV}-c_{\rm UV}) = 24(a_{\rm IR}-c_{\rm IR}) - \dim_{\mathbb{H}} \cM_H^{\rm UV} + \dim_{\mathbb{H}} \cM_H^{\rm IR}~.
\end{equation}
With $\dim_{\mathbb{H}}\cM_H^{\rm UV}-\dim_{\mathbb{H}}\cM_H^{\rm IR}=3$ and $24(a_{\rm UV} - c_{\rm UV}) = -7$, the central charges of the Higgsed theory must satisfy $24(a_{\rm IR} - c_{\rm IR}) = -4$. 

The central charge $c_{\rm IR}$ can be computed independently. Nilpotent Higgsing and the subsequent RG flow to the infrared is implemented at the level of the VOA by quantum Drinfel'd--Sokolov reduction, and for the reduction of an affine Kac--Moody algebra $V_k(\gf)$ the central charge $c = -12 c_{\rm IR}$ of the reduced VOA is given by \cite{Kac_2003}
\begin{equation} \label{centDS}
    -12 c_{\rm IR} = \dim \gf_0 - \frac{1}{2} \dim \gf_{1/2} - \frac{12}{(k+h^\vee_\gf)}|\rho - (k+h^\vee_\gf)x|^2~.
\end{equation}
Here $\rho$ is the Weyl vector of $\gf$, $x = \Lambda(t_0)$ is the semisimple element of the $\sl(2)$-triple associated to the embedding $\Lambda: \sl(2) \hookrightarrow \gf$, and $\gf_i$ is the subspace of degree $i$ with respect to the adjoint action of $x$,
\begin{equation}
    \gf_i = \{y \in \gf : [x, y] = iy \}~.
\end{equation}

For $\gf = \sp(6)$ and the minimal embedding, the semisimple element $x$ and the Weyl vector $\rho$ (realised as an element of the Cartan subalgebra via the Killing form) are
\begin{equation}
\begin{split}
    &x = \mathrm{diag}\left( \frac{1}{2}, 0, 0, -\frac{1}{2}, 0, 0 \right)~,\\
    &\rho = \mathrm{diag}\left(\frac{3}{2}, 1, \frac{1}{2}, -\frac{3}{2}, -1, -\frac{1}{2} \right)~.
\end{split}
\end{equation}
From these one finds\footnote{Note that the bilinear form $(\cdot|\cdot)$ is given by $(x|y) = \mathrm{tr}(xy)$, and is related to the Killing form by $\kappa_\gf(x,y) = 2 h^\vee_\gf (x|y)$.}
\begin{equation}
    (\rho|\rho) = 7~, \qquad (\rho|x) = \frac{3}{2}~, \qquad (x|x) = \frac{1}{2}~,
\end{equation}
and counting the roots of $\sp(6)$ with $(\alpha|x) = 1/2$ and $(\alpha|x) = 0$ gives
\begin{equation}
    \dim \gf_0 = 11~, \qquad \dim \gf_{1/2} = 4~.
\end{equation}
Substituting into \eqref{centDS} yields $12 c_{\rm IR} = 26$, and the Higgs branch anomaly matching condition $24(a_{\rm IR} - c_{\rm IR}) = -4$ then fixes $24 a_{\rm IR} = 48$. Both values are those of $\mathrm{AD}(\cf^{\rm anom}_{2},0)$ (see Table \ref{CcADc2}). 

\subsection{VOA and Higgs branch} 

As a final check, we compute the Schur index of the Higgsed theory, \ie, the vacuum character of the quantum Drinfel'd--Sokolov reduction of $V_{-\frac83}(\sp(6))$ with respect to $\mathds{O}_{[2,1^4]}$. The computation can be carried out using standard techniques for Higgsing the Schur index \cite{Gaiotto:2012xa,Beem:2014rza,Nishinaka:2018zwq} (see \cite{Frenkel:1992ju,Kac_2003,Arakawa:2010ni} for the vertex algebraic treatment), and is particularly tractable thanks to the simple form of the parent character. The result can again be given in plethystic exponential form
\begin{equation}
    \cI_{\cT_{\rm Higgs}}={\rm P.E.}\left[\frac{1}{1-q}\sum_{i\in 3\mathbb{N}}\left(10q^{1+i}+4q^{3/2+i}-4q^{5/2+i}-10q^{3+i}\right)\right]~,
\end{equation}
which matches the form conjectured for the Schur index of this theory in \cite{Kaidi:2021tgr}.

Taking the proposal as given now, at the level of vertex algebras this means that the VOA of $\mathrm{AD}(\cf^{\rm anom}_{2},0)$ is the quantum Drinfel'd--Sokolov reduction of the affine Kac--Moody algebra $V_{-\frac{8}{3}}(\sp(6))$ with respect to $\mathds{O}_{[2,1^4]}$. By the results of \cite{Arakawa:2021ogm}, this reduction can be identified with the \emph{finite extension}
\begin{equation} \label{eq:VOAADc2}
     \mathbb{V}[\mathrm{AD}(\cf^{\rm anom}_{2},0)] \cong V_{-\frac{13}6}(\sp(4))\oplus L_{-\frac{13}6}(\sp(4);\varpi_1)~.
\end{equation}
In other words, the associated VOA of $\mathrm{AD}(\cf^{\rm anom}_{2},0)$ is not the coprincipal admissible (but non-boundary) affine Kac--Moody algebra $V_{-\frac{13}6}(\sp(4))$ itself, but rather its finite extension by a module in the fundamental representation of $\sp(4)$.

Correspondingly, the Higgs branch of $\mathrm{AD}(\cf^{\rm anom}_{2},0)$ is then the Slodowy intersection
\begin{equation}
    \cM_H(\mathrm{AD}(\cf^{\rm anom}_{2},0)) = \overline{\mathds{O}}_{[3^2]} \cap \cS_{[2,1^4]}~,
\end{equation}
whereas the associated variety of $V_{-\frac{13}{6}}(\sp(4))$ is the full nilpotent cone of $\sp(4)$. The two varieties are compared at the level of their Hasse diagrams in Figure \ref{HasseC20}, which makes their relationship apparent: the associated variety $X_{V_{-\frac{13}{6}}(\sp(4))}$ is a $\mathbb{Z}_2$ quotient of the Higgs branch $\cM_H(\mathrm{AD}(\cf^{\rm anom}_{2},0))$, the Kleinian singularity $A_3$ being a $\mathbb{Z}_2$ quotient of the Kleinian singularity $A_1$. Equivalently, the Higgs branch of $\mathrm{AD}(\cf^{\rm anom}_{2},0)$ is a double cover of $X_{V_{-\frac{13}{6}}(\sp(4))}$.

An interesting further consequence of this picture is that the $\mathrm{AD}(\cf^{\rm anom}_{2},0)$ SCFT enjoys a $\mathbb{Z}_2$ symmetry, and gauging this symmetry (we denote the gauged theory by $\mathrm{AD}(\cf^{\rm anom}_{2},0)^{\mathbb{Z}_2}$) has as its associated VOA the affine Kac--Moody algebra itself,
\begin{equation}
    \mathbb{V}[\mathrm{AD}(\cf^{\rm anom}_{2},0)^{\mathbb{Z}_2}] = V_{-\frac{13}{6}}(\sp(4))~.
\end{equation}
This is somewhat notable, as $-\frac{13}{6}$ is not a \emph{boundary} admissible level, whereas boundary admissibility is a recurring feature of affine Kac--Moody VOAs arising from four-dimensional unitary theories \cite{ArabiArdehali:2025fad,Beem:2026lkq}; we will return to this point in Section \ref{sec:discussion}.

\input{figures/Hasse_C2m0}

None of the above features of $\mathrm{AD}(\cf^{\rm anom}_{2},0)$ are accidents of the simplest example. Rather they hold uniformly throughout the anomalous family, as we show in Section \ref{sec:anomalHiggs}, where we relate the Higgs branch of $\mathrm{AD}(\cf^{\rm anom}_{N},m)$ to the associated variety $X_{V_{k^{\rm anom}_{N,m}}(\sp(2N))}$ for all $(N,m)$.

%% file: figures/Hasse_C2m0.tex

\begin{figure}[t]
\begin{tikzpicture}[decoration={markings,
mark=at position .5 with {\arrow{>}}}, gauge/.style={circle, draw=blue, fill=blue, minimum size=7mm, inner sep=0pt},
    flavour/.style={rectangle, draw=red, fill=red, minimum size=6mm, inner sep=0pt},
    arrow/.style={-{Latex[width=2mm]}, thick}]
\begin{scope}[scale=1.5,xshift=1.5cm]
\node[scale=.8] (p0a) at (0,1.5) {$[3^2]$};
\node[scale=.8] (t1a) at (-0.2,1) {$A_1$};
\node[scale=.8] (p1) at (0,.5) {$[2^3]$};
\node[scale=.8] (t1b) at (-.2,0) {$\af_1$};
\node[scale=.8] (p2) at (0,-.5) {$[2^2,1^2]$};
\node[scale=.8] (t1b) at (-.2,-1) {$\cf_2$};
\node[scale=.8] (p0b) at (0,-1.5) {$[2,1^4]$};
\draw[blue] (p0a) -- (p1);
\draw[blue] (p1) -- (p2);
\draw[blue] (p2) -- (p0b);

\end{scope}
\begin{scope}[scale=1.5,xshift=5.5cm]
\node[scale=.8] (p0a) at (0,1.5) {$[4]$};
\node[scale=.8] (t1a) at (-0.2,1) {$A_3$};
\node[scale=.8] (p1) at (0,.5) {$[2^2]$};
\node[scale=.8] (t1b) at (-.2,0) {$\af_1$};
\node[scale=.8] (p2) at (0,-.5) {$[2,1^2]$};
\node[scale=.8] (t1b) at (-.2,-1) {$\cf_2$};
\node[scale=.8] (p0b) at (0,-1.5) {$[1^4]$};
\draw[blue] (p0a) -- (p1);
\draw[blue] (p1) -- (p2);
\draw[blue] (p2) -- (p0b);

\end{scope}
\end{tikzpicture}
{\caption{\label{HasseC20} Hasse diagrams for the Higgs branch moduli space of $\mathrm{AD}(\cf^{\rm anom}_{2},0)$ (left) and the associated variety $X_{V_{-\frac{13}{6}}(\sp(4))}$ (right).}}
\end{figure}
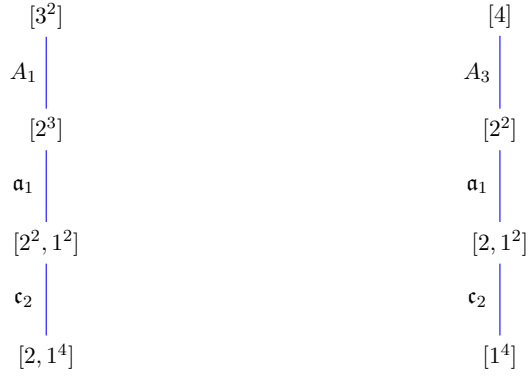

%% file: sections/sec_3_nonanomalous.tex

\section{\label{sec:nonanomalous}VOAs and Higgs branches for \texorpdfstring{$\mathrm{AD}(\cf_{N},m)$}{AD(c(N),m} theories}

The associated VOAs of the non-anomalous $\mathrm{AD}(\cf_{N},m)$ theories without exactly marginal couplings were conjectured in \cite{Wang:2018gvb} to be precisely the affine Kac--Moody VOAs implied by their flavour symmetries, 
\begin{equation}
    \mathbb{V}[\mathrm{AD}(\cf_{N},m)] \cong V_{k_{N,m}}(\sp(2N))~,
\end{equation}
where $k_{N,m}$ is the principal boundary admissible level from \eqref{ADlevels}. As noted previously, the level and the $c$ central charge of $\mathrm{AD}(\cf_{N},m)$ are consistent with this conjecture. In this section we perform a further test of the proposal by showing that the $a$ central charge computed from the high-temperature limit of the Schur index matches the value obtained from the Coulomb branch spectrum. We further review the implications for the Higgs branches of these theories, and recall the collapsing-level relations between members of the non-anomalous family under nilpotent Higgsing. 
 
\subsection{Character of the VOA and modularity}

Due to its being boundary admissible, the vacuum character of $V_{k_{N,m}}(\sp(2N))$ is particularly simple \cite{kac2016remark}. In particular, the unflavoured character reduces to the following simple $\eta$-quotient, 
\begin{equation}
    \chi_0(V_{k_{N,m}}(\sp(2N)))=\left(\frac{\eta(q^{3+2m})}{\eta(q)}\right)^{2N^2 + N}~.
\end{equation}
These characters satisfy finite-order modular linear differential equations (MLDEs), and it is interesting to observe that the order of the MLDE within this family is completely determined by $m$ and is independent of $N$ (for $N$ sufficiently large). For example, it is four for the $m=0$ series and eight for the $m=1$ series, as we illustrate below. These specific modular properties are a special case of a more general phenomenon, which will be discussed systematically in the upcoming work \cite{Etaquotientstoappear}.  

\subsubsection{The $m = 0$ series}

Admissibility of the flavour level in this case imposes the restriction $N \equiv 0,1 \pmod{3}$. The first case, $N = 1$, is the VOA $V_{-\frac{4}{3}}(\sl(2))$ of the $(A_1, A_3)$ Argyres--Douglas SCFT, whose vacuum character was observed to satisfy a second-order MLDE in \cite{Beem:2017ooy}. The $N = 3$ case was treated in the previous section.

In general, the vacuum character for $V_{k_{N,0}}(\sp(2N))$ is annihilated by a fourth-order untwisted MLDO,
\begin{equation}
    \cD_{\mathrm{AD}(\cf_{N},0)}\coloneqq D_q^{(4)}+a_1(N)\E_4(q)D_q^{(2)}+a_2(N)\E_6(q) D_q^{(1)}+a_3(N)(\E_4(q))^2~,
\end{equation}
where $a_1(N)$, $a_2(N)$, and $a_3(N)$ are quartic, sextic, and octic polynomials in $N$: 
\begin{equation}
\begin{split}
    &a_1(N) = -\frac{10}{3} \left( 4N^4 + 4N^3 - 11N^2 - 6N + 18\right)~, \\
    &a_2(N) = \frac{140}{27}\left(8N^6 + 12N^5-66N^4 - 71N^3 + 72N^2 + 45N - 81\right)~, \\
    &a_3(N) = -\frac{25}{27} \left(N^2 (2N+1)^2(2 N^2 + N - 24) (2 N^2 + N - 12) \right)~.
\end{split}
\end{equation}
This reproduces the operator of Section \ref{sec:finite_extensions} at $N=3$. The conformal weights of the remaining components of the corresponding vector-valued modular form are
\begin{equation}
    h_i = \left(\frac{-N(2N+1)}{9}, \frac{-(N-1)(2N+3)}{9}, \frac{-(N+2)(2N-3)}{9}\right)~,
\end{equation}
so $h_{\mathrm min} = \frac{-N(2N+1)}{9}$. Then by the relation $h_{\mathrm min} = 2a_{4d}-\frac{5}{2}c_{4d}$, one finds $a_{4d} = \frac{11}{72}N(2N+1)$, which indeed matches with the $a$ central charge of the $\mathrm{AD}(\cf_{N},0)$ theories in \eqref{acent}.

\subsubsection{The $m = 1$ series}

For $m=1$, the vacuum character of $V_{k_{N,1}}(\sp(2N))$ satisfies an untwisted eighth-order MLDE, again irrespective of $N$. The general expression not being particularly illuminating, here we give the MLDOs for $N=3$ and $N=5$, \ie, for $V_{-\frac{16}{5}}(\cf_3)$ and $V_{-\frac{24}{5}}(\cf_5)$,
\begin{equation}
\begin{split}
    \cD_{\mathrm{AD}(\cf_{3},1)}&\coloneqq D_q^{(8)}-4616\E_4(q)D_q^{(6)}+\frac{102872}{5}\E_6(q) D_q^{(5)}+\frac{87987184}{25}(\E_4(q))^2 D_q^{(4)} \\ &+ \frac{427898464}{25} \E_4(q) \E_6(q) D_q^{(3)} -1195741904 (\E_6(q))^2 D_q^{(2)} \\ &- \frac{1615297152}{5} (\E_4(q))^3 D_q^{(2)} + \frac{13043712192}{5} (\E_4(q))^2 \E_6(q) D_q^{(1)} \\ &+ \frac{12098274048}{5} (\E_6(q))^2 \E_4(q) + 740710656 (\E_4(q))^4~, 
\end{split}
\end{equation}
\begin{equation}
\begin{split}
    \cD_{{\mathrm AD}(\cf_{5},1)}&\coloneqq D_q^{(8)}-33992\E_4(q)D_q^{(6)}+4518584\E_6(q) D_q^{(5)}-\frac{560504336}{25}(\E_4(q))^2 D_q^{(4)} \\ &- \frac{134583919008}{25} \E_4(q) \E_6(q) D_q^{(3)} -\frac{1959359532368}{25} (\E_6(q))^2 D_q^{(2)} \\ &+ \frac{363965027712}{5} (\E_4(q))^3 D_q^{(2)} + \frac{13344210871488}{5} (\E_4(q))^2 \E_6(q) D_q^{(1)} \\ &-33013523499264 (\E_6(q))^2 \E_4(q) -10106180663040 (\E_4(q))^4~. 
\end{split}
\end{equation}
As for the $m = 0$ series, the minimal conformal weight in the corresponding vector-valued modular form takes a closed form $h_{\rm min} = -\frac{1}{5}N(2N+1)$, and consequently so does the $a$ central charge $a_{4d} = \frac{19}{60}N(2N+1)$, which again matches \eqref{acent}.

We note that these closed-form results can be understood (and extended to all $m$) without explicitly determining the MLDE. The modular properties for \emph{principal admissible} affine Kac--Moody algebras are known from the work of Kac--Wakimoto \cite{Kac:1988qc, Kac:1989zz}, and in particular the \emph{asymptotic growth}, or effective central charge, of the vacuum character is given by (see Appendix \ref{app:asymptotic})
\begin{equation}
    \mathbf{g}_{V_{k_{N,m}}(\sp(2N))} = \frac{2N(2N+1)(m+1)}{3+2m}~.
\end{equation}
The asymptotic growth is related to the four-dimensional central charges by
\begin{equation}
    \mathbf{g}_V = 48(c_{4d} - a_{4d})~,
\end{equation}
so one finds
\begin{equation}
    24 a_{4d} = \frac{N(2N+1)(m+1)(8m+11)}{3+2m}~.
\end{equation}
This is precisely the $a$ central charge of $\mathrm{AD}(\cf_{N},m)$ obtained from the Coulomb branch spectrum.

Turning to the Higgs branch, the conjecture of \cite{Beem:2017ooy} identifies this with the associated variety of the corresponding VOA,
\begin{equation}
    \cM_H(\mathrm{AD}(\cf_{N},m)) = X_{V_{k_{N,m}}(\sp(2N))}~.
\end{equation}
Since $k_{N,m}$ is admissible, the associated variety is a nilpotent orbit closure \cite{Arakawa:2010ni}, \ie, $X_{V_{k_{N,m}}(\sp(2N))} \cong \overline{\mathds{O}}_q$, where $\mathds{O}_q$ is the nilpotent orbit given in \eqref{Oqnil} in Appendix \ref{app:KM_admissible}.

\subsection{Collapsing levels}

The notion of a collapsing level for an affine Kac--Moody VOA was introduced in \cite{adamovic2018conformal} (see also \cite{adamovic2020application} for further applications). A level $k$ is said to be collapsing for the nilpotent orbit of the nilpotent element $f\in\gf$ if the (simple quotient of the) corresponding $\mathcal{W}$-algebra is generated by the remaining affine currents, \ie, if one has the isomorphism
\begin{equation}
    \cW_k(\gf, f) \cong V_{k^\natural}(\gf^\natural)~,
\end{equation}
where $\gf^\natural \subset \gf$ is the centraliser subalgebra of the $\sl(2)$ triple $(e,f,h)$ in $\gf$. 

The collapsing levels for $V_k(\sp(2N))$ for $k$ principal boundary admissible have already appeared in the physics literature \cite{Xie:2019yds} and in the mathematics literature \cite{Arakawa:2021ogm}. Writing $q\coloneqq 3+2m$ for the denominator of the level (which is always odd for the theories in question), one has the following isomorphism of vertex operator algebras,
\begin{equation} \label{Wcollap}
       \mathcal{W}_{k'}(\sp(2(N+nq)),\mathds{O}_{[q^{2n},1^{2N}]}) \cong  V_{k}(\sp(2N))
    \qquad \mathrm{with} \ n \in \mathbb{Z}_{\geqslant 0}~, 
\end{equation}
where
\begin{equation}
    k = -(N+1) + \frac{N+1}{q} \ \ \ \mathrm{and} \ \ \ k' = -(N+nq+1) + \frac{N+nq+1}{q}~.
\end{equation}
These are precisely the levels $k_{N,m}$ and $k_{N+nq,m}$ of \eqref{ADlevels}, so the collapse statement implies that the VOA of $\mathrm{AD}(\cf_{N},m)$ is the quantum Drinfel'd--Sokolov reduction of the VOA of $\mathrm{AD}(\cf_{N+nq},m)$ with respect to the nilpotent orbit $\mathds{O}_{[q^{2n},1^{2N}]}$. Consequently, the Higgs branches of the family of non-anomalous theories are related to each other by nilpotent Higgsing,
\begin{equation} \label{HiggsNonanom}
    \cM_H(\mathrm{AD}(\cf_{N},m)) \cong \cM_H(\mathrm{AD}(\cf_{N+nq},m)) \cap \cS_{[q^{2n}, 1^{2N}]}
    \qquad  \mathrm{with} \ n \in \mathbb{Z}_{\geqslant 0}~, 
\end{equation}
where $\cS_{[q^{2n}, 1^{2N}]}$ is the Slodowy slice to the nilpotent orbit $\mathds{O}_{[q^{2n}, 1^{2N}]}$. Thus the non-anomalous theory $\mathrm{AD}(\cf_{N},m)$ admits multiple realisations as nilpotent Higgsings of other members of the non-anomalous family.

%% file: sections/sec_4_anomalous_VOA.tex

\section{\label{sec:anomalVOA}Higgsing and the \texorpdfstring{$\mathrm{AD}(\cf^{\rm anom}_{N},m)$}{AD(c-anom(N),m)} theories}

Having reviewed the VOAs and Higgs branches of the non-anomalous $\mathrm{AD}(\cf_{N},m)$ theories, we now turn to the anomalous $\mathrm{AD}(\cf^{\rm anom}_{N},m)$ theories, \ie, the twisted $A_{2N}$ Argyres--Douglas SCFTs. Recall from \eqref{ADlevels} that the level of the Kac--Moody subalgebra of the associated VOA is $k^{\rm anom}_{N,m}=-h^\vee_{\cf_N}+\frac{1}{2}\frac{2N+1}{3+2m}$, which is coprincipal admissible, but not boundary admissible, whenever $(2N+1, 3+2m) = 1$ (see Appendix \ref{app:KM_admissible}). Throughout the remainder of this paper we restrict to these values of $(N,m)$; physically, these are the anomalous theories without exactly marginal couplings.

In Section \ref{sec:finite_extensions} we found that $\mathrm{AD}(\cf^{\rm anom}_2, 0)$ is realised as a nilpotent Higgsing of $\mathrm{AD}(\cf_{3},0)$, so that its VOA is a quantum Drinfel'd--Sokolov reduction of that of $\mathrm{AD}(\cf_{3},0)$ and is a finite extension of an affine Kac--Moody VOA. Taking this simplest example as representative, we propose the following generalisation.

\smallskip

\begin{conjecture}\label{conj:anom-higgsing}
Higgsing the non-anomalous $\mathrm{AD}(\cf_{N+m+1},m)$ SCFT with respect to the nilpotent orbit $\mathds{O}_{[2+2m,1^{2N}]}$ gives rise to the anomalous $\mathrm{AD}(\cf^{\rm anom}_{N},m)$ SCFT in the IR.
\end{conjecture} 

\smallskip

Since the Higgsing is with respect to the nilpotent orbit $\mathds{O}_{[2+2m, 1^{2N}]}$, the infrared theory has $\sp(2N)$ flavour symmetry, as required. In the next subsection, we describe several anomaly matching checks of this proposal. We then turn to the VOA of the Higgsed theory and to alternative realisations of these SCFTs.

\subsection{Anomaly matching}

\subsubsection{Witten anomaly matching}

The UV theory does not have a global Witten anomaly, so the Witten anomaly of the IR SCFT must cancel against the contribution from the Nambu--Goldstone multiplets, as in Section \ref{subsubsec:witten-matching-simplest}. The only novel ingredient in this more general case is the decomposition of the adjoint representation of $\sp(2(N+m+1))$ under the $\sl(2)$ subalgebra associated to the orbit $\mathds{O}_{[2+2m,1^{2N}]}$ and the commuting, unbroken $\sp(2N)$ flavour symmetry, which determines the contribution of the Nambu--Goldstone modes to the Witten anomaly. The fundamental representation of $\sp(2(N+m+1))$ decomposes as:
\begin{equation}
    \mathbf{2(N+m+1)} \longrightarrow \left(\mathbf{1}_{\sl(2)} \otimes \mathbf{2N}_{\sp(2N)}\right) \oplus \left(\mathbf{(2+2m)}_{\sl(2)} \otimes \mathbf{1}_{\sp(2N)}\right)~.
\end{equation}
From this, we find that the adjoint representation of $\sp(2(N+m+1))$ decomposes as:
\begin{equation} \label{adjdec}
\begin{split}
    \mathbf{adj} \longrightarrow &\left(\mathbf{(2+2m)}_{\sl(2)} \otimes \mathbf{2N}_{\sp(2N)}\right) \oplus \left(\mathbf{1}_{\sl(2)} \otimes \mathbf{adj}_{\sp(2N)} \right) \\  &\bigoplus_{k=0}^{m} \left(\mathbf{(4m+3-4k)}_{\sl(2)} \otimes \mathbf{1}\right)~.
\end{split}
\end{equation}
The Nambu--Goldstone multiplets are associated with the tangent space to the nilpotent orbit of the moment map expectation value, \ie, with the broken generators $X_a$ satisfying
\begin{equation}
    [\Lambda(t_+), X_a] \neq 0~,
\end{equation}
which are the non-highest weight components under the embedded $\sl(2)$. From \eqref{adjdec}, there are $2m+1$ such components in the fundamental representation of $\sp(2N)$ and $2(m+1)^2$ components in the trivial representation, for a total number $n_{\rm NG}$ of Nambu--Goldstone multiplets given by
\begin{equation}
    n_{\rm NG} = N(2m+1) + (m+1)^2~.
\end{equation}
Their Weyl fermions contribute $(2m+1) \bmod 2 = 1$ to the $\mathbb{Z}_2$-valued Witten anomaly, so the IR SCFT must itself carry a Witten anomaly, in agreement with Conjecture \ref{conj:anom-higgsing}.

\subsubsection{Central charge anomaly matching}
The flavour level of the Higgsed theory follows by anomaly matching between the $\sp(2(N+m+1))$ flavour symmetry of the UV theory (with level $-2k_{N+m+1,m}$) and the $\sp(2N)$ flavour symmetry of the IR theory,
\begin{equation}
   k_{\rm IR} = I_{\cf_{N} \hookrightarrow \cf_{N+m+1}}k_{N+m+1,m} - k_{\rm NG}~,
\end{equation}
where $k_{\rm NG} = -\sum_{i} T_2(\mathfrak{R}_i)/2$ is the contribution of the Nambu--Goldstone multiplets in the representation $\mathfrak{R}_i$ of the unbroken $\sp(2N)$, and the embedding index appearing above is one. The $2m+1$ components in the fundamental representation found above give the total contribution $k_{\rm NG} = -\frac{2m+1}{2}$, and hence
\begin{equation}
    k_{\rm IR} = -(N+1) + \frac{2N+1}{2(3+2m)} = k^{\rm anom}_{N,m}~,
\end{equation}
which matches the flavour level of the anomalous $\mathrm{AD}(\cf^{\rm anom}_{N},m)$ theory. 

Matching of the $\Tr\,r^3$ anomaly gives the relation
\begin{equation} \label{HBanommatch}
    24(a_{\rm UV}-c_{\rm UV}) = 24(a_{\rm IR}-c_{\rm IR}) - \dim_{\mathbb{H}} \cM_H^{\rm UV} + \dim_{\mathbb{H}} \cM_H^{\rm IR}~.
\end{equation}
Since the Higgsing is with respect to $\mathds{O}_{[2+2m,1^{2N}]}$, the difference between the quaternionic dimensions of the UV and IR Higgs branches is the quaternionic dimension of this nilpotent orbit: 
\begin{equation}
    \dim_{\mathbb{H}} \cM_H^{\rm UV} - \dim_{\mathbb{H}} \cM_H^{\rm IR} = \dim\mathds{O}_{[2+2m,1^{2N}]} = (N+m+1)^2 - N(N+1)~,
\end{equation}
while the central charges of the non-anomalous $\mathrm{AD}(\cf_{N+m+1},m)$ theory give
\begin{equation}
    24(a_{\rm UV} - c_{\rm UV}) = -\frac{(N+m+1)(2N+2m+3)(m+1)}{3+2m}~.
\end{equation}
It follows that for the IR theory,
\begin{equation} \label{aminuscIR}
    24(a_{\rm IR}-c_{\rm IR}) = -\frac{N(2N + 2Nm + m + 2)}{3+2 m}~,
\end{equation}
which exactly matches $24(a-c)$ of the anomalous $\mathrm{AD}(\cf^{\mathrm{anom}}_{N},m)$ theory according to \eqref{ccent} and \eqref{acent}. 

Finally, as in Section \ref{sec:finite_extensions}, the central charge $c_{\rm Higgs}$ of the Higgsed theory can be computed from \eqref{centDS}. All quantities entering \eqref{centDS} can be computed using the symplectic pyramids corresponding to partitions labelling the nilpotent orbits of $\sp(2(N+m+1))$ \cite{Arakawa:2021ogm}; one finds
\begin{equation}
\begin{split}
    (\rho|\rho) &= \frac{1}{12}(N+m+1)(N+m+2)(2N+2m+3)~,\\
    (x|x) &= \frac{1}{6}(m+1)(2m+1)(2m+3)~,\\
    (\rho|x) &= \frac{1}{12}(m+1)(6N + 6Nm + 4m^2 + 11m+6)~,\\
    \dim\,\gf_0 &= (N+m+1)+2N^2~, \qquad \dim\,\gf_{1/2} = 2 N~,
\end{split}
\end{equation}
and therefore,
\begin{equation}
    12c_{\rm Higgs} = N(4N + 4Nm + 4m + 5)~,
\end{equation}
which is precisely the central charge of $\mathrm{AD}(\cf^{\mathrm{anom}}_{N},m)$. Combined with \eqref{aminuscIR}, this reproduces the correct $a$ central charge as well.

\subsection{Vertex operator algebra}
Since nilpotent Higgsing is implemented as quantum Drinfel'd--Sokolov reduction at the level of the associated VOA, Conjecture \ref{conj:anom-higgsing} implies that
\begin{equation} \label{eq:twA2Nrealis}
 \mathbb{V}[\mathrm{AD}(\cf^{\mathrm{anom}}_{N},m)] = H^0_{DS,\mathds{O}_{[2+2m, 1^{2N}]}}(V_{k_{N+m+1,m}}(\sp(2(N+m+1))))~.
\end{equation}
By the results of \cite{Arakawa:2021ogm}, this $\mathcal{W}$-algebra is a finite extension of the \emph{coprincipal} admissible (but not boundary admissible) affine Kac--Moody algebra $V_{k^{\rm anom}_{N,m}}(\sp(2N))$ by a module with conformal highest weights in the fundamental representation of $\sp(2N)$,
\begin{equation} \label{finiteextsp2N}
    \mathbb{V}[\mathrm{AD}(\cf^{\mathrm{anom}}_{N},m)] \cong V_{k^{\rm anom}_{N,m}}(\sp(2N)) \oplus L_{k^{\rm anom}_{N,m}}(\sp(2N);\varpi_1)~.
\end{equation}
The conformal dimension of the module depends only on $m$, 
\begin{equation}
    h_m = \frac{C_2(2N)}{k^{\rm anom}_{N,m} + h^\vee_{\sp(2N)}} = \frac{\frac{2N+1}{4}}{\frac{2N+1}{2(3+2m)}} = \frac{3 + 2m}{2}~.
\end{equation}   
For $m = -1$ this reproduces the free hypermultiplet result \eqref{eq:freehyperfiniteext} recalled in the introduction, and \eqref{finiteextsp2N} generalises this result to the full family with $m \geqslant 0$.

Equation \eqref{finiteextsp2N} is consistent with all the anomaly matching checks above. The level of the current subalgebra is $k^{\rm anom}_{N,m}$, as found from 't~Hooft anomaly matching, and the asymptotic growth of the vacuum character of the \emph{coprincipal admissible} affine Kac--Moody algebra $V_{k^{\rm anom}_{N,m}}(\sp(2N))$ (see Appendix \ref{app:asymptotic}),
\begin{equation}
    \mathbf{g}_{V_{k^{\rm anom}_{N,m}}(\sp(2N))} = \frac{2N(2N + 2Nm+2+m)}{3+2m}~,
\end{equation}
reproduces $48(c-a)$ of the anomalous theory, in agreement with \eqref{aminuscIR}. The latter agreement is to be expected, since a finite extension does not change the asymptotic growth, so that the $a$ and $c$ central charges are insensitive to the extension. The Higgs branch, on the other hand, may be affected by such an extension. The Higgs branch reconstruction conjecture implies the identification
\begin{equation} \label{anomhiggs}
     \mathcal{M}_H(\mathrm{AD}(\cf^{\mathrm{anom}}_{N},m)) \cong X_{V_{k_{N+m+1,m}}(\sp(2(N+m+1)))} \cap \mathcal{S}_{[2+2m,1^{2N}]}~,
\end{equation}
which need not coincide with the associated variety of $V_{k^{\rm anom}_{N,m}}(\sp(2N))$ itself (as we saw in Section \ref{sec:finite_extensions} and will analyse in general in Section \ref{sec:anomalHiggs}). Before turning to Higgs branches, we take a short diversion and study the VOAs of the $m=0$ and $m=1$ series directly via a bootstrap analysis.

\subsection{Bootstrapping the VOA}

In this section, we produce the VOAs of the $\mathrm{AD}(\cf^{\rm anom}_{N},m)$ theories for $m=0$ and $m=1$ by bootstrapping the OPEs parametrically in $N$.\footnote{All calculations in this section were performed using the \texttt{OPEDefs} package of \cite{Thielemans:1994er}.} Since each VOA is a finite extension of $V_{k^{\rm anom}_{N,m}}(\sp(2N))$ by a (simple) fundamental Weyl module of conformal weight $h_m=\frac{3+2m}{2}$, the strong generators are the affine currents $J_{\alpha\beta}(z)$ in the adjoint of $\sp(2N)$ and the highest-weight fields of the module $W_\g(z)$ where $\g$ is a fundamental index. The singular OPEs for the affine currents $J_{\alpha \beta}$ are as usual,
\begin{equation}
    J_{\alpha \beta}(z) J_{\gamma \delta}(w) \sim \frac{k \Om_{\a (\g} \Om_{\d) \b}}{4(z-w)^2} + \frac{\Om_{(\a (\g} J_{\d) \b)}(w)}{(z-w)}~,
\end{equation}
where $\Om$ is the $2N \times 2N$ antisymmetric symplectic form
\begin{equation}
    \Om = \left( \begin{matrix}
        0 & I_N \\
        -I_N & 0
    \end{matrix} \right)~,
\end{equation}
and the OPE of the currents with the $W_\g(z)$ is similarly standard,
\begin{equation}
     J_{\alpha \beta}(z) W_\g (w) \sim \frac{\Om_{(\a |\g|} W_{\b)}(w)}{2(z-w)}~.
\end{equation}
The general form of the self-OPE $W\times W$ of the module generators depends on $m$. Below we solve the bootstrap problem for this self-OPE for the $m=0$ and $m=1$ cases.

\subsubsection{The $m = 0$ series}

The most general singular self-OPE for $W$ compatible with $\sp(2N)$ covariance is
\begin{equation}\label{eq:m0_ansatz}
    W_\a(z) W_\b(w) \sim \frac{c_1 \Om_{\a \b}}{(z-w)^3} + \frac{c_2 J_{\a \b}(w)}{(z-w)^2} + \frac{c_3 J'_{\a \b}(w) + c_4 \Om^{\g \d} J_{\a \g} J_{\b \d}(w) + c_5 \Om_{\a \b} T(w)}{z-w}~,
\end{equation}
where composite operators are defined by nested normal ordering, a convention we will use throughout this section. Additionally, where $T$ appears in \eqref{eq:m0_ansatz} (and in \eqref{eq:selfOPEmeq1} below), it is shorthand for the Segal--Sugawara composite built from the affine currents. The $JWW$ Jacobi identities uniquely fix the structure constants in terms of the level $k$, and the $WWW$ Jacobi identities then admit exactly two solutions for the level,
\begin{equation} \label{flavlevelbootstrap}
    k = -\frac{2N+3}{2} \qquad \mathrm{or} \qquad k = -\frac{4N+5}{6} = k^{\mathrm{anom}}_{N,0}~.
\end{equation}
The second is, as indicated, the level of the $\mathrm{AD}(\cf^{\rm anom}_N,0)$ VOA. The first satisfies $k < -h^\vee_{\sp(2N)}$ and, under the assumption that the stress tensor (if it exists) is Sugawara, is ruled out as the VOA of a four-dimensional unitary theory by graded unitarity \cite{Beem:2018duj}. We will return to this vertex algebra in the Remark below. For the solution $k = k^{\mathrm{anom}}_{N,0}$, the structure constants at general $N$ (imposing the normalisation $c_1=1$) are given by
\begin{equation}
\begin{alignedat}{2}
    &c_2 = \frac{12}{4N+5}, \qquad &&c_3 = \frac{12(2N+1)}{(N-1)(4N+5)},\\ 
    &c_4 = -\frac{72}{(N-1)(4N+5)}, \qquad &&c_5 = -\frac{9}{(N-1)(4N+5)}~. 
\end{alignedat}
\end{equation}
The constants $c_3$, $c_4$, and $c_5$ are ill-defined when $N = 1$, and indeed the bootstrap problem has no solution with $k = k^{\mathrm{anom}}_{1,0} = -\frac{3}{2}$. Instead here there is a one-parameter family of solutions with $k=-\frac{5}{2}$ given by
\begin{equation}
    c_2 = \frac{4}{5}, \qquad c_3 = \frac{6}{5} + 4 c_5, \qquad c_4 = -\frac{8}{5} - 8 c_5~.
\end{equation}
This can be understood because $\sp(2) \cong \sl(2)$ has only one irreducible representation in the antisymmetric tensor product of two fundamentals (the trivial representation). Consequently, the $c_4$ and $c_5$ terms appearing in the simple pole of the $WW$ OPE are redundant. For our purposes, the restriction $N > 1$ is not a limitation as the $N=1$, $m=0$ case is not admissible.

\medskip

\begin{remark}
    For the first solution of \eqref{flavlevelbootstrap}, the Sugawara vector $T_{\rm sug}$ has the usual Virasoro self-OPE, but the $L_{-1}$ mode of $T_{\rm sug}$ does not act as the derivative/translation operator on the additional generator $W_\alpha$, so this bootstrap solution is \emph{not} a vertex operator algebra, \emph{even though} it contains a Virasoro subalgebra. It can be understood concretely as the quantum Drinfel'd--Sokolov reduction of the critical-level affine Kac--Moody algebra $V_{-h^\vee_{\sp(2N+2)}}(\sp(2N+2))$ with respect to the minimal nilpotent orbit $\mathds{O}_{\mathrm{min}} \coloneqq \mathds{O}_{[2,1^{2N}]}$. Indeed, the $m=0$ case of the decomposition given in \eqref{adjdec} shows that the strong generators of this reduction are $\sp(2N)$ affine currents of weight $h = 1$ and generators $W_\alpha$ with weight $h=\frac32$ in the fundamental representation of $\sp(2N)$. There is no genuine stress tensor because the parent critical-level algebra has none. The level of the $\sp(2N)$ affine currents is determined to be exactly
\begin{equation}
    k_{DS} = -h^\vee_{\sp(2N+2)}+\frac{1}{2} = -(N+2)+\frac{1}{2} = - \frac{2N+3}{2}~,
\end{equation}
matching that of the first solution of the bootstrap problem in \eqref{flavlevelbootstrap}, thus identifying that solution with the minimal critical-level $\mathcal{W}$-algebra for $\sp(2N+2)$.
\end{remark}

\subsubsection{The $m = 1$ series}
For the $\mathrm{AD}(\cf^{\rm anom}_{N},1)$ series the module generator $W_\alpha$ has weight $h = 5/2$, and the most general $W\times W$ self-OPE compatible with $\sp(2N)$ covariance is given by (with the same notational conventions as above),
\begingroup
\allowdisplaybreaks
\begin{align}\label{eq:selfOPEmeq1}
    W_{\alpha}(z) W_{\beta}(w)&\sim \frac{c_1 \Omega_{\alpha\beta}}{(z-w)^5}+ \frac{c_2 J_{\alpha\beta}(w)}{(z-w)^4} + \frac{c_3 J'_{\alpha\beta}(w)- c_4 \Omega^{\gamma\delta} J_{\alpha\gamma} J_{\beta\delta}(w)+ c_5 \Omega_{\alpha\beta} T(w)}{(z-w)^3}\nonumber\\
    &+ \frac{1}{(z-w)^2} \left(c_6 J''_{\alpha\beta}(w) + c_7 \Omega^{\gamma\delta}J_{(\alpha|\gamma|} J'_{\beta)\delta}(w) + c_8 \Omega^{\gamma\delta} J_{[\alpha|\gamma|} J'_{\beta]\delta}(w) \right.\nonumber\\
    &+ \left. c_9 \Omega_{\alpha\beta} \Omega^{\gamma\epsilon} \Omega^{\delta\theta}J_{\gamma\delta} J'_{\epsilon\theta}(w) \right. + \left. c_{10}J_{\alpha\beta} T(w) + c_{11} \Omega^{\gamma\epsilon} \Omega^{\delta\theta}J_{(\alpha|\gamma|} J_{\beta)\delta} J_{\epsilon\theta}(w)\right)\nonumber\\
    &+ \frac{1}{(z-w)} \left(c_{12} J'''_{\alpha\beta}(w) + c_{13} \Omega^{\gamma\delta}J_{(\alpha|\gamma|} J''_{\beta)\delta}(w) + c_{14} \Omega^{\gamma\delta} J_{[\alpha|\gamma|} J''_{\beta]\delta}(w) \right.\nonumber\\
    & + c_{15} \Omega_{\alpha\beta}\Omega^{\gamma\epsilon} \Omega^{\delta\theta}J_{\gamma\delta} J''_{\epsilon\theta}(w) + c_{16} \Omega_{\alpha\beta} \Omega^{\gamma\epsilon} \Omega^{\delta\theta} J'_{\gamma\delta} J'_{\epsilon\theta}(w) - c_{17} \Omega^{\gamma\delta}J'_{[\alpha|\gamma|} J'_{\beta]\delta}(w)\nonumber\\
    &+ c_{18} J'_{\alpha\beta} T(w) + c_{19} J_{\alpha\beta} T'(w) + c_{20} \Omega^{\gamma\epsilon} \Omega^{\delta\theta}J_{(\alpha|\gamma|} J_{\beta)\delta} J'_{\epsilon\theta}(w)\nonumber\\
    &+ c_{21} \Omega^{\gamma\epsilon} \Omega^{\delta\theta}J_{\epsilon\theta} J_{(\alpha|\gamma|} J'_{\beta)\delta}(w) + c_{22} \Omega^{\gamma\epsilon} \Omega^{\delta\theta}J_{\epsilon\theta} J_{[\alpha|\gamma|} J'_{\beta]\delta}(w)\nonumber\\
    &+ c_{23} \Omega_{\alpha\beta}(\Omega^{\gamma \delta}\Omega^{\epsilon \theta} \Omega^{\rho_1 \eta_1}\Omega^{\rho_2 \eta_2})J_{\gamma \epsilon} J_{\delta \theta} J_{\rho_1 \rho_2} J_{\eta_1 \eta_2}(w)\,\nonumber\\
    &+ c_{24} \Omega^{\gamma\delta}J_{[\alpha|\gamma|} J_{\beta]\delta} T(w) + c_{25} \Omega^{\gamma\epsilon}\Omega^{\delta\eta}\Omega^{\theta\rho}J_{[\alpha|\gamma|} J_{\beta]\delta} J_{\epsilon\theta} J_{\eta\rho}(w)\nonumber\\
    &\left. + c_{26} \Omega_{\alpha\beta}\Omega^{\gamma\epsilon}\Omega^{\delta\eta_1}\Omega^{\theta\eta_2}\Omega^{\rho_1\rho_2}J_{\gamma\delta} J_{\epsilon\theta} J_{\eta_1\rho_1} J_{\eta_2\rho_2}(w)\right)~.
\end{align}
\endgroup
The cases of $N = 1$ and $N = 2$ are special due to the isomorphisms $\sp(2) \cong \sl(2)$ and $\sp(4) \cong \so(5)$, which render the bootstrap problem underdetermined as written; we discuss them separately below. For $N >2$, the $JWW$ Jacobi identities fix the structure constants in terms of the level $k$, and again the $WWW$ Jacobi identities admit exactly two solutions, now with levels
\begin{equation} \label{flavlevelbootstrapmeq1}
    k = -\frac{2N+3}{2} \qquad \mathrm{or} \qquad k = -\frac{8N+9}{10} = k^{\mathrm{anom}}_{N,1}~.
\end{equation}
As in the $m = 0$ case, the first case is ruled out by graded unitarity, while the second is precisely the level of the $\mathrm{AD}(\cf^{\rm anom}_{N},1)$ theory. For this level, the structure constants of the $WW$ OPE at general $N$ (with the normalisation $c_1 = 1$) are
\begin{alignat}{2}
    c_{2} &= \frac{20}{8N+9}~, &\qquad c_{3} &= \frac{40\,(2N+1)}{(3N-1)(8N+9)}~, \nn\\
    c_{4} &= c_{8} = \frac{200}{(3N-1)(8N+9)}~, &\qquad c_{5} &= -\frac{25}{(3N-1)(8N+9)}~, \nn\\
    c_{6} &= -\frac{10\,(2N+1)}{(4N+7)(8N+9)}~, &\qquad c_{7} &= -\frac{800\,(2N+1)}{D_N}~, \nn\\
    c_{9} &= \frac{500}{(2N+1)(3N-1)(8N+9)}~, &\qquad c_{10} &= -\frac{500}{D_N}~, \nn\\
    c_{11} &= c_{21} = -\frac{4000}{D_N}~, &\qquad c_{12} &= \frac{5\,(23N+19)}{6\,(3N-1)(8N+9)}~, \nn\\
    c_{13} &= \frac{200\,(N+3)}{D_N}~, &\qquad c_{14} &= -\frac{100\,(2N+1)(7N+6)}{(N-2)\,D_N}~, \\
    c_{15} &= -\frac{1250\,(N^2+2N+2)}{(N-2)(2N+1)\,D_N}~, &\qquad c_{16} &= \frac{125\,(28N^2+33N-3)}{(N-2)(2N+1)\,D_N}~, \nn\\
    c_{17} &= -\frac{200\,(2N+1)(13N+14)}{(N-2)\,D_N}~, &\qquad c_{18} &= c_{19} = -\frac{250}{D_N}~, \nn\\
    c_{20} &= -\frac{2000}{D_N}~, &\qquad c_{22} &= \frac{6000\,(2N+1)}{(N-2)\,D_N}~, \nn\\
    c_{23} &= \frac{125000}{(N-2)(2N+1)^2\,D_N}~, &\qquad c_{24} &= -\frac{5000}{(N-2)\,D_N}~, \nn\\
    c_{25} &= -\frac{40000}{(N-2)\,D_N}~, &\qquad c_{26} &= -\frac{50000}{(N-2)(2N+1)\,D_N}~,\nn
\end{alignat}
where $D_N \coloneqq (3N-1)(4N+7)(8N+9)$. These are well defined for all $N \in \mathbb{N}\setminus\{2\}$, so $k = k^{\rm anom}_{2,1}$ is not a solution of the bootstrap problem; since $k^{\rm anom}_{2,1}$ is not an admissible level, this is not a limitation of the results.

For $N = 1$, the bootstrap Ansatz has redundant parameters and reduces to the bootstrap problem for the rank-two $H_0$ theory solved in \cite{Beem:2019snk}. As first observed there, one in fact finds three allowed values for the level: $k = -\frac{5}{2}$ and $k = -\frac{17}{10}$ (the $N = 1$ specialisations of \eqref{flavlevelbootstrapmeq1}), and also $k = -\frac{7}{4}$ (a solution about which we have nothing to say here). The VOA of the $\mathrm{AD}(\cf^{\rm anom}_{1},1)$ SCFT is then precisely that of the rank-two $H_0$ theory (and indeed $\mathrm{AD}(\cf^{\rm anom}_{1},1)$ should be identified with the rank-two $H_0$ theory itself). 

When $N = 2$, the isomorphism $\mathfrak{sp}(4) \cong \mathfrak{so}(5)$ implies a single linear relation amongst the zero-derivative terms in the simple pole of the $WW$ OPE. Consequently, the bootstrap Ansatz has a free parameter and solving the Jacobi identities leads to a one-parameter family of solutions for the unique level $k = -\frac{7}{2}$ (the $N = 2$ specialisation of the first solution in \eqref{flavlevelbootstrapmeq1}). 

The uniform structure of the first bootstrap solution appearing in \eqref{flavlevelbootstrap} and \eqref{flavlevelbootstrapmeq1} suggests that $k = - \frac{2N+3}{2}$ will solve the corresponding bootstrap problem for any $m \geqslant 0$, and that the corresponding vertex algebra can be understood as a critical-level $\mathcal{W}$-algebra. Generalising the Remark above, we propose the following.

\medskip

\begin{conjecture} \label{conj:bootstrapsolngenm}
    The quantum Drinfel'd--Sokolov reduction of the critical-level affine Kac--Moody algebra $V_{-h^\vee}(\sp(2N+2m+2))$ with respect to the nilpotent orbit $\mathds{O}_{\mathrm{nil}} = \mathds{O}_{[2+2m, 1^{2N}]}$ is a vertex algebra (but not a vertex operator algebra) strongly generated by $\sp(2N)$ affine currents at level $k=-\frac{2N+3}{2}$ and additional strong generators in the fundamental of $\sp(2N)$ with $h=\frac{3+2m}{2}$.
\end{conjecture}

\medskip

For this to hold beyond $m=0$, it must be the case that certain generators of the corresponding universal $\mathcal{W}$-algebra are absent from the simple quotient of the critical-level algebra. Concretely, the strong generators of the universal algebra are $\sp(2N)$ affine currents at level $k=-\frac{2N+3}{2}$, the fundamental generators $W_\alpha$ of weight $\frac{3+2m}{2}$, and additional singlet generators of weight $h=2,4,\ldots,2m+2$. It seems very plausible that the extra singlet generators are all null (as they should descend from generators of the Feigin--Frenkel centre of the parent critical-level affine algebra, which are set to zero in the simple quotient), though we have not attempted to prove this. It is an amusing fact that if the above Conjecture holds, then it means that these algebras \emph{are} realised as associated vertex algebras living on codimension-four defects in the six-dimensional $(2,0)$ theory \cite{Beem:2014kka}.

We now turn to some equivalent realisations of the $\mathrm{AD}(\cf^{\rm anom}_{N},m)$ SCFTs within the landscape of generalised Argyres--Douglas SCFTs with $B/C/D$-type flavour symmetries, which lead to a number of simplifications of that landscape (see \cite{Beem:2023ofp} for similar simplifications in type $A$).

\subsection{Equivalent realisations}

Conjecture \ref{conj:anom-higgsing} realises the anomalous $\mathrm{AD}(\cf^{\rm anom}_N,m)$ theory as a nilpotent Higgsing of the non-anomalous $\mathrm{AD}(\mathfrak{c}_{N+m+1},m)$ theory and, correspondingly, realises its VOA as the quantum Drinfel'd--Sokolov reduction \eqref{eq:twA2Nrealis}. This turns out to be one of several such realisations. In this subsection we describe these other realisations and outline how they fit together. 

Two further ``class I'' families of generalised Argyres--Douglas theories enter the picture: those of untwisted type $D$ and those of twisted type $A_{\rm odd}$, which we denote by $\mathrm{AD}(\mathfrak{d}_{N},m)$ and $\mathrm{AD}(\mathfrak{b}_N, m)$, respectively. Their associated VOAs have been conjectured to be the affine Kac--Moody VOAs $V_{-h^\vee+\frac{h^\vee}{q}}(\so(2N))$ and $V_{-h^\vee + \frac{h^\vee}{q}}(\so(2N+1))$ \cite{Xie:2016evu,Wang:2018gvb}, where, as before, we write $q=3+2m$. 

The proposed relations between these various theories are summarised by the diagram of vertex algebras in Figure \ref{fig:W-algebra-web}, where each arrow corresponds to quantum Drinfel'd--Sokolov reduction with respect to the displayed nilpotent orbit and all dual Coxeter numbers $h^\vee$ and Coxeter numbers $h$ refer to the displayed Lie algebra:

\input{figures/Equivalences_VOA}

The two reductions departing from the untwisted $D$-type VOA and arriving at the boundary admissible $B$- and $C$-type VOAs are both collapsing level isomorphisms that have been established previously in the VOA literature, though the latter does not appear to be widely known. It is proved in the thesis \cite{Riedler2019} and reads
\begin{equation} \label{eq:riedleriso}
    \mathcal{W}_{-h^\vee+\frac{h^\vee}{q}}\left(\so(2N(q-1))~, \mathds{O}_{[(q-1)^{2N}]}\right) \cong V_{-h^\vee+\frac{h^\vee}{q}}(\sp(2N))~.
\end{equation}
In particular, this isomorphism implies a correction to \cite[Conjecture 9.20]{Arakawa:2021ogm}. On the other hand, the upper-left arrow from $D$-type to $B$-type is the $\so$-type collapsing level isomorphism proved in \cite{Arakawa:2021ogm},
\begin{equation} \label{eq:collapsingso}
    \mathcal{W}_{-h^\vee+\frac{h^\vee}{q}}\left(\so(nq+2N+1), \mathds{O}_{[q^n,1^{2N+1}]}\right) \cong V_{-h^\vee+\frac{h^\vee}{q}}(\so(2N+1))~,
\end{equation}
which holds for $n$ and $q$ odd, specialised to $n = q-2$.

The two additional arrows arriving at the VOA of the anomalous theory are, to the best of our knowledge, new. Both can be established by the methods of \cite{Arakawa:2021ogm}, with orthogonal pyramids playing the role of the symplectic pyramids used there. The upper right arrow is the isomorphism
\begin{equation}\label{eq:AeveninAodd}
    \mathcal{W}_{-h^\vee+\frac{h^\vee}{q}}\left(\so(2N(q-1)+1),\mathds{O}_{[(q-1)^{2N},1]}\right) ~\cong~
    \begin{array}{c}V_{-h^\vee+\frac{h+1}{2q}}(\sp(2N))\\ \oplus \\ L_{-h^\vee+\frac{h+1}{2q}}(\sp(2N); \varpi_1)\end{array}~,
\end{equation}
which realises the VOA of the anomalous theory as a reduction of the VOA of the twisted $A_{\rm odd}$ theory $\mathrm{AD}(\mathfrak{b}_{N(2m+2)},m)$. For $N=2$ and $m=0$ this reduces to a variant of the $\mathcal{W}$-algebra isomorphism observed in \cite{Li:2022njl}, but the general case does not seem to have previously appeared in the literature. 

\input{figures/Equivalences_AD}

The horizontal arrow encodes the realisation of that same VOA as a reduction of the untwisted $D$-type VOA in a single step. The relevant orbit can be read off straightforwardly from the top route around the diagram. The first reduction along that route, with respect to $[q^{q-2},1^{2N(q-1)+1}]$, leaves the $\so(2N(q-1)+1)$ current algebra, and the second acts within it, replacing the parts $1^{2N(q-1)+1}$ by $[(q-1)^{2N},1]$; the composite is therefore the reduction with respect to $\mathds{O}_\lambda$ where $\lambda = [q^{q-2}, (q-1)^{2N},1]$, and the corresponding isomorphism,
\begin{equation}\label{eq:oneshotiso}
    \mathcal{W}_{-h^\vee+\frac{h^\vee}{q}}\left(\so((2N+q-1)(q-1)),\mathds{O}_{[q^{q-2}, (q-1)^{2N},1]}\right) ~\cong~\begin{array}{c}V_{-h^\vee+\frac{h+1}{2q}}(\sp(2N))\\ \oplus \\ L_{-h^\vee+\frac{h+1}{2q}}(\sp(2N); \varpi_1)\end{array}~,
\end{equation}
can be proved in the same way as \eqref{eq:AeveninAodd}.\footnote{The identification of the sequential Drinfel'd--Sokolov reduction along the top or bottom edges of Figure \ref{fig:W-algebra-web} with the horizontal arrow is a type of \emph{quantum Hamiltonian reduction in stages} \cite{Genra:2025vfm}. The particular identifications described here are outside of the cases covered by the results proved \emph{loc. cit.} with the exception of the case $q=3$. Furthermore, reduction in stages is proved for universal $\mathcal{W}$-algebras rather than their simple quotients \emph{loc. cit.}.}

The identifications in Figure \ref{fig:W-algebra-web} can be understood directly at the level of associated varieties/Higgs branches. The associated variety of the quantum Drinfel'd--Sokolov reduction of an affine Kac--Moody VOA with respect to a nilpotent orbit $\mathds{O}_\rho$ is the intersection of the associated variety of the affine VOA with the Slodowy slice to $\mathds{O}_\rho$ \cite{Arakawa:2010ni}. The associated varieties of the $\mathcal{W}$-algebras in the figure are therefore Slodowy intersections $\cS_{\sigma,\rho} \coloneqq \overline{\mathds{O}}_{\sigma}\cap\cS_{\rho}$, where $\overline{\mathds{O}}_{\sigma}$ is the associated variety of the parent affine VOA. As all parent levels in the diagram are boundary admissible with denominator $q$, in each case $\sigma$ is the largest partition of the relevant type with no part greater than $q$ (see Appendix \ref{app:KM_admissible}). We will denote these partitions by $\sigma_D$, $\sigma_B$, and $\sigma_C$ for the $\so({\rm even})$, $\so({\rm odd})$, and $\sp$ cases, respectively, with the dependence on $q$ being left implicit. Geometrically, the arrows of Figure \ref{fig:W-algebra-web} thus correspond to the following isomorphisms of Slodowy intersections,
\begin{equation}\label{eq:slodowy-web}
\begin{split}
    \cS^{\so((2N+q-1)(q-1))}_{\sigma_D,\,[q^{q-2},1^{2N(q-1)+1}]} &\cong \overline{\mathds{O}}_{\sigma_B}~,\\
    \cS^{\so((2N+q-1)(q-1))}_{\sigma_D,\,[(q-1)^{2N+q-1}]} &\cong \overline{\mathds{O}}_{\sigma_C}~,\\
    \cS^{\so((2N+q-1)(q-1))}_{\sigma_D,\,[q^{q-2},(q-1)^{2N},1]} &\cong \cS^{\so(2N(q-1)+1)}_{\sigma_B,\,[(q-1)^{2N},1]} \cong \cS^{\sp(2N+q-1)}_{\sigma_C,\,[q-1,1^{2N}]}~.
\end{split}
\end{equation}
The last of these presents the Higgs branch \eqref{anomhiggs} of the anomalous theory in three ways.

All of these isomorphisms follow from classical facts about Slodowy intersections. The first is the row removal rule of Kraft and Procesi \cite{kraft1982geometry}, which states that if the partitions $\sigma$ and $\rho$ share their first $r$ parts, then removing those parts from both leaves the corresponding Slodowy intersection unchanged. Since $\sigma_B$ is obtained from $\sigma_D$ by removing $q-2$ parts of length $q$, removing these rows accounts for the first and third isomorphisms in \eqref{eq:slodowy-web}. The second fact is a rectangular symmetry relating Slodowy intersections of symplectic and orthogonal type, established in \cite{li2018quiver}. If we interpret the parts of a partition $\rho$ of $2n$ as the column lengths of a Young diagram placed in a rectangle with $q$ rows and $2n$ columns, and writing $\rho^\vee$ for the column lengths of the complementary diagram (a partition of $2n(q-1)$), then one has 
\begin{equation}
    \cS^{\sp(2n)}_{\sigma_C,\rho}\cong\cS^{\so(2n(q-1))}_{\sigma_D,\rho^\vee}~.
\end{equation}
For the case $2n = 2N+q-1$ and $\rho = [1^{2N+q-1}]$, the complement is $\rho^\vee=[(q-1)^{2N+q-1}]$, which gives the second isomorphism in \eqref{eq:slodowy-web}. For $\rho = [q-1,1^{2N}]$ the complement is $\rho^\vee = [q^{q-2},(q-1)^{2N},1]$ giving the final isomorphism.

We expect the vertex algebra/Slodowy-intersection identifications encoded in Figure \ref{fig:W-algebra-web} to be upgraded to analogous identifications of generalised Argyres--Douglas theories related by nilpotent Higgsing with respect to the same orbits, giving the web of Higgs branch flows among ``class I'' theories of types $B$, $C$, and $D$ shown in Figure \ref{fig:equiv}. In particular, we propose that the ``class I'' twisted $A_{\rm odd}$, twisted $A_{\rm even}$, and twisted $D$ Argyres--Douglas SCFTs all admit alternative realisations as nilpotent Higgsings of parent ``class I'' untwisted $D$-type Argyres--Douglas SCFTs. 

Another instance of such an equivalence involves a particular ``class I'' twisted $E_6$ Argyres--Douglas SCFT. The twisted $E_6$ Argyres--Douglas SCFT has a flavour symmetry of $\ff_4$ with a principal boundary-admissible flavour level \cite{Wang:2018gvb}. Analogous to the previous conventions, we denote this theory by $\mathrm{AD}(\mathfrak{f}_4,m)$. Largely based on the Sugawara relation for the central charge, its associated VOA has been conjectured to be $V_{-9+\frac{9}{3+2m}}(\ff_4)$ for $m \geqslant 0$ (and $m\neq 0\pmod{3}$ for admissibility). This, combined with the following $\cW$-algebra isomorphism established in \cite{Arakawa:2021ogm},
\begin{equation}
    \mathcal{W}_{-9+\frac{9}{7}}(\ff_4, C_3) \cong \mathcal{W}_{-9+\frac{9}{7}}(\ff_4, B_3) \cong V_{-2 + \frac{2}{7}}(\cf_1)~,
\end{equation}
suggests that the non-anomalous $\mathrm{AD}(\cf_{1},2)$ SCFT admits an alternative realisation as a specific nilpotent Higgsing of the ``class I'' twisted $E_6$ theory $\mathrm{AD}(\mathfrak{f}_4,2)$. 

These identifications lead to a considerable consolidation of the landscape of Argyres--Douglas SCFTs with non-simply laced flavour symmetry. A mathematical interpretation of similar dualities among type $A$ Argyres--Douglas SCFTs at the level of Hitchin systems was recently discussed in \cite{Doucot:2026brk}, and it would be interesting to know whether a comparable analysis can be generalised to the type $B/C/D$ dualities proposed here.

%% file: figures/Equivalences_VOA.tex

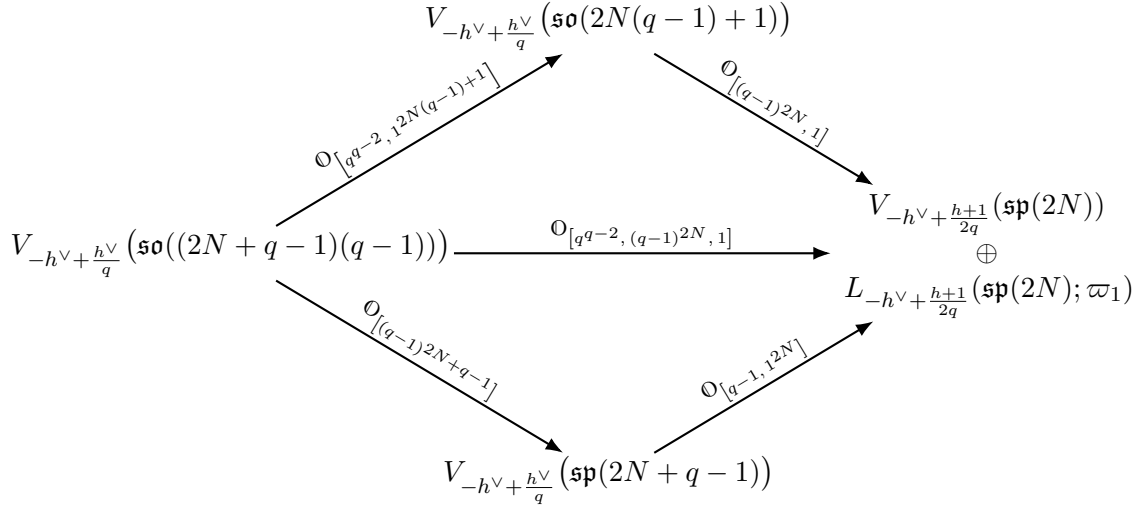
\begin{figure} 
\centering
\begin{tikzpicture}[>=Latex, every node/.style={inner sep=1pt}]

\node (L) at (-5,0) {$V_{-h^\vee+\frac{h^\vee}{q}}\bigl(\so((2N+q-1)(q-1))\bigr)$};
\node (T) at (0,3.0) {$V_{-h^\vee+\frac{h^\vee}{q}}\bigl(\so(2N(q-1)+1)\bigr)$};
\node (R) at (5,0) {$\begin{array}{c} V_{-h^\vee+\frac{h+1}{2q}}(\sp(2N)) \\[2pt]
                              {}\oplus\\{}\, L_{-h^\vee+\frac{h+1}{2q}}(\sp(2N);\varpi_1) \end{array}$};
\node (B) at (0,-3.0) {$V_{-h^\vee+\frac{h^\vee}{q}}\bigl(\sp(2N+q-1)\bigr)$};

\draw[->,thick]
  (L) -- node[above,sloped]
  {$\scriptstyle\mathds{O}_{\left[(q-1)^{2N+q-1}\right]}$} (B);

\draw[->,thick]
  (L) -- node[above,sloped]
  {$\scriptstyle\mathds{O}_{\left[q^{q-2},\,1^{2N(q-1)+1}\right]}$} (T);

\draw[->,thick]
  (L) -- node[above]
  {$\scriptstyle\mathds{O}_{\left[q^{q-2},\,(q-1)^{2N},\,1\right]}$} (R);

\draw[->,thick]
  (T) -- node[above,sloped]
  {$\scriptstyle\mathds{O}_{\left[(q-1)^{2N},\,1\right]}$} (R);

\draw[->,thick]
  (B) -- node[above, sloped]
  {$\scriptstyle\mathds{O}_{\left[q-1,\,1^{2N}\right]}$} (R);

\end{tikzpicture}
\caption{\label{fig:W-algebra-web}$\mathcal{W}$-algebra isomorphisms giving a variety of realisations of the VOA of the anomalous Argyres--Douglas family in terms of Drinfel'd--Sokolov reductions of current algebras in types $B$, $C$, and $D$.}
\end{figure}

%% file: figures/Equivalences_AD.tex

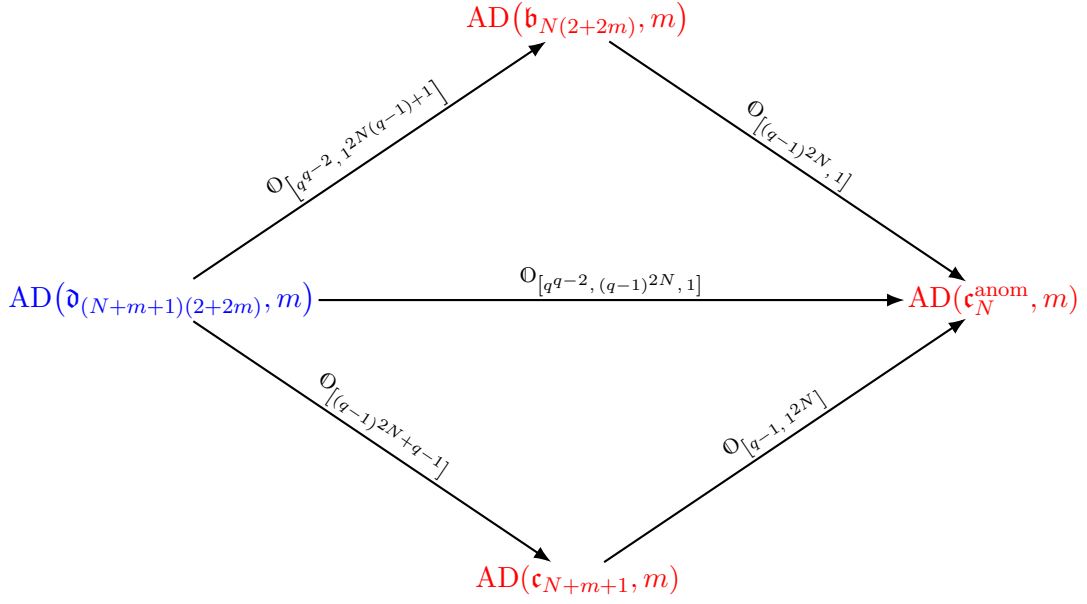
\begin{figure} 
\centering
\begin{tikzpicture}[>=Latex, every node/.style={inner sep=1pt}]

\node (L) at (-5.5,0) {\textcolor{blue}{$\mathrm{AD}\!\left(\df_{(N+m+1)(2+2m)},m\right)$}};
\node (T) at (0,3.7) {\textcolor{red}{$\mathrm{AD}\!\left(\bff_{N(2+2m)},m\right)$}};
\node (R) at (5.5,0) {\textcolor{red}{$\mathrm{AD}\!\left(\cf^{\rm anom}_{N},m\right)$}};
\node (B) at (0,-3.7) {\textcolor{red}{$\mathrm{AD}\!\left(\cf_{N+m+1},m\right)$}};

\draw[->,thick]
  (L) -- node[above,sloped]
  {$\scriptstyle\mathds{O}_{\left[(q-1)^{2N+q-1}\right]}$} (B);

\draw[->,thick]
  (L) -- node[above,sloped]
  {$\scriptstyle\mathds{O}_{\left[q^{q-2},\,1^{2N(q-1)+1}\right]}$} (T);

\draw[->,thick]
  (L) -- node[above]
  {$\scriptstyle\mathds{O}_{\left[q^{q-2},\,(q-1)^{2N},\,1\right]}$} (R);

\draw[->,thick]
  (T) -- node[above,sloped]
  {$\scriptstyle\mathds{O}_{\left[(q-1)^{2N},\,1\right]}$} (R);

\draw[->,thick]
  (B) -- node[above, sloped]
  {$\scriptstyle\mathds{O}_{\left[q-1,\,1^{2N}\right]}$} (R);

\end{tikzpicture}
\caption{\label{fig:equiv} Equivalences in the type $B/C/D$ landscape of generalised Argyres--Douglas SCFTs. The theories with a twisted class $\cS$ construction and non-simply laced flavour symmetry (represented in red) have alternative realisations as specific nilpotent Higgsings of the untwisted $D$-type Argyres--Douglas SCFTs (represented in blue).}
\end{figure}

%% file: sections/sec_5_higgs_branches.tex

\section{Higgs branches of the \texorpdfstring{$\mathrm{AD}(\cf^{\rm anom}_{N},m)$}{AD(c-anom,N,m} theories} \label{sec:anomalHiggs}

In this section we study the Higgs branches of the anomalous $\mathrm{AD}(\cf^{\rm anom}_{N},m)$ theories. We take the Higgs branch to be the associated variety of the full VOA $\mathbb{V}[\mathrm{AD}(\cf^{\rm anom}_{N},m)]$, and our main consideration will be how it is related to the associated variety of the affine current subalgebra $V_{k^{\rm anom}_{N,m}}(\sp(2N))$, which is a nilpotent orbit closure in $\sp(2N)$. For the non-anomalous theories the two coincide, the VOA being just the current algebra itself, but for the anomalous theories the finite extension can, and does, affect the geometry. For clarity, throughout this section we refer to the associated variety of the full VOA as the Higgs branch, and to $X_{V_{k^{\rm anom}_{N,m}}(\sp(2N))}$ as the corresponding orbit closure (or just orbit closure). Of the multiple realisations of the anomalous theories given in the previous section, we will use that of Conjecture \ref{conj:anom-higgsing}. Our main result is the following.

\smallskip

\begin{prop}\label{prop:HBmain}
The Higgs branch of the anomalous $\mathrm{AD}(\cf^{\rm anom}_{N},m)$ theory is isomorphic to the associated variety $X_{V_{k^{\rm anom}_{N,m}}(\sp(2N))}$ of its affine current subalgebra when $N\leqslant m+1$, and is a ramified double cover of $X_{V_{k^{\rm anom}_{N,m}}(\sp(2N))}$ when $N\geqslant m+2$.
\end{prop}

\smallskip

\noindent A visual representation of the isomorphism/double-cover regimes for small values of $(N,m)$ is provided in Figure \ref{GenNmdiag}. We establish the above result in Section \ref{subsec:HBgeomproof} using classical results for Slodowy intersections as well as the description of Slodowy intersections as orthosymplectic quiver varieties \cite{Hanany:2019tji}. In Section \ref{ramifHB}, we investigate the ramification structure of the double cover and illustrate the results in terms of Hasse diagrams of the two varieties in a number of examples. Finally, in Section \ref{subsec:HBfromVOA}, we reconsider the whole picture from the point of view of the associated vertex algebra, with respect to which the double-cover structure reveals itself as a doubling of the vacuum module under certain Drinfel'd--Sokolov reductions. We further discuss the low-energy effective theory at generic points of the Higgs branch in certain families of examples.

\input{figures/Grid}

\subsection{\label{subsec:HBgeomproof}The Higgs branch \emph{vis \`{a} vis} the corresponding orbit closure}

The proof of Proposition \ref{prop:HBmain} involves some details of the partitions labelling the associated varieties of $\sp(2N)$ affine Kac--Moody algebras at principal and coprincipal admissible levels, which we first summarise. We recall from \eqref{anomhiggs} that the Higgs branch is the Slodowy intersection
\begin{equation}\label{eq:anom_higgs_again}
    \cM_H(\mathrm{AD}(\cf^{\rm anom}_{N},m)) \cong X_{V_{k_{N+m+1,m}}(\sp(2N+q-1))} \cap \cS_{[q-1,1^{2N}]}~,
\end{equation}
where $k_{N+m+1,m}$ is the principal boundary admissible level with denominator $q=3+2m$. The associated variety of the parent current algebra appearing here is a nilpotent orbit closure,
\begin{equation}\label{nonanXv}
    X_{V_{k_{N+m+1,m}}(\sp(2N+q-1))}~\cong~\overline{\mathds{O}^{\sp(2N+q-1)}_q}~,
\end{equation}
where $\mathds{O}^{\sp(2N+q-1)}_q$ is the nilpotent orbit of $\sp(2N+q-1)$ corresponding to a partition of the form
\begin{align} \label{Oq}
    \mathds{O}_q =
    \begin{cases}
        [q^n, s]&  0\leqslant s \leqslant q-1~,\qquad n,s\ \mathrm{even}~,\\
        [q^n, q-1, s]&  2 \leqslant s \leqslant q-1~,\qquad n,s\ \mathrm{even}~.
    \end{cases}
\end{align}
Note that this is never the principal nilpotent orbit. The associated variety of the coprincipal admissible current subalgebra $V_{k^{\rm anom}_{N,m}}(\sp(2N))$ is instead
\begin{align} \label{anomXv}
    X_{V_{k^{\rm anom}_{N,m}}(\sp(2N))}  =
    \begin{cases}
    \overline{\mathds{O}^{\sp(2N)}_{\mathrm{prin}}}~,&\qquad  2N \leqslant q+1~,\\
    \overline{{}^L\mathds{O}^{\sp(2N)}_q}~,&\qquad 2N \geqslant q+1~,
    \end{cases}
\end{align}
where $\mathds{O}^{\sp(2N)}_{\mathrm{prin}}$ is the principal nilpotent orbit of $\sp(2N)$ and $^L\mathds{O}^{\sp(2N)}_q$ is the nilpotent orbit of $\sp(2N)$ corresponding to a partition of the form
\begin{align}\label{LOq}
    ^L\mathds{O}_q =
    \begin{cases}
        [q+1, q^{n}, s]& \quad 0\leqslant s \leqslant q-1~, \qquad n, s \ \mathrm{even}~,\\
        [q+1, q^{n}, q-1, s]& \quad 2 \leqslant s \leqslant q-1~, \qquad n, s \ \mathrm{even}~.
    \end{cases}
\end{align}
For $2N=q+1$ the two cases in \eqref{anomXv} coincide, since then $^L\mathds{O}_q=[q+1]=[2N]$ is the principal orbit.

\paragraph{The case $N\leqslant m+1$ ($2N\leqslant q-1$).} In this regime, the associated variety of the parent current algebra is labelled by a partition of $2N+q-1\leqslant 2q-2$, which contains no part equal to $q$ (as $q$ is odd, such parts would be in pairs), so that $\mathds{O}_q=[q-1,2N]$. The Higgs branch is then
\begin{equation} \label{eq:HBNleqmp1}
\begin{split}
     \mathcal{M}_H(\mathrm{AD}(\cf^{\mathrm{anom}}_{N},m)) &\cong \overline{\mathds{O}}_{[q-1,2N]} \cap \mathcal{S}_{[q-1,1^{2N}]} \\
     &\cong \overline{\mathds{O}}_{[2N]}~.
\end{split}
\end{equation}
The second line follows from the row removal rule of Kraft and Procesi \cite{kraft1982geometry}: the partitions $[q-1,2N]$ and $[q-1,1^{2N}]$ share their first part, and removing it leaves the full nilpotent cone of $\sp(2N)$. On the other hand, when $2N\leqslant q-1$ the orbit closure $X_{V_{k^{\rm anom}_{N,m}}(\sp(2N))}$ is the full nilpotent cone by \eqref{anomXv}, and we arrive at the first part of Proposition \ref{prop:HBmain}.

\paragraph{The case $N\geqslant m+2$ ($2N\geqslant q+1$).} In this regime, the form of the partition labelling $X_{V_{k^{\rm anom}_{N,m}}(\sp(2N))}$ is correlated with that labelling the associated variety of the parent current algebra. Comparing \eqref{Oq} and \eqref{LOq} for partitions of $2N+q-1$ and $2N$ respectively, one finds\footnote{The details of the computation have been presented in Appendix \ref{app:HBramifdetails}.}
\begin{equation}\label{partitionmatch}
\begin{alignedat}{3}
    &X_{V_{k^{\rm anom}_{N,m}}(\sp(2N))} = \overline{\mathds{O}}_{[q+1, q^n, s]} &&\quad\Longleftrightarrow\quad &&X_{V_{k_{N+m+1,m}}(\sp(2N+q-1))} = \overline{\mathds{O}}_{[q^{n+2}, s]} \\
    &X_{V_{k^{\rm anom}_{N,m}}(\sp(2N))} = \overline{\mathds{O}}_{[q+1, q^n, q-1, s]} &&\quad\Longleftrightarrow\quad &&X_{V_{k_{N+m+1,m}}(\sp(2N+q-1))} = \overline{\mathds{O}}_{[q^{n+2}, q-1, s]}~,
\end{alignedat}
\end{equation}
where $n$ and $s$ are even. In short, the partition labelling the associated variety of the parent current algebra is obtained from that of $X_{V_{k^{\rm anom}_{N,m}}(\sp(2N))}$ by replacing the part $q+1$ with two parts equal to $q$.

Both the Higgs branch and the corresponding orbit closure can be realised as linear orthosymplectic quiver varieties, the form of which is determined by the quiver subtraction algorithm of \cite{Hanany:2019tji}.\footnote{We thank Julius Grimminger for bringing this to our attention and for helpful discussions on the topic.} The algorithm is reviewed and the computation carried out in Appendix \ref{app:HBramifdetails}; here we simply record the outcome. The two cases in \eqref{partitionmatch} proceed in exact analogy with one another, so we specialise to the first here, for which the Higgs branch is the Slodowy intersection $\cS_{\sigma,\rho}=\overline{\mathds{O}}_\sigma\cap\cS_\rho$ with $\sigma=[q^{n+2},s]$ and $\rho=[q-1,1^{2N}]$. The labels $N_i$ of the gauge nodes of the relevant orthosymplectic quiver are then given by
\begin{equation} \label{gaugeHiggs}
    N_i \equiv
    \begin{cases}
        2N -i(n+2) & \quad 1\leqslant i\leqslant s~, \\
        2N -i(n+1) - s & \quad s+1 \leqslant i \leqslant q-1~,
    \end{cases}
\end{equation}
so there are $q-1$ alternating $B/D$ and $C$ gauge nodes, while the labels of the flavour nodes are
\begin{equation}
    \mathbf{N}_{f_b} = \{\underbrace{2N, 0, 0, \ldots, \textcolor{red}{1}}_{q-1}\}~.
\end{equation}
\input{figures/Quiver_AD}
The quiver describing $\cM_H(\mathrm{AD}(\cf^{\rm anom}_{N},m))$ is therefore the linear orthosymplectic quiver of Figure \ref{QuivGenAD}, with a $C_N\cong{\rm Sp}(2N)$ flavour node attached to the first gauge node and a $B_0\cong{\rm O}(1)\cong\mathbb{Z}_2$ flavour node attached to the last.

By \eqref{partitionmatch}, the corresponding orbit closure $X_{V_{k^{\rm anom}_{N,m}}(\sp(2N))}$ is in this case $\overline{\mathds{O}}_{[q+1,q^n,s]}$, and the gauge nodes of the corresponding quiver are
\begin{equation} \label{gaugeAssoc}
    N_i =
    \begin{cases}
        2N -i(n+2) &\qquad 1\leqslant i\leqslant s~, \\
        2N -i(n+1) - s &\qquad s+1 \leqslant i \leqslant q-1~, \\
        \textcolor{red}{1} &\qquad\textcolor{red}{i = q}~,
    \end{cases}
\end{equation}
with flavour nodes
\begin{equation}
    \mathbf{N}_{f_b} = \{\underbrace{2N, 0, 0, \ldots, 0}_{q}\}~.
\end{equation}
This quiver, shown in Figure \ref{QuivGenCN}, has $q$ gauge nodes, the last of which is a $B_0$ node, and a single $C_N$ flavour node.

\input{figures/Quiver_CN}

The two quivers differ only in that the $B_0$ flavour node of the Higgs branch quiver is replaced with a $B_0$ gauge node in the corresponding orbit closure. In other words, the Higgs branch $\cM_H(\mathrm{AD}(\cf^{\rm anom}_{N},m))$ enjoys an additional $\mathbb{Z}_2$ flavour symmetry, and quotienting by this symmetry produces the associated variety of the current subalgebra,
\begin{equation} \label{doublecoverHB}
\fbox{$\begin{aligned}
    X_{V_{k^{\rm anom}_{N,m}}(\sp(2N))} = \frac{\mathcal{M}_H(\mathrm{AD}(\cf^{\mathrm{anom}}_{N},m))}{\mathbb{Z}_2}~.
    \end{aligned}$}
\end{equation}
Consequently, for $N\geqslant m+2$ the Higgs branch is a double cover of the corresponding orbit closure, and the argument for Proposition \ref{prop:HBmain} is complete. Note that \eqref{doublecoverHB} is the geometric counterpart, at the level of associated varieties, of the fact that there is a $\mathbb{Z}_2$ automorphism of the associated VOA $\mathbb{V}[\mathrm{AD}(\cf^{\rm anom}_{N},m)]$ that acts as minus one on the extending module and as the identity on the affine Kac--Moody subalgebra $V_{k^{\rm anom}_{N,m}}(\sp(2N))$, so the fixed point sub-VOA is just the current algebra.

\subsection{\label{ramifHB}Ramification structure of the double cover}

Proposition \ref{prop:HBmain} can be refined in the regime $N\geqslant m+2$ to determine the precise ramification structure of the double cover. In particular, since the $\mathbb{Z}_2$ action commutes with $\sp(2N)$, the cover is $\sp(2N)$-equivariant and we can ask unambiguously at the level of $\sp(2N)$ orbits in the corresponding orbit closure whether the preimage in the Higgs branch is one or two copies of the orbit.

The answer is as follows: the cover is unramified (two copies in the preimage) over orbits whose partition contains a part equal to $q+1$, and is ramified (one copy in the preimage) over the orbits all of whose parts are at most $q$. We will first derive this result by a direct analysis of the geometry of the relevant Slodowy intersection; an illuminating re-derivation from vertex algebra considerations will be presented in Section \ref{subsubsec:vac_count}.

Let $U\cong\mathbb{C}^{2N+q-1}$ be the defining representation of $\sp(2N+q-1)$ and let $(e,h,f)$ be an $\sl(2)$-triple corresponding to the partition $[q-1,1^{2N}]$. With respect to the action of this $\sl(2)$, we can decompose $U=V\oplus W$, where $V\cong\mathbb{C}^{q-1}$ is an irreducible representation on which $e$ acts as a single Jordan block, and $W\cong\mathbb{C}^{2N}$ is invariant. The centraliser of the triple in ${\rm Sp}(U)$ is then ${\rm O}(1)\times{\rm Sp}(W)$, where the ${\rm O}(1)\cong\mathbb{Z}_2$ acts by the involution $\tau=(-\mathbbm{1}_V)\oplus \mathbbm{1}_W$.

The Slodowy slice $\mathcal{S}_{[q-1,1^{2N}]}$ (before intersecting with any nilpotent orbit closure) is by definition given by
\begin{equation}
    \mathcal{S}_{e} = e + \sp(U)^f~,
\end{equation}
where $\sp(U)^f$ is the centraliser of $f$. From the decomposition of $U$ into $V$ and $W$ we have the decomposition
\begin{equation}
    \sp(U) \cong \sp(V)\oplus \sp(W)\oplus \left(V\otimes W\right)
\end{equation}
as a $\sp(V)\times\sp(W)$ module. The centraliser $\sp(U)^f$ of $f$ decomposes accordingly as
\begin{equation}
    \sp(U)^f=\sp(V)^f\oplus\sp(W)\oplus W~,
\end{equation}
where $W$ is embedded in $V\otimes W$ as the space of $\sl(2)$ lowest-weight vectors. Consequently, an arbitrary point $x$ in the Slodowy slice can be written as
\begin{equation}\label{eq:slicepoint}
    x = e + \nu + \omega + w~,\qquad \nu\in\sp(V)^f~,\quad \omega\in \sp(W)~,\quad w\in W~.
\end{equation}
The group ${\rm Sp}(W)={\rm Sp}(2N)$ acts on $\omega$ and $w$ in the adjoint and fundamental representations, respectively, and trivially on $\nu$. The involution $\tau$ acts as $w\mapsto -w$ and trivially on $\nu$ and $\omega$. Thus the $\mathbb{Z}_2$ fixed points are precisely those for which $w=0$, which are points of the form $x = e + \nu + \omega$. Furthermore, since $w=0$, $e$ is a single Jordan block, and $\nu,\omega$ are valued in $\sp(V)^f$ and $\sp(W)$ respectively, the points $x$ are block diagonal.

To recover the Higgs branch, we further intersect with the relevant nilpotent orbit closure in \eqref{partitionmatch}, which means restricting to points which are nilpotent in $\sp(U)$ and for which no Jordan block is of size greater than $q$. Nilpotency requires that $\nu=0$,\footnote{Note that $x$ is block diagonal and $e+\nu$ is nilpotent in $\sp(U)$ while $\omega$ is nilpotent in $\sp(W)$. But $e+\nu$ lies in the Slodowy slice to $e$ in $\sp(V)$ and $e$ is regular nilpotent in $\sp(V)$ so this Slodowy slice only intersects the relevant nilpotent orbit closure in \eqref{partitionmatch} at $e$, leading to the desired result.} so we see that the fixed locus is exactly the nilpotent elements of $\sp(W)$ with no Jordan block of size greater than $q$, establishing the claim above. 

We also note that if $\omega$ has Jordan type given by the partition $\mu$, then the point $x = e \oplus \omega$ lies in the leaf of the Higgs branch labelled by the partition $[q-1]\cup\mu$. It can be the case that the dimension of that leaf is greater than that of $\mathds{O}_\mu\subset\sp(W)$, in which case the fixed locus is a proper subvariety of that leaf of the Higgs branch. Away from the fixed locus the covering map is a local isomorphism, so the transverse singularity of the orbit closure along an unramified orbit (one with the form $[q+1,\sigma]$) will match the transverse singularity of the Higgs branch along the preimage of that orbit.

\subsubsection{Examples}

It is illustrative to observe the structure described at the level of the Hasse diagrams of the two varieties. Here we will consider several cases for small values of $(N,m)$. The relevant data for these theories are collected in Table \ref{tab:HBexamples}.

\begin{table}[t]
\centering
\renewcommand{\arraystretch}{1.3}
\begin{tabular}{c|c|c|c}
    & $\cM_H$ & $X_{V_{k^{\rm anom}_{N,m}}(\sp(2N))}$ & branch locus \\
    \hline\hline
    $\mathrm{AD}(\cf^{\rm anom}_{2},0)$ & $\overline{\mathds{O}}_{[3^2]}\cap\cS_{[2,1^4]}$ & $\overline{\mathds{O}}_{[4]}$ & $\overline{\mathds{O}}_{[2^2]}$ \\
    $\mathrm{AD}(\cf^{\rm anom}_{3},0)$ & $\overline{\mathds{O}}_{[3^2,2]}\cap\cS_{[2,1^6]}$ & $\overline{\mathds{O}}_{[4,2]}$ & $\overline{\mathds{O}}_{[3^2]}$ \\
    $\mathrm{AD}(\cf^{\rm anom}_{5},0)$ & $\overline{\mathds{O}}_{[3^4]}\cap\cS_{[2,1^{10}]}$ & $\overline{\mathds{O}}_{[4,3^2]}$ & $\overline{\mathds{O}}_{[3^2,2^2]}$ \\
    $\mathrm{AD}(\cf^{\rm anom}_{1},1)$ & $\overline{\mathds{O}}_{[4,2]}\cap\cS_{[4,1^2]}$ & $\overline{\mathds{O}}_{[2]}$ & --- \\
    $\mathrm{AD}(\cf^{\rm anom}_{3},1)$ & $\overline{\mathds{O}}_{[5^2]}\cap\cS_{[4,1^6]}$ & $\overline{\mathds{O}}_{[6]}$ & $\overline{\mathds{O}}_{[4,2]}$ \\
\end{tabular}
\caption{\label{tab:HBexamples}Higgs branches and the corresponding orbit closures for the first few admissible anomalous theories with $m=0$ and $m=1$. The Higgs branch is the Slodowy intersection \eqref{anomhiggs} in $\sp(2(N+m+1))$, and the associated variety of the current subalgebra is the corresponding orbit closure in $\sp(2N)$. The last column is the closure of the largest orbit of $\sp(2N)$ with no part exceeding $q$, along which the double cover is branched. The theory $\mathrm{AD}(\cf^{\rm anom}_{1},1)$ lies in the regime $N\leqslant m+1$, so in that case the two varieties coincide and there is no double cover relation.}
\end{table}

The simplest case, $\mathrm{AD}(\cf^{\rm anom}_{2},0)$, already appeared in Figure \ref{HasseC20} of Section \ref{sec:finite_extensions}. The Hasse diagrams of the Higgs branch and of the corresponding orbit closure agree except at their top transitions, where the $A_1$ singularity of the former is replaced by the $A_3$ singularity of the latter. The only orbit of $\overline{\mathds{O}}_{[4]}$ with a part $q+1=4$ is the open orbit, over which the cover is unramified, and the cover is branched along $\overline{\mathds{O}}_{[2^2]}$. The $A_3$ slice of the orbit closure is thus understood as the $\mathbb{Z}_2$ quotient of the $A_1$ slice in the Higgs branch.

The theory $\mathrm{AD}(\cf^{\rm anom}_{1},1)$, which is the rank-two $H_0$ theory with Higgs branch $\mathbb{C}^2/\mathbb{Z}_2$, is the only example in Table \ref{tab:HBexamples} in the regime $N\leqslant m+1$. There the Hasse diagrams of the two (isomorphic) varieties consist of just a single $A_1$ transition.

The theory $\mathrm{AD}(\cf^{\rm anom}_{3},0)$ is the first example in which the orbit closure has an unramified orbit other than the open one, namely $[4,1^2]$, and also the first in which the fixed locus is not a union of leaves. Its Higgs branch is $\overline{\mathds{O}}_{[3^2,2]}\cap\cS_{[2,1^6]}$, the corresponding orbit closure is $\overline{\mathds{O}}_{[4,2]}$, the branch locus is $\overline{\mathds{O}}_{[3^2]}$, and the two Hasse diagrams are compared in Figure \ref{HasseC30}. Note that the orbit closure has two seven-dimensional orbits, $[4,1^2]$ and $[3^2]$, both with transverse singularity $\af_1$, whereas the Higgs branch has a single seven-dimensional leaf, $[3^2,1^2]$, featuring the same transverse singularity. This leaf cannot have any $\mathbb{Z}_2$ fixed points because its partition has no part equal to two ($q-1$), and therefore maps locally isomorphically onto the unramified orbit $[4,1^2]$. The fixed points over the $[3^2]$ orbit are of the form $e\oplus\mathds{O}_{[3^2]}$ and therefore sit within the leaf $[3^2,2]$ of the Higgs branch (the open leaf). We can further note that the two components of the $\af_1\cup\af_1$ transition of the Higgs branch are exchanged by the $\mathbb{Z}_2$ automorphism, so in the orbit closure one instead sees  the $\af_1$ transition from $[2^3]$ to $[3^2]$.

\input{figures/Hasse_C3m0}

The theory $\mathrm{AD}(\cf^{\rm anom}_{3},1)$ is an example of the threshold case $N=m+2$, so the orbit closure $X_{V_{k^{\rm anom}_{3,1}}(\sp(6))}$ is the full nilpotent cone of $\sp(6)$, while the Higgs branch is $\overline{\mathds{O}}_{[5^2]}\cap\cS_{[4,1^6]}$ and the branch locus is $\overline{\mathds{O}}_{[4,2]}$, so the cover is only unramified over the open orbit. The fixed locus $e\oplus\mathds{O}_{[4,2]}$ lies in the leaf $[4^2,2]$ of the Higgs branch and has the same dimension; the $D_4$ singularity of the nilpotent cone along $[4,2]$ is thus the $\mathbb{Z}_2$ quotient of the $A_3$ singularity of the Higgs branch along $[4^2,2]$. The two Hasse diagrams are compared in Figure \ref{HasseC31}.

\input{figures/Hasse_C3m1}

The most intricate of our examples is $\mathrm{AD}(\cf^{\rm anom}_{5},0)$, whose Higgs branch is $\overline{\mathds{O}}_{[3^4]}\cap\cS_{[2,1^{10}]}$ and for which the corresponding orbit closure is $\overline{\mathds{O}}_{[4,3^2]}$ (see Figure \ref{HasseC50}). The branch locus is $\overline{\mathds{O}}_{[3^2,2^2]}$, and the cover is unramified over the five orbits with a part equal to four $(q+1)$; these are the orbits along the top right edge of the diagram. The fixed locus over the largest orbits $[3^2,2^2]$, $[3^2,2,1^2]$, and $[3^2,1^4]$ of the branch locus are smoothly embedded into the $[3^2,2^3]$, $[3^2,2^2,1^2]$, and $[3^2,2,1^4]$ leaves of the Higgs branch, respectively, each of which is one quaternionic dimension larger, which explains the three extra leaves in the orbit closure. Furthermore, the transverse singularity of the orbit closure along the unramified orbits, $\overline{\mathds{O}}_{[4,3^2]} \cap \mathcal{S}_{[4,1^6]}$, is isomorphic to the transverse singularity of the Higgs branch along the corresponding preimage $\overline{\mathds{O}}_{[3^4]} \cap \mathcal{S}_{[3^2,1^6]}$.

\input{figures/Hasse_C5m0}

\subsection{\label{subsec:HBfromVOA}The vertex algebra perspective}

The double cover of Proposition \ref{prop:HBmain}, and its ramification, can also be understood in vertex algebraic terms. Nilpotent Higgsing of a theory with respect to a nilpotent orbit $\mathds{O}$ of its flavour symmetry is implemented on the VOA by Drinfel'd--Sokolov reduction with respect to $\mathds{O}$, and the reduced vertex algebra is the VOA of the infrared theory at the point on the Higgs branch determined by the moment map expectation value. The sharp version of this statement presumes that the moment map expectation value determines a unique vacuum (where all other Higgs branch chiral ring generators are set to zero). Equivalently, it presumes that the fibre product of the Higgs branch with the Slodowy slice to $\mathds{O}$ is a single conical geometry. This fails for the anomalous theories in the regime $N\geqslant m+2$ for nilpotents in the doubly covered strata, as for these there are two equally good vacua sharing the same $\sp(2N)$ moment map value and the fibre product is two cones, one anchored on each vacuum. Drinfel'd--Sokolov reduction does not distinguish between the two vacua, and instead should produce a vertex algebra that, as a module for the reduction $\mathcal{W}_{\mathds{O}}$ of the current algebra, looks like a direct sum of two copies of the vacuum module. Indeed, we can identify the number of copies of the $\mathcal{W}_{\mathds{O}}$ vacuum module in the reduced algebra with the number of preimages in the Higgs branch of a generic point of $\mathds{O}$, thus giving a vertex algebra window into the ramification structure of the Higgs branch.

\subsubsection{\label{subsubsec:vac_count}Vacuum module counting}

Let $\mathds{O}$ be a nilpotent orbit of $\sp(2N)$ contained in the orbit closure $X_{V_{k^{\rm anom}_{N,m}}(\sp(2N))}$. Since the VOA of the anomalous theory is the finite extension \eqref{finiteextsp2N} of the current algebra by the module $L_{k^{\rm anom}_{N,m}}(\sp(2N);\varpi_1)$, the reduced vertex algebra is an extension of the algebra $\mathcal{W}_{\mathds{O}}\coloneqq H^0_{DS,\mathds{O}}(V_{k^{\rm anom}_{N,m}}(\sp(2N)))$ by the $\mathcal{W}_{\mathds{O}}$-module $M_{\mathds{O}}$ obtained by reducing the fundamental module,
\begin{equation}\label{eq:redext}
    V_{k^{\rm anom}_{N,m}}(\sp(2N)) \oplus L_{k^{\rm anom}_{N,m}}(\sp(2N);\varpi_1) \xrightarrow{~~\mathds{O}~~} \mathcal{W}_{\mathds{O}} \oplus M_{\mathds{O}}~.
\end{equation}
The relevant question is then whether $M_{\mathds{O}}$ is isomorphic to the vacuum module of $\mathcal{W}_{\mathds{O}}$.

We first note that for a physically allowed reduction, the reduced module $M_{\mathds{O}}$ is nontrivial (the fundamental module is not in the kernel of the reduction functor).\footnote{By \eqref{eq:redext} and \eqref{AsDatqDS}, the quantum dimension $\mathrm{qdim}_{\mathcal{W}_{\mathds{O}}} M_{\mathds{O}}$ is non-zero.} Whether the reduced module contains an additional vacuum-like vector can then be decided by inspecting the conformal weights of its generators. The states descending from the generators $W_\alpha$ of the fundamental module form a (possibly overcomplete) set of strong generators of $M_{\mathds{O}}$, and their conformal weights are obtained from the conformal weight $h_m=\tfrac12(3+2m)$ of $W_\alpha$ by the standard shift of the reduced conformal vector,\footnote{This is standard \cite{Kac_2003}; see \cite{Beem:2014rza} for a treatment in the SCFT/VOA context.}
\begin{equation} \label{confwtDS}
    h^i_{m,\rm{DS}} = h_m - \lambda_i \qquad i=1,2,\ldots, 2N~,
\end{equation}
where the $\lambda_i$ are the eigenvalues in the fundamental representation of the semisimple element $\Lambda(t_0)$ of the $\sl(2)$-triple defining the reduction. For a nilpotent orbit of $\sp(2N)$ with partition $[d_1,d_2,\dots,d_k]$, where $d_i$ are the sizes of the Jordan blocks of the nilpotent element, $\Lambda(t_0)$ takes the form
\begin{equation} \label{JordanBlock}
    \Lambda(t_0) = \tfrac{1}{2}\mathrm{diag}\left(D_{d_1}, \dots, D_{d_k}\right)~, \qquad D_{d} = \mathrm{diag}\left(d-1, d-3, \dots, 1-d\right)~,
\end{equation}
so the conformal weights \eqref{confwtDS} are
\begin{equation}
    h_{i,j} = \frac{3+2m}{2} - \frac{1}{2}(d_i - (2j+1))~,
\end{equation}
with $1\leqslant i\leqslant k$ and $0\leqslant j\leqslant d_i-1$. The smallest of these is $h_{1,0}=\tfrac12(q+1-d_1)$, which is determined entirely by the size $d_1$ of the largest Jordan block. Since $d_1\leqslant q+1$ for every orbit in the orbit closure, the list of weights includes zero if and only if the partition labelling $\mathds{O}$ has a part equal to $q+1$; otherwise all of the weights are strictly positive.

When the partition of $\mathds{O}$ has no part equal to $q+1$, the module $M_{\mathds{O}}$ has strictly positive conformal weights and the reduced vertex algebra \eqref{eq:redext} has a single vacuum vector. This is the case when the moment map expectation value determines a unique vacuum (points on $\mathds{O}$ have a single preimage in the Higgs branch). For the anomalous theories in the regime $N \leqslant m+1$, the Higgs branch and the corresponding orbit closure are both given by the full nilpotent cone $\overline{\mathds{O}}_{[2N]}$. Since all nilpotent orbits $\mathds{O} \subset \overline{\mathds{O}}_{[2N]}$ have parts strictly less than $q+1$, the module $M_{\mathds{O}}$ is \emph{never} a second copy of the vacuum module in these cases.

When the partition of $\mathds{O}$ contains a part $q+1$, the module $M_{\mathds{O}}$ contains a state $\mathbf{1}'$ of conformal weight zero, which is a second vacuum-like vector. In this case the reduced vertex algebra takes a simple form. Its weight-zero subspace is spanned by $\mathbf{1}$ and $\mathbf{1}'$ and is a commutative algebra under the normally ordered product. In fact, this weight-zero subspace realises the group algebra of $\mathbb{Z}_2$, and the reduced vertex algebra is
\begin{equation}\label{eq:twovacua}
    \mathcal{W}_{\mathds{O}} \oplus M_{\mathds{O}} \cong \mathcal{W}_{\mathds{O}} \otimes \mathbb{C}[\mathbb{Z}_2]~,
\end{equation}
the tensor product of the $\mathcal{W}$-algebra with the group algebra of $\mathbb{Z}_2$ regarded as a commutative vertex algebra concentrated in conformal weight zero. Going to an idempotent basis $\tfrac12(\mathbf{1}\pm\mathbf{1}')$ realises this as the direct sum of two copies of $\mathcal{W}_{\mathds{O}}$ which are exchanged by the $\mathbb{Z}_2$ symmetry, and its associated variety is two disjoint copies of $X_{\mathcal{W}_{\mathds{O}}}$. Thus, we recover the previous geometric description of the ramification of the double cover for the Higgs branches of the anomalous theories in the regime $N \geqslant m+2$. 

\subsubsection{\label{subsubsec:EFT_ex}Low-energy effective theories}

We now apply the preceding discussion to the maximal physically allowed Drinfel'd--Sokolov reduction, which probes the low-energy effective theory at a generic point on the Higgs branch. In certain families, the resulting vertex algebra can be identified explicitly. To this end, we note that it can be shown by direct computation that the relation
\begin{equation} \label{freehypers}
    -24(a - c) = \mathrm{dim}_{\mathbb{H}}(\mathcal{M}_H(\mathrm{AD}(\cf^{\mathrm{anom}}_{N},m)))~,
\end{equation}
holds if and only if $N = (0,m+2)~\mathrm{mod}~(3+2m)$. Hence, the low-energy effective theory at a generic point on the Higgs branch of the anomalous theories is just free hypermultiplets precisely when $N = (0,m+2)~\mathrm{mod}~(3+2m)$. An immediate consequence is that the low-energy effective theory is always interacting in the regime $N \leqslant m+1$, whereas for the AD$(\cf_N^{\rm anom},0)$ SCFTs it is just free hypermultiplets. Precisely in the cases satisfying \eqref{freehypers}, using the result in \eqref{eq:twovacua} and an established $\mathcal{W}$-algebra isomorphism \cite{Arakawa:2021ogm}, one finds
\begin{equation}
    V_{k^{\rm anom}_{N,m}}(\sp(2N)) \oplus L_{k^{\rm anom}_{N,m}}(\sp(2N);\varpi_1) \xrightarrow{\ \mathds{O}_{\rm max}\ } \mathbb{C} \oplus \mathbb{C}~,
\end{equation}
where $\mathds{O}_{\rm max}$ is the nilpotent orbit corresponding to the maximal physically allowed reduction. 

The interacting effective theory at a generic point on the Higgs branch can be identified explicitly when $N = (m+3)~\mathrm{mod}~(3+2m)$. Leveraging a known $\mathcal{W}$-algebra isomorphism from \cite{Arakawa:2021ogm}, the result \eqref{eq:twovacua} implies
\begin{equation} \label{eq:EFT_Vir}
    V_{k^{\rm anom}_{N,m}}(\sp(2N)) \oplus L_{k^{\rm anom}_{N,m}}(\sp(2N);\varpi_1) \xrightarrow{\ \mathds{O}_{\rm max}\ } \mathrm{Vir}_{2,3+2m} \oplus \mathrm{Vir}_{2,3+2m}~.
\end{equation}
Therefore, the low-energy effective theory at a generic point of the Higgs branch of the AD$(\cf_N^{\rm anom},m)$ theory is the $(A_1, A_{2m})$ Argyres--Douglas SCFT when $N = (m+3)~\mathrm{mod}~(3+2m)$. The direct sum decomposition in \eqref{eq:EFT_Vir} reflects the presence of two vacua, as in Section \ref{subsubsec:vac_count}.

%% file: figures/Grid.tex

\begin{figure} 
\begin{center}
\begin{tikzpicture}

\def\gridwidth{10}
\def\gridheight{10}  
\def\cellsize{1}   
\def\linewidth{0.5mm}  

\def\crossboxes{2/2, 5/2, 8/2, 3/3, 8/3, 4/4, 2/5, 5/5, 8/5, 6/6, 7/7, 2/8, 3/8, 5/8, 8/8, 9/9} 

\foreach \x in {2,3,4,5,6,7,8,9,10} {
    \foreach \y in {1,2,3,4,5,6,7,8,9} {
        \pgfmathsetmacro{\isCross}{0}
        \foreach \cx/\cy in \crossboxes {
            \ifnum\x=\cx \ifnum\y=\cy
                \pgfmathsetmacro{\isCross}{1}
            \fi \fi
        }

        \ifnum\x > \y
            \fill[red!50] (\x-1,9-\y) rectangle (\x,10-\y);
        \else
            \fill[yellow!50] (\x-1,9-\y) rectangle (\x,10-\y);
        \fi
    }
}

\draw[line width=\linewidth] (0,0) grid (\gridwidth,\gridheight);

\foreach \cx/\cy in \crossboxes {
    \draw[black, thick] (\cx-1+0.2, 9-\cy+0.2) -- (\cx-0.2, 9-\cy+0.8);
    \draw[black, thick] (\cx-1+0.2, 9-\cy+0.8) -- (\cx-0.2, 9-\cy+0.2);
}

\foreach \i in {1,2,3,4,5,6,7,8,9} {
    \node at (\i+0.5, 9.5) {\textbf{\i}};
}
\foreach \i in {-1,0,1,2,3,4,5,6,7} {
    \node at (0.5, 7.5-\i) {\textbf{\i}};
}

\draw[black, thick] (0.15,9.85) -- (0.85,9.15);
\node at (0.7, 9.7) {$N$};
\node at (0.3, 9.3) {$m$};

\end{tikzpicture}
\end{center}
{\caption{\label{GenNmdiag} The Higgs branch of $\mathrm{AD}(\cf^{\rm anom}_{N},m)$ is the double cover of the associated variety $X_{V_{k^{\rm anom}_{N,m}}(\sp(2N))}$ for the values of $(N,m)$ represented by uncrossed red boxes, and is exactly equal to the associated variety $X_{V_{k^{\rm anom}_{N,m}}(\sp(2N))}$ for values of $(N,m)$ corresponding to the yellow, uncrossed boxes. The crossed boxes correspond to non-admissible levels, and we do not comment on the Higgs branch for these values of $(N,m)$.}}
\end{figure}

%% file: figures/Quiver_AD.tex

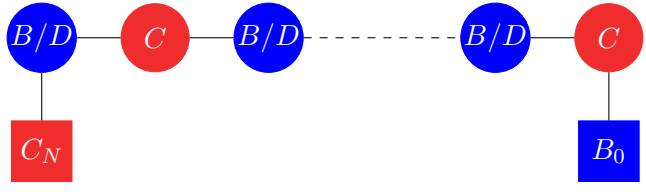
\begin{figure}[t]
\begin{tikzpicture}[
    gaugeblue/.style={
        circle, draw=blue, fill=blue,
        minimum size=9mm, inner sep=0pt
    },
    gaugered/.style={
        circle, draw=mutedred, fill=mutedred,
        minimum size=9mm, inner sep=0pt
    },
    flavourred/.style={
        rectangle, draw=mutedred, fill=mutedred,
        minimum size=8mm, inner sep=0pt
    },
    flavourblue/.style={
        rectangle, draw=blue, fill=blue,
        minimum size=8mm, inner sep=0pt
    },
    arrow/.style={-{Latex[width=2mm]}, thick}
]
\begin{scope}[xshift = 2cm]
\node[gaugeblue] (BN1) at (0,0) {$\textcolor{white}{B/D}$};
\node[gaugered] (CN2) at (1.5,0) {$\textcolor{white}{C}$};
\node[gaugeblue] (BN3) at (3,0) {$\textcolor{white}{B/D}$};
\node[gaugeblue] (BDq2) at (6,0) {$\textcolor{white}{B/D}$};
\node[gaugered] (CNq) at (7.5,0) {$\textcolor{white}{C}$};

\node[flavourred] (CN) at (0,-1.5) {$\textcolor{white}{C_N}$};
\node[flavourblue] (B0) at (7.5,-1.5) {$\textcolor{white}{B_0}$};

\draw[black] (BN1) -- (CN2);    
\draw[black] (CN2) -- (BN3);    
\draw[dashed] (BN3) -- (BDq2);  
\draw[black] (BDq2) -- (CNq);   
\draw[black] (BN1) -- (CN);     
\draw[black] (CNq) -- (B0);     
\end{scope}
\end{tikzpicture}
{\caption{\label{QuivGenAD} The orthosymplectic quiver for the Higgs branch of the $\mathrm{AD}(\cf^{\rm anom}_{N},m)$ theory. The $q-1$ gauge nodes have $\mathbf{N} = \{N_1, N_2, \dots, N_{q-1}\}$, where $B/D$ nodes carry ${\rm O}(N_i)$ gauge groups and $C$ nodes carry ${\rm USp}(N_i)$ gauge groups.}}
\end{figure}

%% file: figures/Quiver_CN.tex

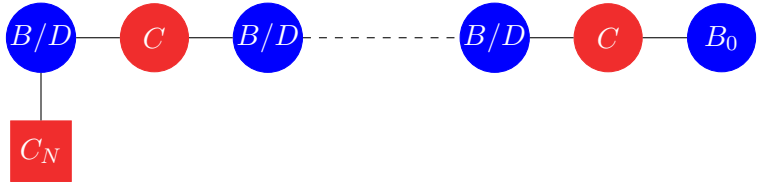
\begin{figure}[t]
\begin{tikzpicture}[
    gaugeblue/.style={
        circle, draw=blue, fill=blue,
        minimum size=9mm, inner sep=0pt
    },
    gaugered/.style={
        circle, draw=mutedred, fill=mutedred,
        minimum size=9mm, inner sep=0pt
    },
    flavourred/.style={
        rectangle, draw=mutedred, fill=mutedred,
        minimum size=8mm, inner sep=0pt
    },
    arrow/.style={-{Latex[width=2mm]}, thick}
]
\begin{scope}[xshift = 2cm]
\node[gaugeblue] (BN1) at (0,0) {$\textcolor{white}{B/D}$};
\node[gaugered] (CN2) at (1.5,0) {$\textcolor{white}{C}$};
\node[gaugeblue] (BN3) at (3,0) {$\textcolor{white}{B/D}$};
\node[gaugeblue] (BDq2) at (6,0) {$\textcolor{white}{B/D}$};
\node[gaugered] (CNq) at (7.5,0) {$\textcolor{white}{C}$};
\node[gaugeblue] (B0) at (9,0) {$\textcolor{white}{B_0}$};

\node[flavourred] (CN) at (0,-1.5) {$\textcolor{white}{C_N}$};

\draw[black] (BN1) -- (CN2);    
\draw[black] (CN2) -- (BN3);    
\draw[dashed] (BN3) -- (BDq2);  
\draw[black] (BDq2) -- (CNq);   
\draw[black] (CNq) -- (B0);     
\draw[black] (BN1) -- (CN);     
\end{scope}
\end{tikzpicture}
\caption{\label{QuivGenCN} The orthosymplectic quiver for the associated variety $X_{V_{k^{\rm anom}_{N,m}}(\sp(2N))}$. The $q$ gauge nodes have $\mathbf{N} = \{N_1, N_2, \dots, N_{q-1}, 1\}$, where the $N_i$ are exactly the same as in Figure \ref{QuivGenAD}. The rightmost gauge node is an ${\rm O}(1)\cong\mathbb{Z}_2$ discrete gauging.}
\end{figure}

%% file: figures/Hasse_C3m0.tex

\begin{figure}
\begin{tikzpicture}[decoration={markings,
mark=at position .5 with {\arrow{>}}},gauge/.style={circle, draw=blue, fill=blue, minimum size=7mm, inner sep=0pt},
    flavour/.style={rectangle, draw=red, fill=red, minimum size=6mm, inner sep=0pt},
    arrow/.style={-{Latex[width=2mm]}, thick}]
\begin{scope}[scale=1.5,xshift=1.5cm]
\node[scale=.8] (p5) at (0,3.5) {$[3^2,2]$};
\node[scale=.8] (t5) at (-0.2,3) {$\af_1$};
\node[scale=.8] (p4) at (0,2.5) {$[3^2,1^2]$};
\node[scale=.8] (t4) at (-.5,2) {$\af_1\cup \af_1$};
\node[scale=.8] (p3) at (0,1.5) {$[2^4]$};
\node[scale=.8] (t3) at (-.2,1) {$\af_1$};
\node[scale=.8] (p2) at (0,.5) {$[2^3,1^2]$};
\node[scale=.8] (t2) at (-.2,0) {$\cf_2$};
\node[scale=.8] (p1) at (0,-.5) {$[2^2,1^4]$};
\node[scale=.8] (t1) at (-.2,-1) {$\cf_3$};
\node[scale=.8] (p0) at (0,-1.5) {$[2,1^6]$};
\draw[blue] (p0) -- (p1);
\draw[blue] (p1) -- (p2);
\draw[blue] (p2) -- (p3);
\draw[blue] (p3) -- (p4);
\draw[blue] (p4) -- (p5);

\end{scope}
\begin{scope}[scale=1.5,xshift=5.5cm]
\node[scale=.8] (p5) at (0,3.5) {$[4,2]$};
\node[scale=.8] (t5a) at (0.7,3) {$\af_1$};
\node[scale=.8] (t5a) at (-0.7,3) {$\af_1$};
\node[scale=.8] (p4b) at (.7,2.5) {$[3^2]$};
\node[scale=.8] (t4b) at (-.7,2) {$\af_1$};
\node[scale=.8] (p4a) at (-.7,2.5) {$[4,1^2]$};
\node[scale=.8] (t4a) at (.7,2) {$\af_1$};
\node[scale=.8] (p3) at (0,1.5) {$[2^3]$};
\node[scale=.8] (t3) at (-.2,1) {$\af_1$};
\node[scale=.8] (p2) at (0,.5) {$[2^2,1^2]$};
\node[scale=.8] (t2) at (-.2,0) {$\cf_2$};
\node[scale=.8] (p1) at (0,-.5) {$[2,1^4]$};
\node[scale=.8] (t1) at (-.2,-1) {$\cf_3$};
\node[scale=.8] (p0) at (0,-1.5) {$[1^6]$};
\draw[blue] (p0) -- (p1);
\draw[blue] (p1) -- (p2);
\draw[blue] (p2) -- (p3);
\draw[blue] (p3) -- (p4a);
\draw[blue] (p3) -- (p4b);
\draw[blue] (p4a) -- (p5);
\draw[blue] (p4b) -- (p5);

\end{scope}
\end{tikzpicture}
{\caption{\label{HasseC30} Hasse diagrams for the Higgs branch moduli space of $\mathrm{AD}(\cf^{\rm anom}_{3},0)$ (left) and the associated variety $X_{V_{k^{\rm anom}_{3,0}}(\sp(6))}$ (right).}}
\end{figure}%

%% file: figures/Hasse_C3m1.tex

\begin{figure}
\begin{tikzpicture}[decoration={markings,
mark=at position .5 with {\arrow{>}}}, gauge/.style={circle, draw=blue, fill=blue, minimum size=6.5mm, inner sep=0pt},
    flavour/.style={rectangle, draw=red, fill=red, minimum size=5.5mm, inner sep=0pt},
    arrow/.style={-{Latex[width=2mm]}, thick}]
\begin{scope}[scale=1.5,xshift=1.5cm]
\node[scale=.8] (p6) at (0,4.5) {$[5,5]$};
\node[scale=.8] (t5a1) at (-0.2,4) {$A_3$};
\node[scale=.8] (p5) at (0,3.5) {$[4,4,2]$};
\node[scale=.8] (t5a) at (0.7,3) {$\af_1$};
\node[scale=.8] (t5a) at (-0.7,3) {$\af_1$};
\node[scale=.8] (p4b) at (.7,2.5) {$[4,4,1,1]$};
\node[scale=.8] (t4b) at (-.7,2) {$\af_1$};
\node[scale=.8] (p4a) at (-.7,2.5) {$[4,3,3]$};
\node[scale=.8] (t4a) at (.7,2) {$\af_1$};
\node[scale=.8] (p3) at (0,1.5) {$[4,2,2,2]$};
\node[scale=.8] (t3) at (-.2,1) {$\af_1$};
\node[scale=.8] (p2) at (0,.5) {$[4,2,2,1,1]$};
\node[scale=.8] (t2) at (-.2,0) {$\cf_2$};
\node[scale=.8] (p1) at (0,-.5) {$[4,2,1^4]$};
\node[scale=.8] (t1) at (-.2,-1) {$\cf_3$};
\node[scale=.8] (p0) at (0,-1.5) {$[4,1^6]$};
\draw[blue] (p0) -- (p1);
\draw[blue] (p1) -- (p2);
\draw[blue] (p2) -- (p3);
\draw[blue] (p3) -- (p4a);
\draw[blue] (p3) -- (p4b);
\draw[blue] (p4a) -- (p5);
\draw[blue] (p4b) -- (p5);
\draw[blue] (p5) -- (p6);

\end{scope}
\begin{scope}[scale=1.5,xshift=5.5cm]
\node[scale=.8] (p6) at (0,4.5) {$[6]$};
\node[scale=.8] (t5a1) at (-0.2,4) {$D_4$};
\node[scale=.8] (p5) at (0,3.5) {$[4,2]$};
\node[scale=.8] (t5a) at (0.7,3) {$\af_1$};
\node[scale=.8] (t5a) at (-0.7,3) {$\af_1$};
\node[scale=.8] (p4b) at (.7,2.5) {$[3^2]$};
\node[scale=.8] (t4b) at (-.7,2) {$\af_1$};
\node[scale=.8] (p4a) at (-.7,2.5) {$[4,1^2]$};
\node[scale=.8] (t4a) at (.7,2) {$\af_1$};
\node[scale=.8] (p3) at (0,1.5) {$[2^3]$};
\node[scale=.8] (t3) at (-.2,1) {$\af_1$};
\node[scale=.8] (p2) at (0,.5) {$[2^2,1^2]$};
\node[scale=.8] (t2) at (-.2,0) {$\cf_2$};
\node[scale=.8] (p1) at (0,-.5) {$[2,1^4]$};
\node[scale=.8] (t1) at (-.2,-1) {$\cf_3$};
\node[scale=.8] (p0) at (0,-1.5) {$[1^6]$};
\draw[blue] (p0) -- (p1);
\draw[blue] (p1) -- (p2);
\draw[blue] (p2) -- (p3);
\draw[blue] (p3) -- (p4a);
\draw[blue] (p3) -- (p4b);
\draw[blue] (p4a) -- (p5);
\draw[blue] (p4b) -- (p5);
\draw[blue] (p5) -- (p6);

\end{scope}
\end{tikzpicture}
{\caption{\label{HasseC31} Hasse diagrams for the Higgs branch moduli space of $\mathrm{AD}(\cf^{\rm anom}_{3},1)$ (left) and the associated variety $X_{V_{k^{\rm anom}_{3,1}}(\sp(6))}$ (right).}}
\end{figure}%

%% file: figures/Hasse_C5m0.tex

\begin{figure}
\begin{tikzpicture}[decoration={markings,
mark=at position .5 with {\arrow{>}}}, gauge/.style={circle, draw=blue, fill=blue, minimum size=7mm, inner sep=0pt},
    flavour/.style={rectangle, draw=red, fill=red, minimum size=6mm, inner sep=0pt},
    arrow/.style={-{Latex[width=2mm]}, thick}]
\begin{scope}[scale=1.5,xshift=1.5cm]
\node[scale=.8] (p10) at (-.7,8.5) {$[3^4]$};
\node[scale=.8] (t10) at (-.1,8) {$\af_1$};
\node[scale=.8] (p9b) at (0,7.5) {$[3^2,2^3]$};
\node[scale=.8] (t9b) at (.6,7) {$\af_1$};
\node[scale=.8] (p8b) at (.7,6.5) {$[3^2,2^2,1^2]$};
\node[scale=.8] (t8b) at (1.2,6) {$\cf_2$};
\node[scale=.8] (t4a) at (-.1,5) {$\df_3$};
\node[scale=.8] (p7b) at (1.4,5.5) {$[3^2,2,1^4]$};
\node[scale=.8] (t7b) at (1.9,5) {$\cf_3$};
\node[scale=.8] (p6c) at (2.1,4.5) {$[3^2,1^6]$};
\node[scale=.8] (t4a) at (.6,4) {$\bff_2$};
\node[scale=.8] (p5a) at (-1.4,3.5) {$[2^6]$};
\node[scale=.8] (t5a) at (-1.4,3) {$\af_1$};
\node[scale=.8] (p4) at (-.7,2.5) {$[2^5,1^2]$};
\node[scale=.8] (t4b) at (-.7,2) {$\cf_2$};
\node[scale=.8] (t4a) at (1.4,2.9) {$\af_1 \cup \af_1$};
\node[scale=.8] (p3) at (0,1.5) {$[2^4,1^4]$};
\node[scale=.8] (t3) at (-.2,1) {$\cf_3$};
\node[scale=.8] (p2) at (0,.5) {$[2^3,1^6]$};
\node[scale=.8] (t2) at (-.2,0) {$\cf_4$};
\node[scale=.8] (p1) at (0,-.5) {$[2^2,1^8]$};
\node[scale=.8] (t1) at (-.2,-1) {$\cf_5$};
\node[scale=.8] (p0) at (0,-1.5) {$[2,1^{10}]$};
\draw[blue] (p0) -- (p1);
\draw[blue] (p1) -- (p2);
\draw[blue] (p2) -- (p3);
\draw[blue] (p3) -- (p4);
\draw[blue] (p3) -- (p6c);
\draw[blue] (p4) -- (p5a);
\draw[blue] (p4) -- (p7b);
\draw[blue] (p5a) -- (p8b);
\draw[blue] (p6c) -- (p7b);
\draw[blue] (p7b) -- (p8b);
\draw[blue] (p8b) -- (p9b);
\draw[blue] (p9b) -- (p10);

\end{scope}
\begin{scope}[scale=1.5,xshift=6.5cm]
\node[scale=.8] (p10) at (-.7,8.5) {$[4,3^2]$};
\node[scale=.8] (t10) at (-.1,8) {$\af_1$};
\node[scale=.8] (p9b) at (0,7.5) {$[4,2^3]$};
\node[scale=.8] (t9b) at (.6,7) {$\af_1$};
\node[scale=.8] (p8b) at (.7,6.5) {$[4,2^2,1^2]$};
\node[scale=.8] (t8b) at (1.2,6) {$\cf_2$};
\node[scale=.8] (p7b) at (1.4,5.5) {$[4,2,1^4]$};
\node[scale=.8] (t7b) at (1.9,5) {$\cf_3$};
\node[scale=.8] (tEx5) at (-.8,6.6) {$\af_1$};
\node[scale=.8] (p7a) at (-1.4,5.5) {$[3^2,2^2]$};
\node[scale=.8] (t7a) at (-1.2,4.9) {$\af_1$};
\node[scale=.8] (p6c) at (2.1,4.5) {$[4,1^6]$};
\node[scale=.8] (tEx4) at (-.1,5.6) {$\af_1$};
\node[scale=.8] (p6b) at (-.7,4.5) {$[3^2,2,1^2]$};
\node[scale=.8] (t6b) at (-.5,3.9) {$\cf_2$};
\node[scale=.8] (tEx3) at (.6,4.6) {$\af_1$};
\node[scale=.8] (p5b) at (0,3.5) {$[3^2,1^4]$};
\node[scale=.8] (t5b) at (0.1,3) {$\af_1\cup\af_1$};
\node[scale=.8] (tEx1) at (-1.2,4.1) {$\bff_2$};
\node[scale=.8] (p5a) at (-1.4,3.5) {$[2^5]$};
\node[scale=.8] (t5a) at (-1.4,3) {$\af_1$};
\node[scale=.8] (p4) at (-.7,2.5) {$[2^4,1^2]$};
\node[scale=.8] (t4b) at (-.7,2) {$\cf_2$};
\node[scale=.8] (t4a) at (1.2,2.9) {$\af_1$};
\node[scale=.8] (p3) at (0,1.5) {$[2^3,1^4]$};
\node[scale=.8] (t3) at (-.2,1) {$\cf_3$};
\node[scale=.8] (p2) at (0,.5) {$[2^2,1^6]$};
\node[scale=.8] (t2) at (-.2,0) {$\cf_4$};
\node[scale=.8] (p1) at (0,-.5) {$[2,1^8]$};
\node[scale=.8] (t1) at (-.2,-1) {$\cf_5$};
\node[scale=.8] (p0) at (0,-1.5) {$[1^{10}]$};
\draw[blue] (p0) -- (p1);
\draw[blue] (p1) -- (p2);
\draw[blue] (p2) -- (p3);
\draw[blue] (p3) -- (p4);
\draw[blue] (p3) -- (p6c);
\draw[blue] (p4) -- (p5a);
\draw[blue] (p4) -- (p5b);
\draw[blue] (p5b) -- (p6b);
\draw[blue] (p5a) -- (p6b);
\draw[blue] (p6b) -- (p7a);
\draw[blue] (p5b) -- (p7b);
\draw[blue] (p6c) -- (p7b);
\draw[blue] (p7b) -- (p8b);
\draw[blue] (p6b) -- (p8b);
\draw[blue] (p8b) -- (p9b);
\draw[blue] (p7a) -- (p9b);
\draw[blue] (p9b) -- (p10);

\end{scope}
\end{tikzpicture}
{\caption{\label{HasseC50} Hasse diagrams for the Higgs branch moduli space of $\mathrm{AD}(\cf^{\rm anom}_{5},0)$ (left) and the associated variety $X_{V_{k^{\rm anom}_{5,0}}(\sp(10))}$ (right).}}
\end{figure}%

%% file: sections/sec_6_discussion.tex

\section{Observations and outlook}\label{sec:discussion}

In this paper, we have studied the Higgs branches and associated vertex operator algebras of the ``class I'' twisted $A_{2N}$ Argyres--Douglas theories, which we have called $\mathrm{AD}(\cf^{\rm anom}_{N},m)$. On the basis of global Witten anomaly matching, 't~Hooft anomaly matching, and Higgs branch anomaly matching, we have proposed that $\mathrm{AD}(\cf^{\rm anom}_{N},m)$ admits an alternative realisation as a specific nilpotent Higgsing of the ``class I'' twisted $D_{N+m+2}$ theory. A consequence of this proposal is that the associated VOA is not the affine Kac--Moody algebra $V_{k^{\rm anom}_{N,m}}(\sp(2N))$ itself, but rather its finite extension by a module of conformal weight $h_m = \tfrac12(3+2m)$ in the fundamental representation of $\sp(2N)$, the level being coprincipal admissible but not boundary admissible. For the $m=0$ and $m=1$ series we bootstrapped these VOAs parametrically in $N$ and found that OPE associativity together with graded unitarity single out precisely the flavour level of the corresponding SCFT. Most of the remaining bootstrap solutions admit an interpretation as critical-level $\mathcal{W}$-algebras; it would be interesting to know whether the isolated solution with $k=-\frac{7}{4}$ of the $N=1$, $m=1$ problem, for which we have no such interpretation, has any physical relevance.

We found that the anomalous VOAs admit several further realisations as Drinfel'd--Sokolov reductions of current algebras of types $B$, $C$, and $D$. These isomorphisms suggest that the ``class I'' twisted $A_{\rm even}$, twisted $A_{\rm odd}$, and twisted $D$-type theories can all be realised as nilpotent Higgsings of ``class I'' \emph{untwisted} $D$-type theories. Finally, we established that the Higgs branch of $\mathrm{AD}(\cf^{\rm anom}_{N},m)$ coincides with the associated variety of its current subalgebra for $N\leqslant m+1$ and is a ramified double cover thereof for $N\geqslant m+2$. Furthermore, we also determined the branch locus of the cover both geometrically and by counting vacuum modules in Drinfel'd--Sokolov reductions of the VOA.

The theories studied in this paper are also of interest from the point of view of graded unitarity \cite{ArabiArdehali:2025fad,Beem:2026lkq}. Boundary admissible levels have been observed to be well suited to arise as the affine Kac--Moody VOAs of unitary four-dimensional $\mathcal{N}=2$ SCFTs. While the VOAs of the anomalous theories are finite extensions of the non-boundary admissible algebras $V_{k^{\rm anom}_{N,m}}(\sp(2N))$, the $\mathbb{Z}_2$ gauging of $\mathrm{AD}(\cf^{\rm anom}_{N},m)$ has as its associated VOA the $\mathbb{Z}_2$-invariant subalgebra, which is precisely the affine Kac--Moody algebra,
\begin{equation}\label{eq:non-boundary-z2-gauging}
    \mathbb{V}[\mathrm{AD}(\cf^{\mathrm{anom}}_{N},m)^{\mathbb{Z}_2}] = V_{k^{\rm anom}_{N,m}}(\sp(2N))~.
\end{equation}
For $N = 1$, this gives examples of unitary four-dimensional $\mathcal{N} = 2$ SCFTs which have non-boundary admissible $\sl(2)$ affine Kac--Moody algebras as their associated VOAs. Naively, this is in tension with the graded unitarity analysis of \cite{ArabiArdehali:2025fad}, but there is no actual conflict because, as we will show presently, the natural $\mathfrak{R}$-filtration on the VOA $V_{k^{\mathrm{anom}}_{N,m}}(\sp(2N))$ inherited from the ungauged parent theory is \emph{not} the standard $\mathfrak{R}$-filtration used in \cite{ArabiArdehali:2025fad}.

Consider $\mathrm{AD}(\cf^{\rm anom}_{1},1)^{\mathbb{Z}_2}$, the $\mathbb{Z}_2$ gauging of the rank-two $H_0$ theory, whose associated VOA is $\mathbb{V} \coloneqq V_{-\frac{17}{10}}(\sl(2))$. Under the assumption of the standard $\sl(2)$ $\mathfrak{R}$-filtration,
\begin{equation} \label{eq:standardsl2}
    \mathfrak{R}_p^{\mathrm{std}} \mathbb{V} = \mathrm{Span}\{L_{-m_1} \ldots L_{-m_j} J^{a_1}_{-n_1} \ldots J^{a_k}_{-n_k} \Omega~,\quad m_i \geqslant 2~, n_i \geqslant 1~,\quad j + k \leqslant p\}~, 
\end{equation}
all flavour levels in the interval $k \in (-\frac{12}{7}, -\frac{8}{5} )$ were ruled out in \cite{ArabiArdehali:2025fad} by graded unitarity at conformal weight seven. The existence of $\mathrm{AD}(\cf^{\rm anom}_{1},1)^{\mathbb{Z}_2}$ as a unitary SCFT therefore requires the $\mathfrak{R}$-filtration on $V_{-\frac{17}{10}}(\sl(2))$ to differ from \eqref{eq:standardsl2}. To see how this comes about we first consider the $\mathfrak{R}$-filtration on the VOA of the rank-two $H_0$ theory itself.

The strong generators of the rank-two $H_0$ VOA are the $\sl(2)$ currents $J_{\alpha\beta}$ and the doublet $W_\alpha$ of conformal weight $\tfrac52$. The standard $\mathfrak{R}$-filtration for this VOA is weight-based and determined by the assignments $R(J_{\alpha\beta}) = R(T) = 1$ and $R(W_\alpha) = \tfrac32$. The first null states of this VOA appear at levels $\tfrac72$ and $5$, and give rise to the following relations,
\begin{align}
    \epsilon^{\beta\gamma} J_{\alpha \beta} W_\gamma &= -\tfrac{3}{10} W'_\alpha~,\label{eq:JWnullstate}\\
    W_\alpha W_\beta &= \tfrac{625}{1683}\,(TTJ_{\alpha\beta})+\tfrac{200}{561}\,\left(\epsilon^{\gamma\delta} T\,J'_{\gamma\alpha}J_{\delta\beta}\right) -\tfrac{125}{1122}\, \epsilon^{\gamma\delta}\epsilon^{\theta\eta} \left(J_{\alpha\gamma} J_{\beta \theta} J''_{\delta \eta} \right)\label{eq:WWnullstate}\\
    &+\tfrac{200}{561}\,\epsilon^{\gamma\delta}\epsilon^{\theta\eta} \left( J'_{\alpha\gamma} J'_{\beta\theta} J_{\delta\eta} \right)+\tfrac{25}{132}\,\left(T''J_{\alpha\beta}\right) +\tfrac{775}{2244}\,\left(T J''_{\alpha\beta}\right) -\tfrac{280}{1683}\, \epsilon^{\gamma\delta} J_{\alpha\gamma}J'''_{\beta\delta}\nonumber\\
    &+\tfrac{235}{3366}\, \epsilon^{\gamma\delta} J'''_{\alpha\gamma} J_{\beta\delta}-\tfrac{5}{22}\, \epsilon^{\gamma\delta} J'_{\alpha\gamma} J''_{\beta\delta} -\tfrac{235}{561}\, \epsilon^{\gamma\delta} J''_{\alpha\gamma} J'_{\beta\delta} -\tfrac{163}{2244}\, J''''_{\alpha\beta} -\tfrac{500}{1683}\, \epsilon_{\alpha\beta}\, TT'~.\nonumber
\end{align}
The relation \eqref{eq:WWnullstate} is such that the presumed $\mathfrak{R}$-filtration induces the standard filtration on the $\mathbb{Z}_2$-invariant subalgebra up to at least level five. The same is true at level six, where one can check that every linear combination of the zero-charge states: $\{ W'_1 W_2, \allowbreak W'_2 W_1, \allowbreak J_{11}W_2 W_2, \allowbreak J_{12} W_1 W_2, \allowbreak J_{22} W_1 W_1\}$, once expressed in terms of normal ordered products of $T$, $J_{\alpha \beta}$ and their derivatives, satisfies $\mathfrak{R}_p \mathbb{V} = \mathfrak{R}^{\mathrm{std}}_p \mathbb{V}$ for $p \leqslant 4$.

At level seven, however, the null relations \eqref{eq:JWnullstate} and \eqref{eq:WWnullstate} give rise to a null descendant of the form
\begin{equation} \label{eq:Lvl7nullrelation}
    W''_{(1} W_{2)} - W'_{(1} W'_{2)} \sim TTT J_{12} + \ldots \eqqcolon \mathcal{N}_{TTTJ}~,
\end{equation}
where $\ldots$ denotes terms with one or more derivatives. The weight-based filtration of the rank-two $H_0$ VOA assigns $\mathcal{N}_{TTTJ} \in \mathfrak{R}_3 \mathbb{V}$, whereas the standard filtration assigns it to $\mathfrak{R}^{\mathrm{std}}_4 \mathbb{V}$ but not $\mathfrak{R}^{\mathrm{std}}_3 \mathbb{V}$. One can further check that the $\mathfrak{R}$-charges of all $\mathbb{Z}_2$-invariant states of weight seven that are linearly independent of $\mathcal{N}_{TTTJ}$ agree with the standard assignment. The net effect of the modified filtration is thus an additional factor of $(-1)^{4-3} = -1$ in the Gram determinant constraint of graded unitarity in the zero-charge sector at level seven. This implies that the interval $k \in (-\frac{12}{7}, -\frac{8}{5})$, which was excluded at level seven under the assumption of the standard $\sl(2)$ $\mathfrak{R}$-filtration, is no longer excluded once the inherited filtration is used, thus rescuing compatibility with the graded unitarity analysis in \cite{ArabiArdehali:2025fad}.

Note that a principal nilpotent Higgsing of the rank-two $H_0$ theory RG flows to a four-dimensional $\mathcal{N}=2$ SCFT with the associated VOA given by the $\text{Vir}_{2,5} \otimes \text{Vir}_{2,5}$ vertex algebra. It can be shown that the diagonal subalgebra of this VOA is precisely $\mathrm{Vir}_{3,10}$, which coincides with the $\mathbb{Z}_2$-invariant subalgebra of the $\text{Vir}_{2,5} \otimes \text{Vir}_{2,5}$ VOA. Consequently, a $\mathbb{Z}_2$ gauging of the Higgsed SCFT has the associated VOA given by the \emph{non-boundary} minimal model $\mathrm{Vir}_{3,10}$. This provides an example in the $C_2$-cofinite setting of a discretely gauged VOA which must be equipped with a non-standard $\mathfrak{R}$-filtration.

More generally, we expect that the non-boundary admissible affine Kac--Moody algebras arising as $\mathbb{Z}_2$ gaugings will inherit similarly ``non-standard'' $\mathfrak{R}$-filtrations from their parent extended VOAs. These discretely gauged generalised Argyres--Douglas SCFTs enjoy corresponding $2$-form $\mathbb{Z}_2$ symmetries, and it is amusing to speculate that there might be a deeper connection between higher-form symmetries of $\mathcal{N}=2$ SCFTs and non-standard modifications of the $\mathfrak{R}$-filtrations of their associated vertex operator algebras.

%% file: sections/appendix_a.tex

\section{\label{app:irregular_puncture_review}Brief review of irregular punctures}

A key role in the characterisation of the physics of the four-dimensional $\cN=2$ SCFTs of class $\cS$ \cite{Witten:1997sc,Gaiotto:2009we} is played by the punctures which appear at marked points of the UV curve $\cC$. These punctures can be broadly divided into two classes: regular (tame) \cite{Gaiotto:2009we,Nanopoulos:2009uw,Chacaltana:2012zy} and irregular (wild) \cite{Cecotti:2011rv,Bonelli:2011aa,Xie:2012hs,Wang:2016yha}. We will assume familiarity of the reader with the former and briefly review the characterisation of irregular punctures. This is mainly to position the class of theories studied in the main text within the wider ecosystem of generalised Argyres--Douglas theories, and the review here is far from comprehensive.

\subsection{\label{subsec:untwisted_punctures}Untwisted irregular punctures}

Irregular punctures are characterised in terms of the induced irregular singularities for the Higgs field $\Phi(z)$ of the Hitchin system associated to the class $\cS$ theory, where $z$ is a local coordinate around the puncture on $\cC$. These are parametrised in the following way,
\begin{equation}\label{HField}
\Phi(z)=\frac{T}{z^{2+\frac k b}}+\cdots~,
\end{equation}
where $b\in\Z_{>0}$ and $k\in\Z$ with $k>-b$. The specification of $k>-b$ ensures that the singularity is irregular with a higher order pole. $T$ is a semisimple element of the simply laced Lie algebra $\jf$ defining the parent $(2,0)$ theory. The Higgs field \eqref{HField} further satisfies the constraint that the monodromy around the singularity be given by the action of an automorphism of $\jf$,
\begin{equation}\label{HFtransf}
    \Phi(e^{2\pi i}z)=\s_g\cdot \Phi(z)~.
\end{equation}
In the untwisted case, $\s_g$ is an inner automorphism and solving this constraint provides the allowed triple $(T,k,b)$. A particular set of such singularities, known as those of regular semisimple type, are those for which $T$ is a regular semisimple element of $\jf$. The four-dimensional theories associated to a genus-zero UV curve with a single irregular puncture as above are denoted in the literature by $J^{(b)}[k]$ \cite{Wang:2016yha}. Alternatively, when such irregular punctures are paired with an additional regular maximal puncture carrying flavour symmetry $\jf$, the corresponding theories are denoted by $(J^{(b)}[k],F)$, where $F$ represents the presence of the extra regular puncture.

The associated vertex operator algebras for the $(J^{(b)}[k])$ and $(J^{(b)}[k],F)$ theories have been convincingly identified in \cite{Xie:2016evu,Xie:2017vaf,Xie:2017aqx}. They were proposed to be a rational $\mathcal{W}$-algebra in the former case and an affine Kac--Moody algebra in the latter case with level:
\begin{equation}
k_F = -h_\jf^\vee + \frac{b}{b+k}~,
\end{equation}
where $h_\jf^\vee$ is the dual Coxeter number of the simply laced Lie algebra $\jf$ and the pairs of numbers $(b,k)$ are those characterising the irregular puncture. The defining Lie algebra $\jf$ of the six-dimensional $(2,0)$ SCFTs has an ADE classification. Thus, compactifying the six-dimensional $(2,0)$ SCFT on a Riemann surface with untwisted irregular punctures and a regular puncture necessarily leads to four-dimensional $\cN = 2$ SCFTs with \emph{simply laced} flavour symmetries. However, four-dimensional $\cN = 2$ SCFTs with \emph{non-simply laced} flavour symmetry can be engineered from the six-dimensional $(2,0)$ SCFT by invoking outer automorphism twists of the Lie algebra $\jf$ as reviewed below.

\subsection{\label{subsec:twisted_irregular}Twisted irregular punctures}

In addition to codimension two defects, the six-dimensional $(2,0)$ SCFT also has (topological) codimension one defects corresponding to discrete symmetries associated to the outer-automorphism group of the ADE Lie algebra $\jf$. An irregular puncture on which such twist lines end is referred to as a twisted irregular puncture. Due to the outer automorphism twist, the constraint satisfied by the Higgs field $\Phi(z)$ of the Hitchin system on looping around the singularity is modified from \eqref{HFtransf} to: 
\begin{equation}
    \Phi(e^{2\pi i}z)=\s_g\cdot\s_o\cdot \Phi(z)~.
\end{equation}
Here $\s_o$ is the outer automorphism of the Lie algebra $\jf$ labelling the twist line, and $\s_g$ is an inner automorphism. 

As with the untwisted case, we restrict to the case where the coefficient $T$ appearing in the irregular singularity of the Higgs field is regular semisimple. The corresponding irregular singularity is then parametrised as follows,
\begin{equation}
\Phi(z)=\frac{T_k}{z^{2+\frac {{}_{\phantom{t}}k_{\phantom{t}}}{{}_{\phantom{t}}b_t}}}+\cdots~,
\end{equation}
where $b_t\in\Z_{>0}$ and $k\in\Z$ with $k>-b_t$. Looping around the puncture $z\to e^{2\pi i}z$ induces a monodromy action on the Higgs field,
\begin{equation}
\Phi\to \omega^k\Phi~,
\end{equation}
where $\w=e^{2 \pi i/b_t}$. This leads to a constraint relating the automorphism of $\jf$ and the parameters labelling the irregular singularity of the Higgs field: 
\begin{equation}\label{autP}
\tilde{\s}_gT_k \coloneqq (\s_g \s_o)T_k=\w^kT_k~.
\end{equation}
As $k$ and $b_t$ are not necessarily coprime, the order of the automorphism in \eqref{autP} is not necessarily $b_t$. 

It is worth mentioning briefly how the twisted irregular punctures of regular semisimple type are classified. Since $T_k$ is regular semisimple, on restricting to the commutant Cartan subalgebra $\hf$, the outer automorphism $\tilde{\s}_g$ induces an element in the twisted Weyl group $W_t(\hf) = W(\hf) \rtimes \mathrm{Out}(\jf)$. This element is also known as the corresponding twisted regular element. Each twisted regular element is uniquely determined by a positive integer $d_t$ such that its regular eigenvector in $\hf$ has eigenvalue $e^{2\pi i/d_t}$. Moreover, the order of the associated twisted regular element is given by $\mathrm{lcm}(d_t, |o|)$ (where $|o|$ is the cardinality of the outer automorphism group). 

Given a Lie algebra and an outer automorphism group, the positive integers $d_t$ have been classified \cite{Springer:1974}. To relate the parameter $d_t$ to the Hitchin field parameters, note that corresponding to the outer automorphism $\tilde{\s}_g$, the associated positive integer $d_t$ is given by
\begin{equation} \label{eq:dtbtrelation}
    d_t = \frac{b_t}{\gcd(b_t,k)}~.
\end{equation}
Clearly, $b_t$ is the maximum possible order for $\tilde{\s}_g$.\footnote{In the mathematics literature, this is referred to as the twisted Coxeter number, and has been determined in \cite{Springer:1974,Kac}.} Furthermore, the parameter $b_t$ in the Hitchin pole is taken to be equal to the order of the associated twisted regular element \ie, $b_t = \mathrm{lcm}(d_t, |o|)$ \cite{Wang:2018gvb}. The allowed values of $k$ are then determined by \eqref{eq:dtbtrelation} and the condition that $k > -b_t$. Consequently, the problem of classifying twisted irregular singularities of regular semisimple type can be translated to classifying all the allowed values for $d_t$.

\begin{table}[t]
\begin{adjustbox}{center,max width=.9\textwidth}
\renewcommand{\arraystretch}{1.3}
$\begin{array}{|c|c|c|}
\hline
\hline
\multicolumn{3}{|c|}{\text{\bf Twisted $D_{N+1}$ Irregular Punctures}}\\
\hline
\hline
d_t &\ \#\ {\rm of\ marginal\ couplings}\,\,\, &\ \#\ {\rm of\ mass\ parameters}\,\,\,\\
\hline
d_t=1 & N-1 & N\\
d_t\in 2\Z+1,\ d_t|N & N/d_t-1 & N/d_t\\
d_t\in 2\Z,\ 2N/d_t \in 2\Z+1 & 2N/d_t & 0\\
d_t\in 2\Z,\ 2N/d_t \in 2\Z & 2N/d_t-1 & 1\\
 \textcolor{red}{\ d_t>2\in 2\Z,\ (2N+2)/d_t \in 2\Z+1} & \textcolor{red}{(2N+2)/d_t-1} & \textcolor{red}{0}\\
\hline
\hline
\end{array}$
\caption{\label{tab:IPunDn}Properties of twisted $D_{N+1}$ irregular punctures. The ``class I'' theories' parameters correspond to $d_t = 2N+2$ in the highlighted entry above.}
\end{adjustbox}
\end{table}

\begin{table}[t]
\begin{adjustbox}{center,max width=.9\textwidth}
\renewcommand{\arraystretch}{1.3}
$\begin{array}{|c|c|c|}
\hline
\hline
\multicolumn{3}{|c|}{\text{\bf Twisted $A_{2N}$ Irregular Punctures}}\\
\hline
\hline
d_t &\ \#\ {\rm of\ marginal\ couplings}\,\,\, &\ \#\ {\rm of\ mass\ parameters}\,\,\,\\
\hline
d_t=1 & N-1 & N\\
d_t\in 2\Z+1,\ d_t|N & N/d_t-1 & N/d_t\\
d_t\in 4\Z,\ d_t|2N & 2N/d_t-1 & 2N/d_t\\
d_t\in 4\Z+2,\ d_t|(4N) & 4N/d_t-1 & 0\\
\textcolor{red}{\ d_t\in 4\Z+2,\ d_t|(4N+2)\ (d_t>2)}\, \, \, & \textcolor{red}{(4N+2)/d_t-1} & \textcolor{red}{0}\\
\hline
\hline
\end{array}$
\caption{\label{tab:IPunA2n}Properties of twisted $A_{2N}$ irregular punctures. The ``class I'' theories' parameters correspond to $d_t = 4N+2$ in the highlighted entry above.}
\end{adjustbox}
\end{table}

All the allowed values for $d_t$ were identified in \cite{Springer:1974} and used in \cite{Wang:2018gvb} to classify twisted irregular punctures of regular semisimple type. Since the theories of interest in this paper are the twisted $D_{N+1}$ and twisted $A_{2N}$ theories, we will mention all the allowed values of $d_t$, the number of exactly marginal couplings, and the number of extra mass parameters corresponding to irregular punctures in Tables \ref{tab:IPunDn} and \ref{tab:IPunA2n}. It is worth noting that the number of exactly marginal couplings and mass parameters is only dependent on $d_t$. We refer the reader to \cite{Wang:2016yha, Wang:2018gvb} for further details and for a complete classification of other twisted Argyres--Douglas SCFTs. 

Within the family of twisted $D_{N+1}$ and twisted $A_{2N}$ Argyres--Douglas SCFTs, we have restricted our attention to the ``class I'' theories, which are the subset of theories without exactly marginal couplings and without extra mass parameters from the irregular punctures. From Tables \ref{tab:IPunDn} and \ref{tab:IPunA2n}, one can read off the parameter $d_t$ for these theories to be $d_t = 2N+2$ and $d_t = 4N+2$ respectively, which is precisely the maximum order for the automorphism $\tilde{\s}_g$  \cite{Springer:1974}. Consequently, the irregular singularity of the Higgs field in the Hitchin system for the twisted $D_{N+1}$ and twisted $A_{2N}$ class $\cS$ theories is parametrised by \eqref{Hitchinparam}. These families of theories have flavour symmetry given by the Langlands dual of the invariant subalgebra under the $\mathbb{Z}_2$ automorphism twist, \ie, the flavour symmetry is given by $\gf = \spf(2N)$ in both cases, with the twisted $A_{2N}$ theory having a global Witten anomaly, whereas the twisted $D_{N+1}$ does not.

%% file: sections/appendix_b.tex

\section{\label{app:KM_admissible}Admissible affine Kac--Moody algebras}

Admissible level affine Kac--Moody algebras appear repeatedly as the associated VOAs of families of four-dimensional $\cN = 2$ SCFTs. In this appendix, we briefly summarise the associated varieties of admissible-level affine Kac--Moody algebras as determined by Arakawa in \cite{Arakawa:2010ni}. 

Let $\gf$ be a semisimple Lie algebra. The extended affine Kac--Moody algebra $\tilde{\gf}$ is defined as the central extension of the loop algebra $\gf[t,t^{-1}]$: 
\begin{equation}
    \tilde{\gf} = \gf[t,t^{-1}]\oplus\mathbb{C}K\oplus\mathbb{C} D~.
\end{equation}
Then given any $k \in \mathbb{C}$, we can define the following vertex algebra:
\begin{equation}
    V^k(\gf) = U(\hat \gf) \otimes_{U(\gf[t] \oplus \mathbb{C}K)} \mathbb{C}_k~,
\end{equation}
where $U(\hat \gf)$ is the universal enveloping algebra of the affine Kac--Moody algebra $\hat{\gf}$, $\gf[t]$ is the loop algebra of $\gf$, and $\mathbb{C}_k$ is the one-dimensional representation of $\gf[t] \oplus \mathbb{C}K$ on which $\gf[t]$ acts by 0 and $K$ acts as multiplication by the scalar $k$. The vertex algebra $V^k(\gf)$ is called the universal affine vertex algebra associated with $\gf$ at level $k$. $V^k(\gf)$ has a unique simple quotient, and we denote this simple quotient by $V_k(\gf)$.

A simple $V_k(\gf)$-module corresponding to the highest weight $\lambda$ of $\gf$ can also be regarded as an irreducible representation of $\tilde{\gf}$ with weight $\lambda + k \Lambda_0$. We denote such a $V_k(\gf)$-module by $M(\lambda)$. This simple module $M(\lambda)$ has conformal dimension given by: 
\begin{equation}
    h_{M(\lambda)} = \frac{(\lambda|\lambda + 2 \rho)}{2(k + h^\vee_\gf)}~,
\end{equation}
where $\rho$ is the Weyl vector, which is the half-sum of positive roots of $\gf$, and $(\cdot|\cdot)$ is the bilinear form defined on the root space of the Lie algebra $\gf$.

The level, $k$, is said to be \emph{principal admissible} if it satisfies:
\begin{align}
    k = - h^\vee_{\gf} + \frac{p}{q}~,\qquad p,q \in \mathbb{Z}_{\geqslant 1}~,\quad(p,q) = 1~,\quad (r^\vee, q) = 1~, \quad p \geqslant h^\vee_{\gf}~,
\end{align}
where $r^\vee$ is the lacing number of $\gf$ and $h_\gf^\vee$ the dual Coxeter number. The level $k$ is further called \emph{principal boundary admissible} if $p = h^\vee_\gf$. On the other hand, the level is said to be \emph{coprincipal admissible} if it satisfies:
\begin{align}
    k = - h^\vee + \frac{p}{r^\vee q}~,\qquad p,q \in \mathbb{Z}_{\geqslant 1}~,\quad(p,q) = 1~,\quad (r^\vee, q) = 1~, \quad p \geqslant h~,
\end{align}
with $h$ the Coxeter number of $\gf$. The vertex algebra $V_k(\gf)$ is called an admissible level affine Kac--Moody algebra if $k$ is either principal admissible or coprincipal admissible.

The associated varieties of admissible-level affine Kac--Moody algebras have been determined in \cite{Arakawa:2010ni}. Summarising, one has
\begin{equation}
    X_{V_k(\gf)} = \begin{cases}
        \overline{\mathcal{N}_q} &\qquad k \ \text{principal admissible}~, \\
        \overline{{}^L\mathcal{N}_q} &\qquad k \ \text{coprincipal admissible}~,
    \end{cases}
\end{equation}
where $\mathcal{N}_q$ and ${}^L\mathcal{N}_q$ are certain nilpotent orbits in $\gf^\ast$ determined by the denominator $q$,
\begin{align}
    \mathcal{N}_q &= \begin{cases}
        \mathds{O}_{\mathrm{prin}}~,&\qquad q \geqslant h^\vee_{\gf} \\
        \mathds{O}_{q}~,&\qquad q < h^\vee_{\gf}
    \end{cases}~, \\ 
    {}^L\mathcal{N}_q &= \begin{cases}
        \mathds{O}_{\mathrm{prin}}~,&\qquad q \geqslant h^\vee_{\gf^\vee} \\
        {}^L\mathds{O}_{q}~,&\qquad q < h^\vee_{\gf^\vee} 
    \end{cases}~,
\end{align}
where $\gf^\vee$ is the Langlands dual of $\gf$, and $\mathds{O}_{\mathrm{prin}}$ is the principal nilpotent orbit of $\gf$.

For $\gf = \sp(2N)$, the partitions corresponding to the nilpotent orbits $\mathds{O}_q$ and ${}^L\mathds{O}_q$ are as follows. When $q$ is even they agree, and are given by
\begin{equation}
    \mathds{O}_q = {}^L\mathds{O}_q = [q,q,\dots,q,s]~,\qquad 0 \leqslant s \leqslant q-1~,\quad s~\mathrm{even}~.
\end{equation}
When $q$ is odd, there are more possibilities depending, roughly, on whether $q$ fits into $2N$ and even or odd number of times:
\begin{align}\label{Oqnil}
    \mathds{O}_q = 
    \begin{cases}
        [\underbrace{q, q, \dots, q}_{\mathrm{even}}, s] &\quad 0\leqslant s \leqslant q-1~, \quad s \ \mathrm{even}~, \\
        [\underbrace{q, q, \dots, q}_{\mathrm{even}}, q-1, s] &\quad 2 \leqslant s \leqslant q-1~, \quad s \ \mathrm{even}~.
    \end{cases}
\end{align}
and 
\begin{align}\label{LOqnil}
    {}^L\mathds{O}_q = 
    \begin{cases}
        [q+1, \underbrace{q, q, \dots, q}_{\mathrm{even}}, s] &\quad 0\leqslant s \leqslant q-1~, \quad s \ \mathrm{even}~, \\
        [q+1, \underbrace{q, q, \dots, q}_{\mathrm{even}}, q-1, s] &\quad 2 \leqslant s \leqslant q-1~, \quad s \ \mathrm{even}~.
    \end{cases}
\end{align}

%% file: sections/appendix_c.tex

\section{\label{app:asymptotic}Asymptotic data}

In this appendix, we review some general results relating to the ``high-temperature'' $q\to1$ limit of the Schur index/VOA vacuum character, which have also been studied under the moniker ``asymptotic data'' in \cite{Arakawa:2021ogm}. A vertex operator algebra $V$ is said to admit an asymptotic datum if the $q\to1$ limit of its vacuum character behaves according to
\begin{equation}
    \chi_V(\tau) \xrightarrow[]{\tau \to 0} \mathbf{A}_V\cdot(-i\tau)^{\mathbf{w}_V} e^{\frac{i\pi}{12\tau}\mathbf{g}_V}~,
\end{equation}
for some constants $\mathbf{A}_V \in \mathbb{R}$, $\mathbf{w}_V \in \mathbb{R}$, and $\mathbf{g}_V \in \mathbb{R}$. If this is the case, then $\mathbf{A}_V$, $\mathbf{w}_V$, and $\mathbf{g}_V$ are called the \emph{asymptotic dimension}, the \emph{asymptotic weight}, and the \emph{asymptotic growth} of the vertex operator algebra $V$, respectively. In the same way, one defines an asymptotic datum ($\mathbf{A}_M$, $\mathbf{w}_M$, $\mathbf{g}_M$) for a $V$-module $M$ when the character of $M$ has the asymptotic behaviour
\begin{equation}
    \chi_M(\tau) \xrightarrow[]{\tau \to 0} \mathbf{A}_M\cdot(-i\tau)^{\mathbf{w}_M} e^{\frac{i\pi}{12\tau}\mathbf{g}_M}~.
\end{equation}
The leading high-temperature limit of the vacuum character of the VOA is then characterised by the asymptotic growth according to
\begin{equation}
    \lim_{\tau \to 0} \log(\chi_V(\tau)) \sim \frac{i \pi \mathbf{g}_V}{12 \tau} + \cdots~.
\end{equation}
For a four-dimensional $\cN = 2$ SCFT, the high-temperature limit of the Schur index is related to the Weyl anomaly coefficients according to
\begin{equation}
    \lim_{\tau \to 0} \log(\mathcal{I}_{\mathrm{Schur}}(q)) \sim \frac{4\pi i (c_{4d} - a_{4d})}{\tau} + \dots~.
\end{equation}
Thus, the asymptotic growth $\mathbf{g}_V$ of the associated VOA determines the $a_{4d}$ anomaly coefficient in terms of the Virasoro central charge.

The quantum dimension (if it exists) of an ordinary module $M$ over a vertex operator algebra $V$ is defined according to
\begin{equation}
    \mathrm{qdim}_{V}M = \lim_{\tau \to 0} \frac{\chi_M(\tau)}{\chi_V(\tau)}~.
\end{equation}
When the asymptotic weight and the asymptotic growth of $V$ and $M$ agree, \ie, $\mathbf{w}_V = \mathbf{w}_M$ and $\mathbf{g}_V = \mathbf{g}_M$, then the quantum dimension of $M$ exists and is given by,
\begin{equation}
    \mathrm{qdim}_VM = \frac{\mathbf{A}_M}{\mathbf{A}_V}~.
\end{equation}

The modular properties of characters of principal admissible affine Kac--Moody algebras and their admissible modules, including the asymptotic data, have been known since the work of Kac--Wakimoto \cite{Kac:1988qc, Kac:1989zz}. For the case of coprincipal admissible affine Kac--Moody algebras and for a review of the former as well, see \cite{Arakawa:2021ogm}. It will be useful for us to record some of these data here. 

For an admissible affine Kac--Moody algebra $V$ and any simple, ordinary module $M$ over it, one has the following. For $k$ \emph{principal admissible},
\begin{equation} \label{asymptdat1}
    \begin{split}
        \mathbf{w}_V &= \mathbf{w}_M = 0~,\\
        \mathbf{g}_V &= \mathbf{g}_M = \left(1 - \frac{h^\vee_{\gf}}{p q} \right) \dim\,\gf~.
    \end{split}
\end{equation}
For $k$ \emph{coprincipal admissible}, 
\begin{equation}\label{asymptdat2}
    \begin{split}
        \mathbf{w}_V &= \mathbf{w}_M = 0~, \\
        \mathbf{g}_V &= \mathbf{g}_M = \left(1 - \frac{r^\vee h^\vee_{\gf^\vee}}{p q} \right) \dim\,\gf~.
    \end{split}
\end{equation}

Asymptotic data for an affine Kac--Moody algebra $V$ or an ordinary $V$-module $M$ can also be used to extract information about the asymptotic data of the quantum Drinfel'd--Sokolov reduction of $V$ or $M$ \cite{Arakawa:2021ogm}. In particular, the reduction of any \emph{admissible} affine Kac--Moody algebra $V$ or any ordinary module $M$ also admits an asymptotic datum and the following hold:
\begin{itemize}
    \item $\mathbf{w}_{H^0_{DS,f}(M)} = \mathbf{w}_{H^0_{DS,f}(V)} = 0$.
    \item $\mathbf{g}_{H^0_{DS,f}(M)} = \mathbf{g}_{H^0_{DS,f}(V)}$.
    \item The quantum dimension of any module $M$ over the vertex algebra $V_k(\gf)$ remains unchanged under reduction with respect to any nilpotent orbit $\mathds{O}_{\rm nil} \subset X_{V_k(\gf)}$. In particular: 
    \begin{equation} \label{AsDatqDS}
        \mathrm{qdim}_{H^0_{DS,f}(V)} H^0_{DS,f}(M) = \mathrm{qdim}_{V}M~,
    \end{equation}
    with $f \in X_V$.
\end{itemize}

%% file: sections/appendix_d.tex

\section{\label{app:HBramifdetails}Quiver variety descriptions of \texorpdfstring{$\cM_H(\mathrm{AD}(\cf^{\rm anom}_{N},m))$}{M-Higgs(AD(c-anom(N),m))} and \texorpdfstring{$X_{V_{k^{\rm anom}_{N,m}}(\sp(2N))}$}{X(V(k-anom)(sp(2N)))}}

In this appendix, we present the details of the derivation of the quiver representations mentioned in Figures \ref{QuivGenAD} and \ref{QuivGenCN} for the Slodowy intersections that appear in Section \ref{sec:anomalHiggs}. These were crucial in deriving the main result \eqref{doublecoverHB} in the regime $N \geqslant m + 2$. 

\subsection{Relating partitions describing \texorpdfstring{$\cM_H(\mathrm{AD}(\cf^{\rm anom}_{N},m))$}{M-Higgs(AD(c-anom(N),m))} and \texorpdfstring{$X_{V_{k^{\rm anom}_{N,m}}(\sp(2N))}$}{X(V(k-anom)(sp(2N)))}}

Before finding the quivers corresponding to the Higgs branch of the $\mathrm{AD}(\cf^{\rm anom}_{N},m)$ theory and the associated variety $X_{V_{k^{\rm anom}_{N,m}}(\sp(2N))}$, it is relevant to first compare the possible partitions whose nilpotent orbit closures correspond to the associated varieties $X_{V_{k^{\rm anom}_{N,m}}(\sp(2N))}$ and $X_{V_{k_{N+m+1,m}}(\sp(2(N+m+1)))}$. More precisely, given the associated variety $X_{V_{k^{\rm anom}_{N,m}}(\sp(2N))}$ as the closure of some $\sp(2N)$ nilpotent orbit of the form \eqref{anomXv}, we will determine the nilpotent orbit closure of $\sp(2(N+m+1))$ of the form \eqref{nonanXv} that corresponds to the associated variety $X_{V_{k_{N+m+1,m}}(\sp(2(N+m+1)))}$.

Note that since we are in the $N \geqslant m+2$ regime, we always have: 
\begin{equation}
\begin{split}
    X_{V_{k^{\rm anom}_{N,m}}(\sp(2N))} &= \overline{{}^L\mathds{O}_q^{\sp(2N)}}~,\\
    X_{V_{k_{N+m+1,m}}(\sp(2(N+m+1)))} &= \overline{\mathds{O}_q^{\sp(2N+q-1)}}~,
\end{split}
\end{equation}
with the partitions corresponding to these nilpotent orbits given by \eqref{LOq} and \eqref{Oq}  respectively.\footnote{We remind the reader that for brevity, we are using the notation $q = 3+2m$.}

Let us first consider the associated variety $X_{V_{k^{\rm anom}_{N,m}}(\sp(2N))} = \overline{{}^L\mathds{O}_q^{\sp(2N)}}$ with the corresponding partition given by ${}^L\mathds{O}_q = [q+1, q^n, s]$ where $n$ and $s$ are even (this is one of the two possible partitions in \eqref{LOq}). Since the length of the partition is $2N$, we have:
\begin{equation} \label{constraint1}
    2N = (n+1)q + 1 + s~.
\end{equation}
We have two possible cases for partitions corresponding to the nilpotent orbit $\mathds{O}_q^{\sp(2N+q-1)}$ such that $X_{V_{k_{N+m+1,m}}(\sp(2(N+m+1)))} = \overline{\mathds{O}_q^{\sp(2N+q-1)}}$ as determined in \eqref{nonanXv}. We will show that only one of these is consistent with the constraint \eqref{constraint1}. 

\paragraph{Case 1:} $\mathds{O}_q = [q^{n'},q-1,s']$ where $n'$ and $s'$ are even. Since this corresponds to a partition of $\sp(2(N+m+1))$, its length must be $2N+q-1$. Thus, in this case, we have:
\begin{equation}
    2N + q-1 = (n'+1)q -1+ s'~.
\end{equation}
Consistency of this with the previous equation requires $(n+2)q + s = (n'+1)q + s'-1$. Since both $s,s'\leqslant q-1$, the only possible solution to this equation is $n' = n+1$ and $s' = s+1$. However, $n, n', s, s'$ are all necessarily even, implying that the two equations above are inconsistent. 

\paragraph{Case 2:} $\mathds{O}_q = [q^{n'},s']$, with again $n'$ and $s'$ even. In this case, we have:  
\begin{equation}
    2N + q-1 = n'q + s'~.
\end{equation}
Consistency of this equation with \eqref{constraint1} implies $n' = n+2$ and $s' = s$. 

This implies that when the associated variety $X_{V_{k^{\rm anom}_{N,m}}(\sp(2N))} = \overline{\mathds{O}}_{[q+1, q^n, s]}$, then the associated variety $X_{V_{k_{N+m+1,m}}(\sp(2(N+m+1)))} = \overline{\mathds{O}}_{[q^2, q^n, s]}$. A similar argument holds for the other possibility $X_{V_{k^{\rm anom}_{N,m}}(\sp(2N))} = \overline{\mathds{O}}_{[q+1, q^n, q-1, s]}$, and one reproduces \eqref{partitionmatch} establishing the following relation between the associated varieties, 
\begin{equation}\label{eq:app_partitionmatch}
\begin{alignedat}{3}
    &X_{V_{k^{\rm anom}_{N,m}}(\sp(2N))} = \overline{\mathds{O}}_{[q+1, q^n, s]} &&~~~\Longleftrightarrow~~~ &&X_{V_{k_{N+m+1,m}}(\sp(2(N+m+1)))} = \overline{\mathds{O}}_{[q^2, q^n, s]} \\
    &X_{V_{k^{\rm anom}_{N,m}}(\sp(2N))} = \overline{\mathds{O}}_{[q+1, q^n, q-1, s]} &&~~~\Longleftrightarrow~~~ &&X_{V_{k_{N+m+1,m}}(\sp(2(N+m+1)))} = \overline{\mathds{O}}_{[q^2, q^n, q-1, s]}~.
\end{alignedat}     
\end{equation} 

\subsection{Quivers for Slodowy intersections of \texorpdfstring{$\sp(2N)$}{sp(2N)}}

We now review some generalities about the quiver variety description of Slodowy intersections. The machinery holds generally for Slodowy intersections in $\gf = \af_N, \bff_N, \df_N$ Lie algebras as well, but we focus on $\gf = \cf_N$ which is relevant for derivations of the main text of the paper. We refer the reader to \cite{Hanany:2019tji} for further details.

Let $\rho$ and $\sigma$ be two partitions of $2N$ which label nilpotent orbits of $\sp(2N)$ such that the nilpotent orbit closure of $\sigma$ contains that of $\rho$. Consider the Slodowy intersection $\cS_{\sigma, \rho}=\overline{\mathds{O}}_{\sigma} \cap \cS_{\rho}$. The associated quiver encoding this Slodowy intersection as a symplectic singularity is given by a linear orthosymplectic quiver with alternating $C$ and $B/D$ gauge and flavour nodes. The quiver representation for this $\sp(2N)$ Slodowy intersection is denoted as $\cM_{C_N}(\sigma, \rho)$, \ie, 
\begin{equation}
    \cS_{\sigma, \rho} = \mathcal{H}[\cM_{C_N}(\sigma, \rho)]~.
\end{equation}
Thus, the associated varieties given by Slodowy intersections in $\gf = \cf_N$ are equivalently encoded in orthosymplectic quiver diagrams. To determine the precise quiver descriptions, note that,
\begin{equation}
    \cM_{C_N}(\sigma, \rho) \coloneqq \cM_{C_N}(\sigma, 0) \ominus \cM_{C_N}(\rho, 0)~, 
\end{equation}
where $\cM_{C_N}(\rho,0)$ (and similarly $\cM_{C_N}(\sigma,0)$) is defined by the linear chain of orthosymplectic $\sp(N_i)$ or $\so(N_i)$ gauge nodes, with decrements given by the transpose partition $\lambda \equiv \rho^T$. More precisely, with the transpose partition $\lambda = [\lambda_1, \lambda_2, \dots, \lambda_k]$ with $\lambda_i \geqslant \lambda_{i+1}$, the array describing the gauge nodes of the quiver $\mathbf{N} = \{N_1, N_2, \dots, N_k\}$ can be determined by $\lambda_i = N_{i-1} - N_{i}$, with $\sum^k_{i=1} \lambda_i = 2N$ and $N_0 \coloneqq 2N$. In addition to the linear chain of gauge nodes, the quiver also has a single flavour node $C_N$. 

The notation $\ominus$ denotes a quiver subtraction, which is defined as follows. Given any partition $\rho$ of $\sp(2N)$, we then have $\mathbf{N} = \{N_1, N_2, \dots, N_k\}$ describing the gauge nodes of the quiver $\cM_{C_N}(\rho, 0)$ and $\mathbf{N}_f = \{2N, 0, \dots, 0\}$ describing the flavour nodes of this quiver. Consequently, the quiver $\cM_{C_N}(\rho, 0)$ can be equivalently represented as $\cM_{C_N}(\mathbf{N},\mathbf{N}_f)$ emphasising the gauge and flavour node data $\{\mathbf{N}, \mathbf{N}_f\}$. The quiver subtraction operation is defined by, 
\begin{equation}
    \cM_{C_N}(\mathbf{N}_a,\mathbf{N}_{f}) \ominus \cM_{C_N}(\mathbf{N}_b,\mathbf{N}_{f}) = \cM_{C_N}(\mathbf{N}_a - \mathbf{N}_b,\mathbf{N}_{f_b})~,
\end{equation}
where $\mathbf{N}_f = \{2N, 0, \dots, 0\}$ for $\sp(2N)$ and 
\begin{equation} \label{Nfb}
    \mathbf{N}_{f_b} \coloneqq \mathbf{N}_f - \mathbf{A}\cdot\mathbf{N}_b~,
\end{equation}
where $\mathbf{A}$ is the Cartan matrix of $\su(k+1)$. Note that the Cartan matrix $\mathbf{A}$ only depends on the length, $k$, of the linear orthosymplectic quiver. Thus, given the partitions $\sigma$ and $\rho$ corresponding to some Slodowy intersection $\cS_{\sigma, \rho}$, one can associate to it a quiver $\cM_{C_N}(\mathbf{N}_a - \mathbf{N}_b,\mathbf{N}_{f_b})$, with the quiver data $\mathbf{N}_a$, $\mathbf{N}_b$, and $\mathbf{N}_{f_b}$ completely determined by the two partitions $\sigma$ and $\rho$ of $\sp(2N)$. We now use this general technology for computing quiver representations for $\sp(2N)$ Slodowy intersections to find the relevant quivers corresponding to the Higgs branch of the anomalous $\mathrm{AD}(\cf^{\rm anom}_{N},m)$ theory and the associated variety $X_{V_{k^{\rm anom}_{N,m}}(\sp(2N))}$.

\subsection{Derivation of the quiver descriptions}

Recall that, as established in Section \ref{sec:anomalVOA}, the Higgs branch of the anomalous $\mathrm{AD}(\cf^{\rm anom}_{N},m)$ theory is the Slodowy intersection, 
\begin{equation}
\begin{split}
     \cM_H(\mathrm{AD}(\cf^{\rm anom}_{N},m)) &\cong X_{V_{k_{N+m+1,m}}(\sp(2(N+m+1)))} \cap \cS_{[q-1,1^{2N}]} \\ 
     &\cong \overline{\mathds{O}}_q \cap \cS_{[q-1,1^{2N}]}~.
\end{split}     
\end{equation}
Hence, with $\mathds{O}_q = \mathds{O}_{[q^2, q^n, s]}$ where $2N + q-1 = (n+2)q + s$, \eqref{eq:app_partitionmatch} implies that, 
\begin{equation}
    X_{V_{k^{\rm anom}_{N,m}}(\sp(2N))} = \overline{\mathds{O}}_{[q+1, q^n, s]}~.
\end{equation}
Both of these affine varieties of interest are Slodowy intersections, and can be written in terms of orthosymplectic quivers $\cM_{C_N}(\sigma, \rho)$. In particular, we have,
\begin{equation} \label{case1HB}
\begin{split}
    \cM_H(\mathrm{AD}(\cf^{\rm anom}_{N},m)) &= \mathcal{H}[\cM_{C_{N+m+1}}(\sigma_1,\rho_1)] \\
    X_{V_{k^{\rm anom}_{N,m}}(\sp(2N))} &= \mathcal{H}[\cM_{C_{N}}(\sigma_2,\rho_2)]~,
\end{split}   
\end{equation}
with the partitions given by $\sigma_1 = [q^2,q^n,s]$, $\rho_1 = [q-1, 1^{2N}]$, and $\sigma_2 = [q+1,q^n,s]$, $\rho_2 = [1^{2N}]$. We note that the structure of the quivers and the arguments that follow are similar for the other possibility in \eqref{eq:app_partitionmatch} when the Higgs branch of the anomalous $\mathrm{AD}(\cf^{\rm anom}_{N},m)$ theory and the associated variety $X_{V_{k^{\rm anom}_{N,m}}(\sp(2N))}$ are instead given by,
\begin{equation}
\begin{split}
    \cM_H(\mathrm{AD}(\cf^{\rm anom}_{N},m)) &= \mathcal{H}[\cM_{C_{N+m+1}}(\sigma'_1,\rho'_1)] \\
    X_{V_{k^{\rm anom}_{N,m}}(\sp(2N))} &= \mathcal{H}[\cM_{C_{N}}(\sigma'_2,\rho'_2)]~,
\end{split}
\end{equation}
with $\sigma'_1 = [q^2,q^n,q-1,s]$, $\rho'_1 = [q-1, 1^{2N}]$, and $\sigma'_2 = [q+1,q^n,q-1,s]$, $\rho'_2 = [1^{2N}]$. Thus, we will only present the details of the computation of the quiver representation for the first case above. To this end, we now turn to finding the quivers $\cM_{C_{N+m+1}}(\sigma_1,\rho_1)$ and $\cM_{C_{N}}(\sigma_2,\rho_2)$.

As reviewed earlier, to find the quiver descriptions of Slodowy intersections, one first needs to determine the arrays $\mathbf{N}$ and $\mathbf{N}_f$ representing the gauge and flavour nodes of the quivers, respectively. From \eqref{case1HB}, we have the partitions corresponding to the Slodowy intersection $\cS_{\sigma_1, \rho_1}$ given by $\sigma_1 = [q^2,q^n,s]$ and $\rho_1 = [q-1, 1^{2N}]$. The transposes of these partitions, $\sigma_1^T$ and $\rho_1^T$, are given by
\begin{equation}
    \sigma_1^T = [(n+3)^s, (n+2)^{q-s}] \qquad \mathrm{and} \qquad \rho_1^T = [2N+1, 1^{q-2}]~.
\end{equation}
Since $\cS_{\sigma_1, \rho_1}$ is a Slodowy intersection in $\sp(2(N+m+1))$, the length of the partitions is $2(N+m+1)$ and we have $N_0 = 2N+q-1$. Consequently, the gauge nodes of the linear orthosymplectic quiver $\cM_{C_{N+m+1}}(\sigma_1,0)$ encoding this Slodowy intersection are given by $\mathbf{N}_a = \{N^a_1, N^a_2, \dots, N^a_{q-1}\}$ with
\begin{equation}
    N^a_i = 
    \begin{cases}
        2N + q-1 - i(n+3) & \quad 1\leqslant i\leqslant s~, \\
        2N + q-1 -i(n+2) - s & \quad s+1 \leqslant i \leqslant q-1~.
    \end{cases}
\end{equation}
Similarly, the gauge nodes of the linear orthosymplectic quiver $\cM_{C_{N+m+1}}(\rho_1,0)$ are given by $\mathbf{N}_b = \{N^b_1, N^b_2, \dots, N^b_{q-1}\}$ with
\begin{equation}
    N^b_i = q - 1 - i \qquad 1 \leqslant i \leqslant q-1~.
\end{equation}
From this, we can read off $\mathbf{N} \coloneqq \mathbf{N}_a - \mathbf{N}_b$ to be, 
\begin{equation}
    N_i \coloneqq N^a_i - N^b_i = 
    \begin{cases}
        2N -i(n+2) & \quad 1\leqslant i\leqslant s~, \\
        2N -i(n+1) - s & \quad  s+1 \leqslant i \leqslant q-1~.
    \end{cases}
\end{equation}
These correspond to the $q-1$ alternating $C$ and $B/D$ gauge nodes in the linear orthosymplectic quiver $\cM_{C_{N+m+1}}(\sigma_1,\rho_1)$ describing the Higgs branch of the anomalous $\mathrm{AD}(\cf^{\rm anom}_{N},m)$ theory. Using $\mathbf{N}_f = \{2N+q-1, 0, 0, \dots, 0\}$ and taking $\mathbf{A}$ to be the Cartan matrix of $\su(q)$, we also find the flavour nodes $\mathbf{N}_{f_b}$ by quiver subtraction to be given by
\begin{equation}
    \mathbf{N}_{f_b} = \{\underbrace{2N, 0, 0, \dots, \textcolor{red}{1}}_{q-1}\}~.
\end{equation}
This completes the derivation of the quiver representation of the Higgs branch of the $\mathrm{AD}(\cf_N^{\rm anom},m)$ theory, leading to Figure \ref{QuivGenAD}.

We now proceed to determining the quiver representation corresponding to the associated variety $X_{V_{k^{\rm anom}_{N,m}}(\sp(2N))}$. From \eqref{case1HB}, the partitions corresponding to the associated variety $X_{V_{k^{\rm anom}_{N,m}}(\sp(2N))}$ are given by $\sigma_2 = [q+1, q^n, s]$ and $\rho_2 = [1^{2N}]$. The transposes of these partitions, $\sigma_2^T$ and $\rho_2^T$, are given by
\begin{equation}
    \sigma_2^T = [(n+2)^s, (n+1)^{q-s}, 1] \qquad \mathrm{and} \qquad \rho_2^T = [2N]~.
\end{equation}
Since these partitions label nilpotent orbits of $\sp(2N)$, we have $N_0 = 2N$. The gauge nodes of the linear orthosymplectic quiver $\cM_{C_{N}}(\sigma_2,0)$ are given by $\mathbf{N}_a = \{N^a_1, N^a_2, \dots, N^a_{q}\}$ with
\begin{equation}
    N^a_i = 
    \begin{cases}
        2N - i(n+2) &\quad 1\leqslant i\leqslant s~, \\
        2N -i(n+1) - s &\quad s+1 \leqslant i \leqslant q~,
    \end{cases}
\end{equation}
and similarly, the gauge nodes of the linear orthosymplectic quiver $\cM_{C_{N}}(\rho_2,0)$ are given by $\mathbf{N}_b = \{N^b_1, N^b_2, \dots, N^b_{q}\}$ with
\begin{equation}
    N^b_i = 0~.
\end{equation}
We again use quiver subtraction to find the gauge nodes $\mathbf{N} \coloneqq \mathbf{N}_a - \mathbf{N}_b$, 
\begin{equation}
    N_i \coloneqq N^a_i - N^b_i = 
    \begin{cases}
        2N -i(n+2) &\quad 1\leqslant i\leqslant s~, \\
        2N -i(n+1) - s &\quad s+1 \leqslant i \leqslant q-1~, \\
        \textcolor{red}{1} &\quad \textcolor{red}{i = q}~.
    \end{cases}
\end{equation}
Hence, in this case, we have $q$ gauge nodes in the linear orthosymplectic quiver $\cM_{C_{N}}(\sigma_2,\rho_2)$ representing the associated variety $X_{V_{k^{\rm anom}_{N,m}}(\sp(2N))}$. Moreover, from $\mathbf{N}_f = \{2N, 0, 0, \dots, 0\}$ and $\mathbf{N}_b = \mathbf{0}$, we find the flavour nodes $\mathbf{N}_{f_b}$ of the quiver $\cM_{C_{N}}(\sigma_2,\rho_2)$ to be
\begin{equation}
    \mathbf{N}_{f_b} = \{\underbrace{2N, 0, 0, \dots, 0}_{q}\}~.
\end{equation}
This completes the derivation of the quiver representation of $X_{V_{k^{\rm anom}_{N,m}}(\sp(2N))}$.